\documentclass[aps,pra,twocolumn,superscriptaddress,10pt,showpacs]{revtex4}
\usepackage[dvips]{graphics}
\usepackage{graphicx}
\usepackage{amsfonts}
\usepackage{amssymb}
\usepackage{amsmath}
\usepackage{subfigure}

\begin{document}

\title{Solitons and vortices in honeycomb defocusing photonic lattices}
\author{K.\ J.\ H.\ Law}
\affiliation{Department of Mathematics and Statistics, University of Massachusetts,
Amherst MA 01003-4515, USA}
\author{H.\ Susanto}
\affiliation{
School of Mathematical Sciences, University of Nottingham, University Park, Nottingham, NG7 2RD, UK
}
\author{P.\ G.\ Kevrekidis}
\affiliation{Department of Mathematics and Statistics, University of Massachusetts,
Amherst MA 01003-4515, USA}

\begin{abstract}
Solitons and necklaces in the first band-gap of a two-dimensional optically
induced honeycomb defocusing photonic lattice are theoretically considered.
It is shown that dipoles, soliton necklaces, and vortex necklaces
exist and may possess regions of stable propagation through a
photorefractive crystal.
Most of the configurations disappear in bifurcations
close to the upper edge of the first band. Solutions associated with
such bifurcations are also
numerically examined, and it is found that they are
often asymmetric and more exotic. The dynamics of the relevant unstable
structures are also examined through direct numerical simulations
revealing either breathing oscillations or, in some cases, destruction
of the original waveform.
\end{abstract}

\maketitle

\section{Introduction}
\label{intro}

In the past few years, there has been a considerable growth of
interest in the examination of the
self-trapping of light in photonic lattices optically induced in
nonlinear photorefractive crystals,
such as
strontium barium niobate (SBN).
This can be attributed to a considerable extent to the fact that
the theoretical inception \cite{efrem} of the relevant phenomena
was rapidly followed by
the experimental realization \cite{moti1,neshevol03,martinprl04},
revealing a considerable wealth of new possibilities.
This setting naturally permits the consideration of the competition
between  effects of nonlinearity
and those of diffraction, therefore enabling the
examination of effects of periodic ``potentials''
on solitary waves. In this context the role of the effective potential
is played by the ordinary polarization of light forming a waveguide array in
which the nonlinear, extra-ordinarily polarized probe beam evolves.

Numerous nonlinear waves and coherent structures have been elucidated
and experimentally realized in this context. In particular,
discrete dipole \cite{yang04}, necklace \cite{neck} solitons and
even stripe patterns \cite{multi}, rotary solitons \cite{rings},
discrete vortices \cite{vortex} or the realization of photonic
quasicrystals \cite{moti22} and Anderson localization \cite{moti23}
are among the recently reported experimental results in the field.
These efforts illustrate the potential that this setting holds for
the examination of localized structures
that may be usable as carriers and conduits for data transmission
and processing in all-optical communication schemes. In parallel
to this more practical aspect, this framework remains
an experimentally tunable playground where
numerous fundamental issues of solitons and nonlinear waves
can be explored.

The above mentioned
interplay of nonlinearity with periodicity is important
not only in the physics of optically induced lattices in photorefractive
crystals, but also in a variety of other contexts in optical and
atomic physics. These involve e.g. on the optical end,
the numerous developments on the experimental and theoretical
investigation of optical waveguide arrays; see e.g.
\cite{review_opt,general_review} for relevant
reviews. In the case of atomic physics, and particularly of
Bose-Einstein condensates, the confinement of
dilute alkali vapors in optical lattice potentials \cite{konotop}
has  offered a similarly far-reaching opportunity to examine many fundamental
phenomena involving (effective) nonlinearity and spatial periodicity.
These include, but are not limited to
modulational instabilities, Bloch
oscillations, Landau-Zener tunneling and gap solitons among others;
see \cite{markus2} for a recent review.

Our present study, motivated by optically induced
lattices in photorefractive SBN crystals, focuses on two-dimensional 
periodic, nonlinear media
with a {\it non-square} lattice. While most of the above studies
have been dedicated to square lattices, only a few have tackled
the coherent structures possible in non-square settings; see
e.g., as relevant examples \cite{yurig,ablo,seg,ol2007a,ol2007b,moti07} and
references therein.
Furthermore, the vast majority of the above-mentioned
studies has centered around
focusing nonlinearities.
At least partly, this is due to technical limitations,
as it is easier to work with
voltages that are in the regime of focusing rather than in that of
the defocusing nonlinearity (in the latter case, sufficiently large
voltage, which is tantamount to large nonlinearity,
may  actually change the sign of the nonlinearity by
inverting the orientation of the permanent polarization of the crystal).
As a result,
coherent structures in the defocusing regime, have only rather
sparsely been examined.  Such an experimental example
is the fundamental and higher order gap solitons excited in the 
vicinity of the edge
of the first Brillouin zone \cite{moti1,moti07}. More complex gap structures
(multipoles and vortices)
are only now starting to be explored in square
lattices \cite{ourol}. In parallel to these experimental
developments, a theoretical framework is starting to emerge
to address such multipole and vortex structures in
square lattices with cubic nonlinearities \cite{ourpre,pgk_dnls},
whose qualitative predictions can however be extended to non-square
settings and the main ones among which will also be compared to
the results presented below.
Our main focus in the present work is on employing
a continuum model to examine the waveforms present in a context
involving a {\it triangular lattice} ({\it honeycomb }) potential and
a {\it saturable defocusing} nonlinearity associated
with appropriate optically induced lattices in SBN crystals.
In particular,
we study in detail
multipole (dipole and hexapole) solitons in such lattices
induced with a self-defocusing nonlinearity.

We numerically
analyze both the existence and the stability of these
structures and follow their dynamics, in the cases where we find them
to be unstable. We also qualitatively
compare our findings with the roadmap provided by the {\it discrete} model
\cite{ourpre}.

Our presentation is structured as follows. In section II,
we present our theoretical model setup.
Dipole solutions with the two excited sites in adjacent wells of the periodic
potential (nearest-neighbor dipoles) are studied in section III.
Subsequently, we do the same for next-nearest-neighbor dipoles
(excited in two diagonal sites, separated by one lattice site)
in section IV and opposite dipoles on either end of the
hexagonal configuration (ie. the excited sites are
separated by two empty wells) in section V. Section VI addresses
the case of more complex structures such as
hexapoles (all six sites from one period of the potential) and
vortices. Finally, in
section VII, we summarize our findings, posing some interesting
questions for future study.

\section{Setup}

\begin{figure}[tbh]
\begin{center}
\includegraphics[width=0.4\textwidth]{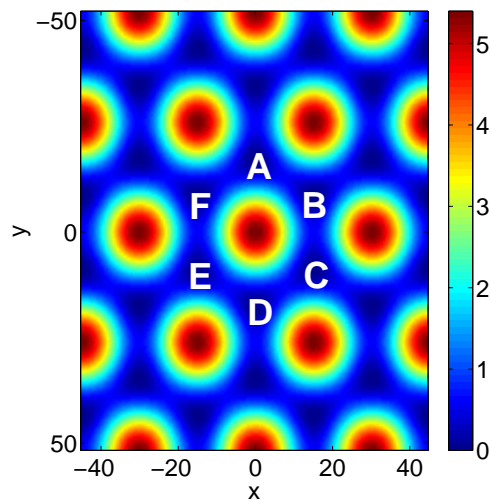}
\end{center}
\caption{(Color online) A spatial (x-y) contour plot of the
ordinary polarization standing
wave [lattice beam in Eq.\ (\ref{eq2})]. In this context,
the light intensity maxima correspond to
the minima of the resulting refractive index lattice  (i.e., honeycomb
lattice), as opposed to the
focusing nonlinearity
lattice field, where they correspond to the maxima (i.e., triangular lattice).
Points $A,\, B,\, C,\, D,\, E,\,$ and $F$
are used for naming the various configurations.
 $A$ is a ``nearest-neighbor''
minimum of $B$ and $F$, a ``next-nearest-neighbor''
of $C$ and $E$, and an ``opposite'' of $D$
(with respect to the local maximum of the lattice).
Because of the symmetry of the setup, this is a
complete characterization of dipole
configurations.  We will refer to the configurations
with the names given above.}
\label{lattice}
\end{figure}

We use the standard partial differential equation
for the amplitude of the electric field $U$ \cite{ol2007a,ol2007b,yang04_3,yang04_4},
in the following form:
\begin{center}
\begin{equation}
-iU_z=[L+N(\textbf{x},|U|^2)]U,\label{eq1}
\end{equation}
\begin{equation}
N(\textbf{x},|U|^2)=\frac{E_0}{1+I(\textbf{x})+|U|^2},
\label{eq1_1}
\end{equation}
\end{center}
where $L=D \nabla^2$ and $\nabla^2$ is the two-dimensional Laplacian,
$U$ is the slowly varying amplitude of the probe beam, and
\begin{equation}
I(\textbf{x})=I_0\left|e^{ik\textbf{b}_1 \textbf{x}}+{e}^{ik\textbf{b}_2\textbf{x}}+{e}^{ik\textbf{b}_3\textbf{x}}\right|^2
\label{eq2}
\end{equation}
is the optical lattice intensity function formed by three laser beams with
$\textbf{b}_1=(1,0)$,
$\textbf{b}_2=(-\frac{1}{2},-\frac{\sqrt{3}}{2})$, and
$\textbf{b}_3=(-\frac{1}{2},\frac{\sqrt{3}}{2})$.
Here $I_0$ is the lattice peak intensity,
$z$ is the propagation distance and
$\textbf{x}=(x, y)$ are transverse distances
(normalized to $z_s=1$ mm and $x_s=y_s=1 \mu$m),
$E_0$ is proportional to the applied DC field voltage,
$D=z_s \lambda/(4\pi n_e x_s y_s)$ is the diffraction coefficient,
$\lambda$ is the wavelength of the laser in a vacuum,
$d$ is the period in the x direction with $k=4\pi/(3d)$
(period in the $y$ direction is $\sqrt{3}d$), and
$n_e$ is the refractive index along the extraordinary axis.
We choose the lattice intensity $I_0 = 0.6$.
A plot of the optical lattice is shown in Fig.\ \ref{lattice} for
illustrative purposes regarding the location where our localized pulses
will be ``inserted''. In addition, we choose other physical parameters
consistently with a typical experimentally accessible setting
\cite{ol2007a,ourol} as

\[
d= 30  \mu\textrm{m},\quad \lambda = 532\ \textrm{nm},\quad n_e = 2.35,\quad E_0 = 8.
\]

The non-dimensional value $D=18.01$,
and we note that this dispersion coefficient is equivalent to
rescaling space by a factor $\sqrt{D}$ as e.g. in \cite{our_sq}.

The numerical simulations are performed in a rectangular
domain corresponding to the
periodicity of the lattice using a rectangular spatial mesh with
$\Delta x \approx 0.75$ and $\Delta y \approx 0.86$
and domain size $ 4 d \times 3 \sqrt{3} d $,
i.e.\ $160\times180$ grid points.
See Fig.\ \ref{lattice} for a schematic of the spatial configurations.

Regarding the typical dynamics of a soliton when it is unstable,
we simulate the z-dependent evolution
using a Runge-Kutta fourth-order scheme with a step $\Delta z=0.01$.

Assuming a stationary state $u(x,y)$ exists, and
letting the propagation constant $\mu$
represent the (nonlinear)
real eigenvalue of the operator of the right-hand-side
of Eq. (\ref{eq1}),
then the corresponding eigenvector $u(x,y)$ is a fixed point of

\begin{equation}
[\mu-L-N(\textbf{x},|u|^2)]u=0.
\label{sta_eq}
\end{equation}

The localized states $u$ of (\ref{sta_eq}) were obtained using the
Newton-GMRES fixed point solver nsoli from \cite{kell03} and
a pseudo arc-length continuation \cite{doedel} was used to follow
each branch and locate the bifurcations which occur at the edge of
the first band.
Since the parameter of interest is $\mu$, diagnostics are
plotted against $\mu$.

We restrict $\mu$ to those values within the first spectral gap of
the linear eigenvalue problem,

\begin{equation}
[\mu-L-N(\textbf{x},0)]u=0.
\label{lin}
\end{equation}

Values of the propagation constant $\mu$ within this forbidden gap in the
spectrum of the linearized problem will
correspond to exponentially localized in space, so-called gap-soliton,
states of the original nonlinear partial differential equation.
Using a standard eigenvalue solver package
implemented through MATLAB, we identify
the spectral gap for our given parameters and gridsize to be
$3.62\lesssim\mu\lesssim4.94$.

The square root of the optical power (or, mathematically, the $L^2$ norm) of
the  localized waves is defined as follows:
\begin{equation}
P = \left[\int_{-\infty}^{\infty}\int_{-\infty}^{\infty}|U|^2\,dx\,dy\right]^{1/2}.
\end{equation}

Introducing a linearization around an exact
stationary solution $u$, and expanding the leading order
perturbation into a eigenfunctions and eigenvalues,
we obtain the following Bogoliubov system
\begin{eqnarray}
&&\left[i\lambda+\mu-L-\frac{\partial (N U)}{\partial U}|_u\right] \tilde{u}
- \frac{\partial (N U)}{\partial U^*}|_u \tilde{u}^* = 0,
\nonumber\\
&&\left[i\lambda-\mu+L+\frac{\partial (N U)^*}{\partial U^*}|_u\right] \tilde{u}^*
+\frac{\partial (N U)^*}{\partial U}|_u \tilde{u} = 0.
\label{lin_eq}
\end{eqnarray}
We solve the above linear eigenvalue problem using
MATLAB's standard eigenvalue solver package.
The symplectic nature of the resulting eigenvalue equations
guarantees that the relevant eigenvalues should come in quartets,
hence an instability is present whenever the solution of the above
linearization problem of Eqs. (\ref{lin_eq}) possesses an eigenvalue with a
non-zero real part.


We now briefly discuss the principal stability conclusions, for
the defocusing case of \cite{ourpre}, which we should expect
to still be valid in the present configuration.
Nearest neighbor excitations in the defocusing case correspond
to nearest neighbor excitations in the focusing case, but
with an additional $\pi$ phase in the relative phase of the
sites added by the so-called staggering transformation \cite{ourpre}.
Therefore, the in-phase nearest neighbor configuration in the defocusing
case corresponds to an out-of-phase such configuration in the focusing case
(and should thus be stable) \cite{pgk_dnls}. On the other hand,
next nearest neighbor out-of-phase defocusing configurations would correspond
to next nearest neighbor out-of-phase focusing configurations and
should also be stable (at least in some parameter regimes). By the
same token, out-of-phase nearest neighbor, and in-phase next nearest neighbor
structures should be unstable. These considerations also indicate
that in-phase opposite dipoles should be stable, while out-of-phase
such dipoles should always be unstable. Finally, vortex-like structures
and in-phase hexapoles should be stable as well.
Notice, however, that as discussed in \cite{ourpre} the multipole structures
characterized as potentially stable above will, in fact, typically possess
imaginary eigenvalues of negative Krein signature (see e.g. \cite{kks}
and references therein). These may lead to oscillatory instabilities
through complex quartets of eigenvalues. These arise by means
of Hamiltonian-Hopf
bifurcations \cite{vdm} emerging from collisions with eigenvalues
of opposite (i.e., positive) Krein signature. These conclusions
will be discussed in connections with our detailed numerical results
in what follows.



\section{Nearest Neighbor Dipole Solitons}

In this section, we report dipole solitons where the two lobes of the
wave are located in two nearest neighbor (N)
lattice sites in the 2D triangular
potential shown in Fig.\ \ref{lattice}. The lobes can have the
same phase or $\pi$ phase difference so we define them as
in-phase (IP) dipoles and out-of-phase (OP) dipoles, respectively.

\subsection{In-Phase Nearest Neighbor Dipole Solitons}

\begin{figure}[tbp!]
\includegraphics[width=0.23\textwidth]{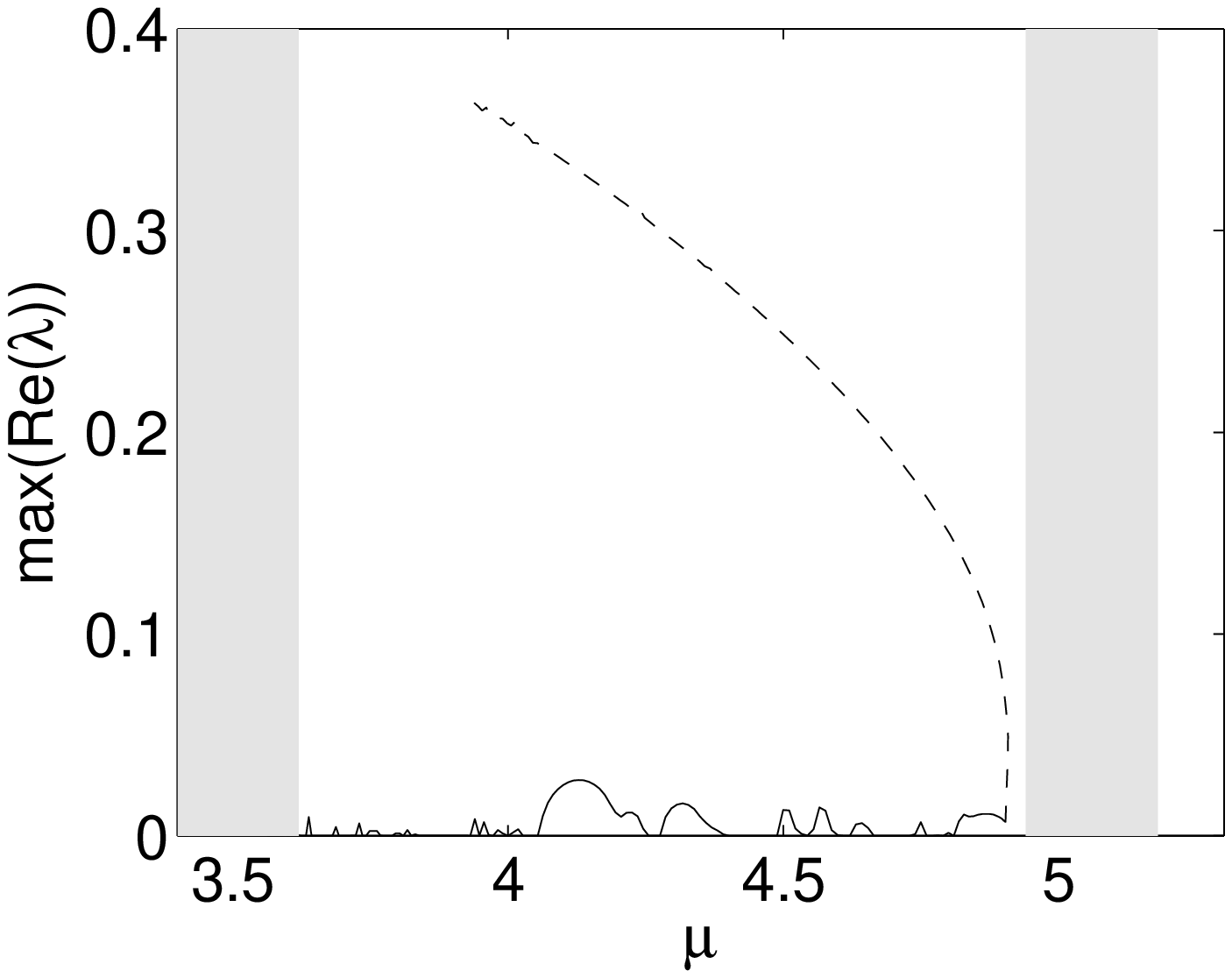}
\includegraphics[width=0.23\textwidth]{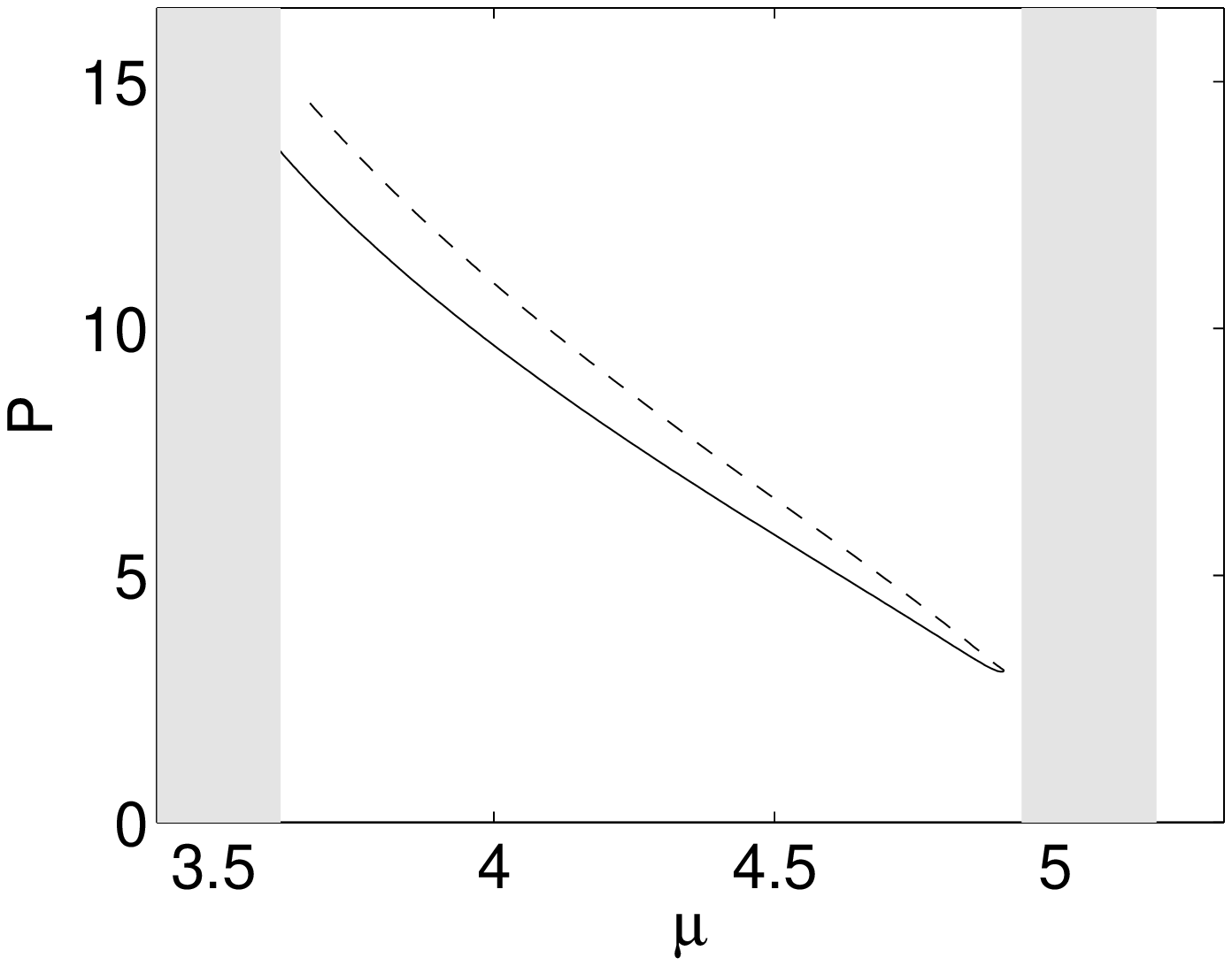}\\
\includegraphics[width=0.23\textwidth]{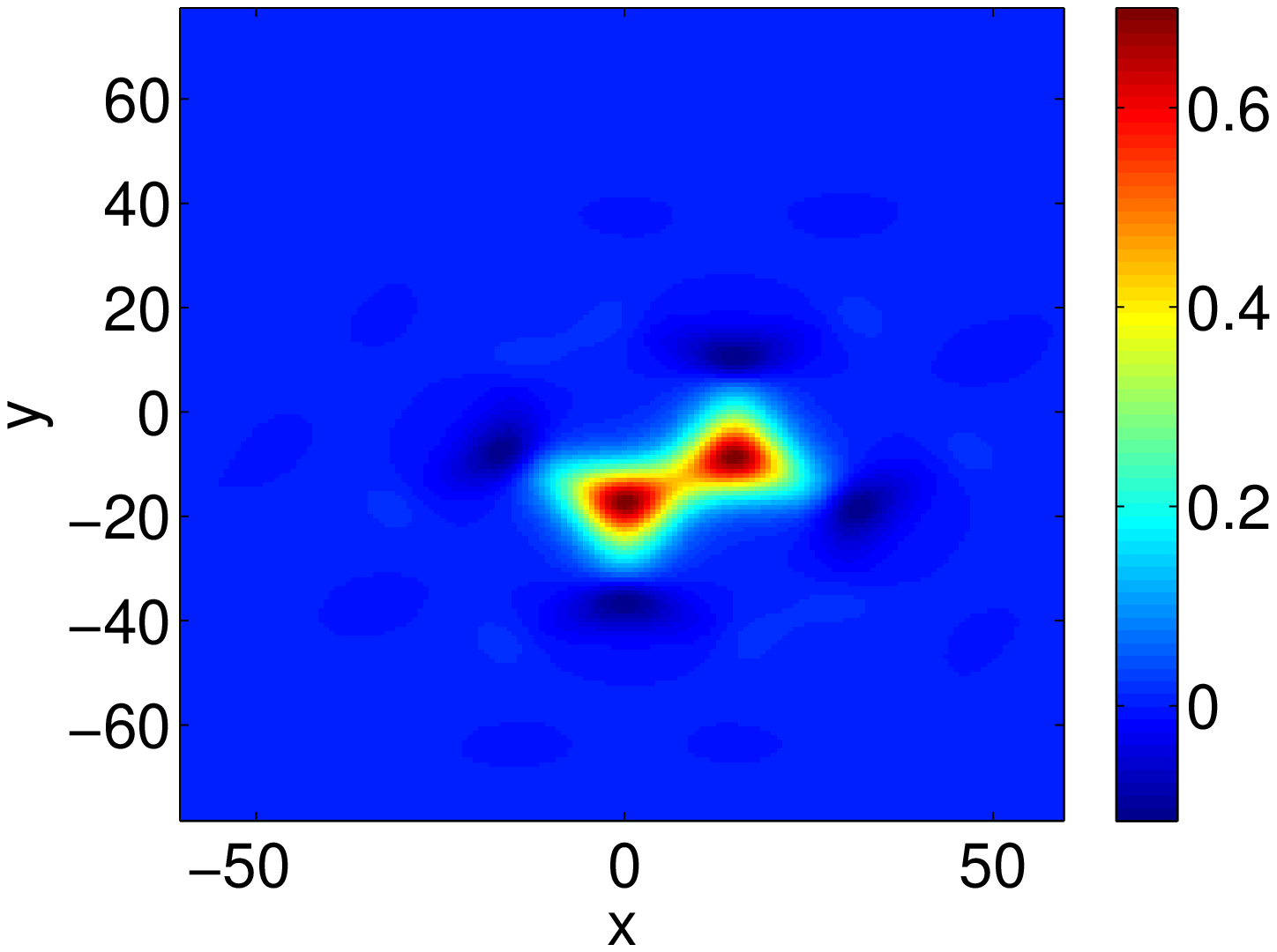}
\includegraphics[width=0.23\textwidth]{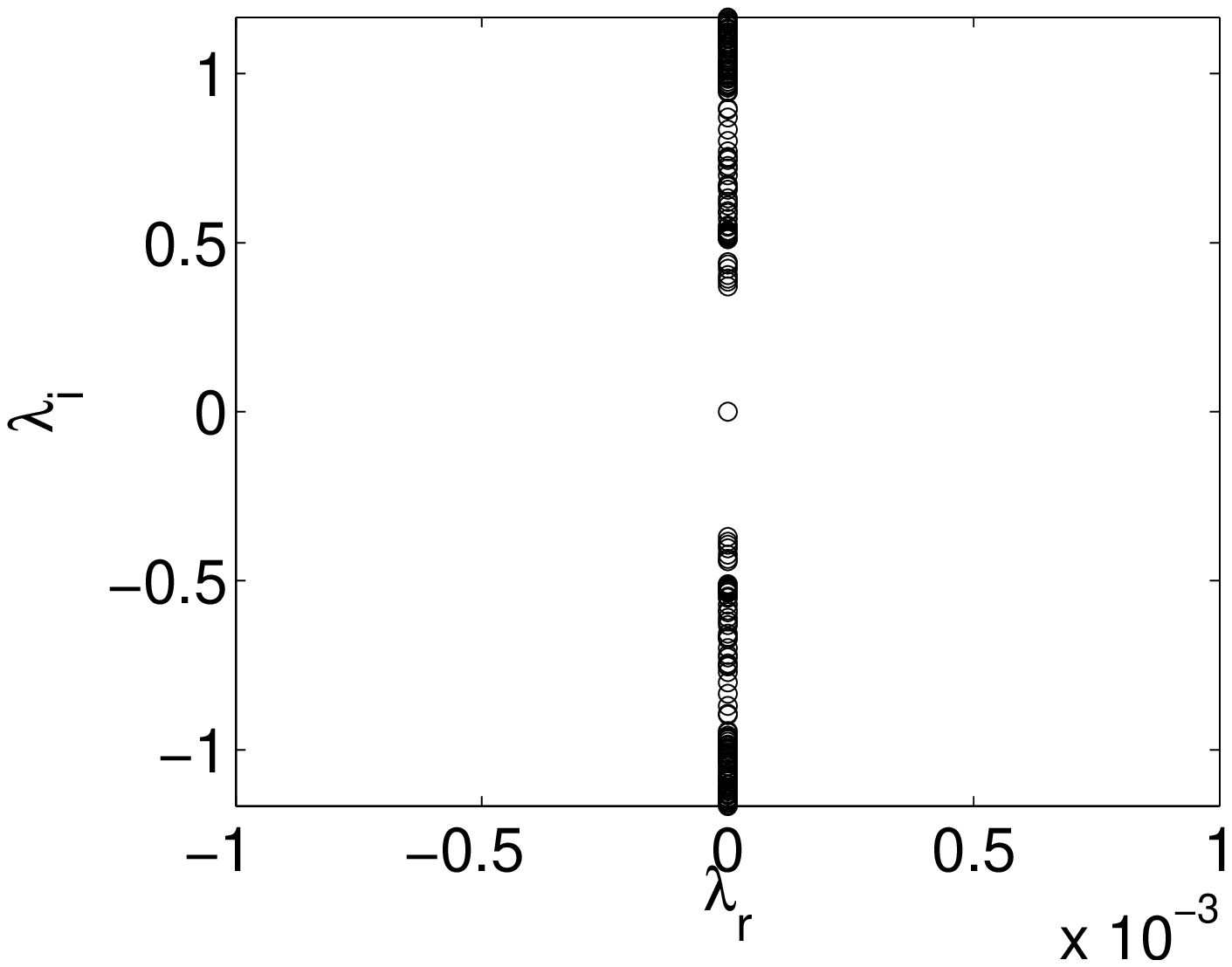}\\
\includegraphics[width=0.23\textwidth]{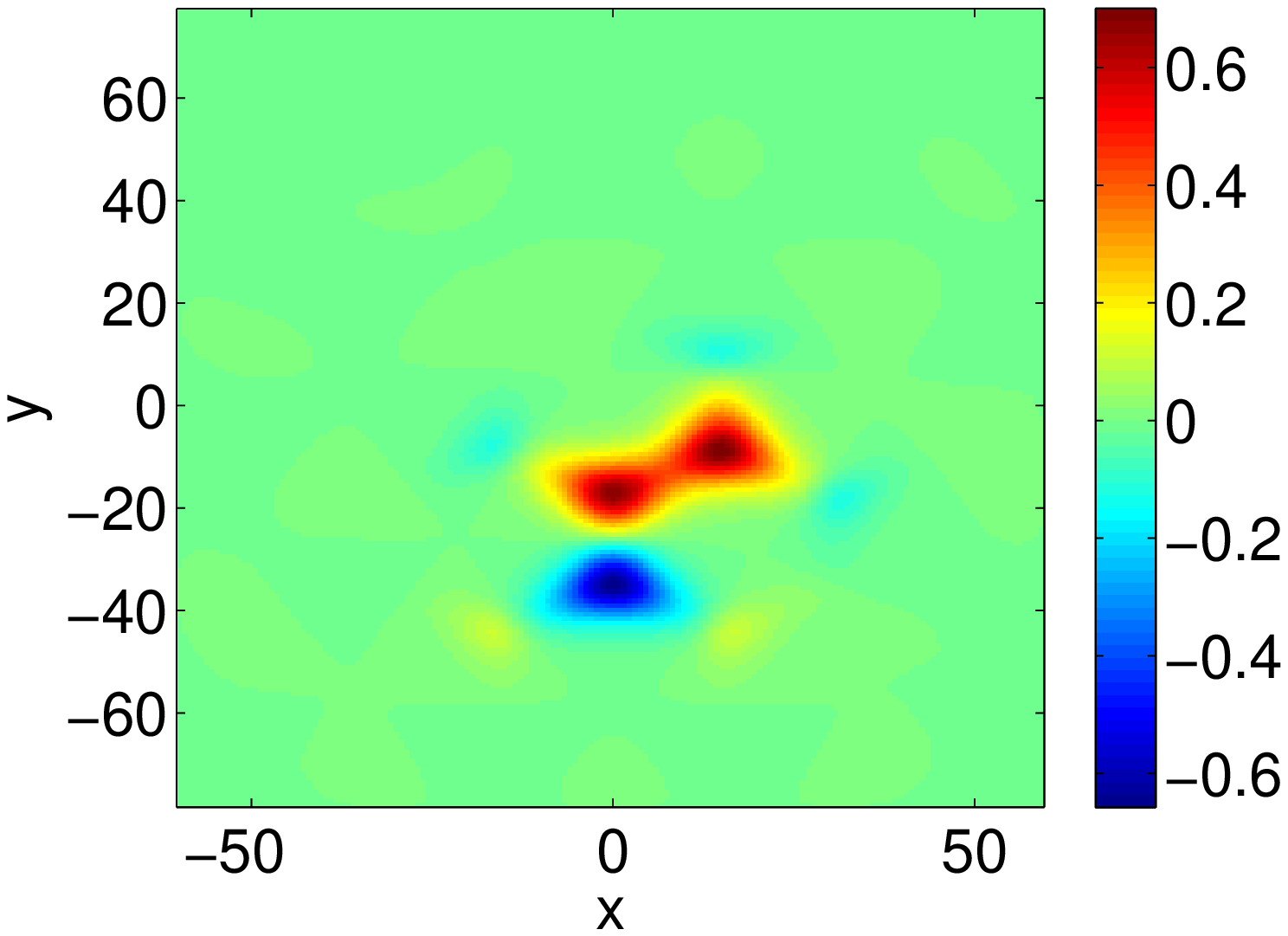}
\includegraphics[width=0.23\textwidth]{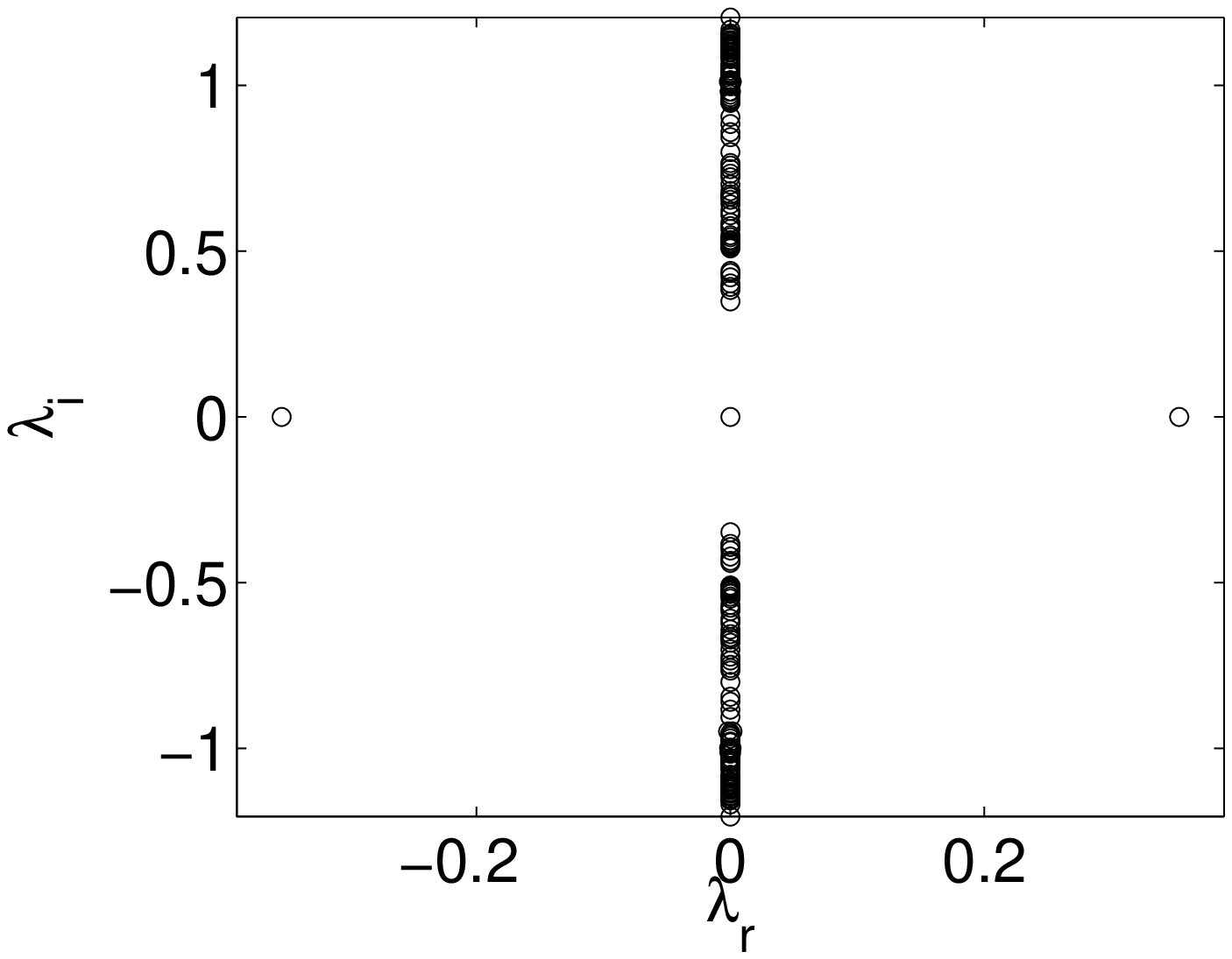}
\caption{(Color online) The top left panel shows the stability of
the dipoles against the propagation constant $\mu$. It is stable
when the spectra is
purely imaginary (i.e., when
$\max({\rm Re}(\lambda))=0$).
The top right panel depicts the power of the dipoles against the
propagation constant. In each of these images the solution branch
is denoted by a solid line. The branch with which the dipole collides
and terminates in a saddle-node bifurcation is shown by a dashed line.
The shaded areas in both of these panels represent the bands
of linear spectrum (\ref{lin}).
The middle left and right panels show the profile
$u$ of the dipole at $\mu=4$ and the corresponding complex spectral
plane $({\rm Re}(\lambda),{\rm Im}(\lambda))$ of
$\lambda={\rm Re}(\lambda) + i {\rm Im}(\lambda)$.
Finally, the bottom panels show the same features for the unstable
saddle solution corresponding to the dashed line.}
\label{IPN}
\end{figure}

\begin{figure}[tbp!]
\begin{center}
\includegraphics[width=0.4\textwidth]{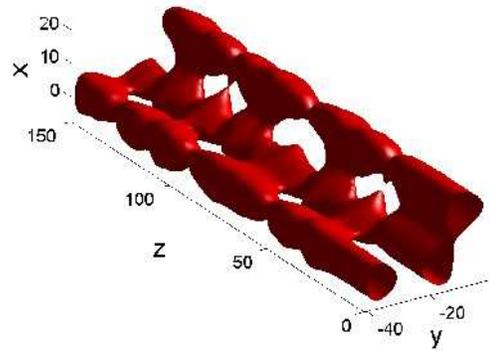}
\end{center}
\caption{The typical time-dependent dynamics of an unstable
configuration along
the upper (dashed line) branch of the existence curve presented in the
top panels of Fig\ \ref{IPN}. Depicted
here is the isosurface of height $0.15$
 of the dynamics of the of the intensity, $|U|^2$, of the
configuration shown in the
bottom panel of Fig.\ \ref{IPN}.}
\label{dyn_IPN}
\end{figure}

We have found IP dipoles in adjacent wells for values of the
propagation constant $\mu$ throughout the entire Bragg reflection gap
for a given $E_0$. We found that the solitons exist for
$\mu$ between 3.62 and 4.94, and that the intensity of the dipoles cannot be
arbitrary low, a result similar to the observed results of the
focusing and defocusing cases for square lattices
\cite{yang04,yang04_4,our_sq}.
The relevant findings are summarized in Fig.\ \ref{IPN}.

The top left panel of Fig.\ \ref{IPN} shows the stability of the dipoles
against the propagation constant $\mu$, by illustrating the
maximal growth rate (maximum real part of all eigenvalues $\lambda$) 
of perturbations.
When $\max({\rm Re}(\lambda))=0$, this implies stability
of the configuration, while the configuration is unstable if
$\max({\rm Re}(\lambda)) \neq 0$ in this Hamiltonian system.
We found that this type of dipoles may be stable for windows throughout
the first Bragg gap, as predicted above, although it
is possible for small oscillatory Hopf instabilities to arise
due to opposite signature eigenvalue collisions.
The dipole configuration disappears in a saddle-node bifurcation
at the edge of the first spectral band, depicted in the
top panels of Fig.\ \ref{IPN},
as $\mu\rightarrow 4.94$, and a real pair of
eigenvalues emerges.  At this point, the
configuration collides with a configuration shown at the bottom
panel of Fig. \ref{IPN} in which the adjacent well next to one of the
populated ones becomes excited out-of-phase with the others.
Consistent with our theoretical expectation from its having an out-of-phase set
of nearest neighbors, the latter configuration
always has a real pair of eigenvalues $\lambda$.

\begin{figure}[tbh]
\begin{center}
\includegraphics[width=0.23\textwidth]{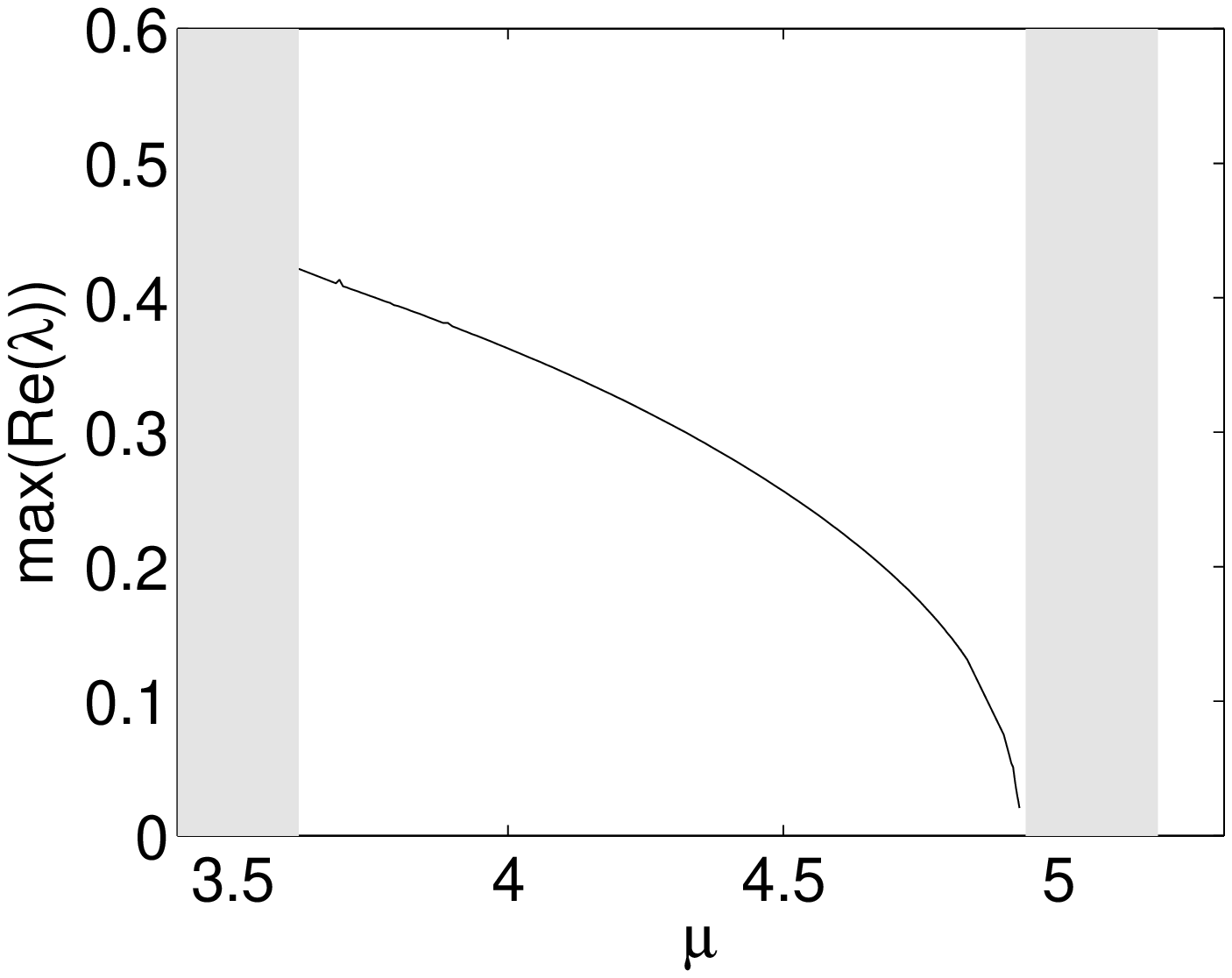}
\includegraphics[width=0.23\textwidth]{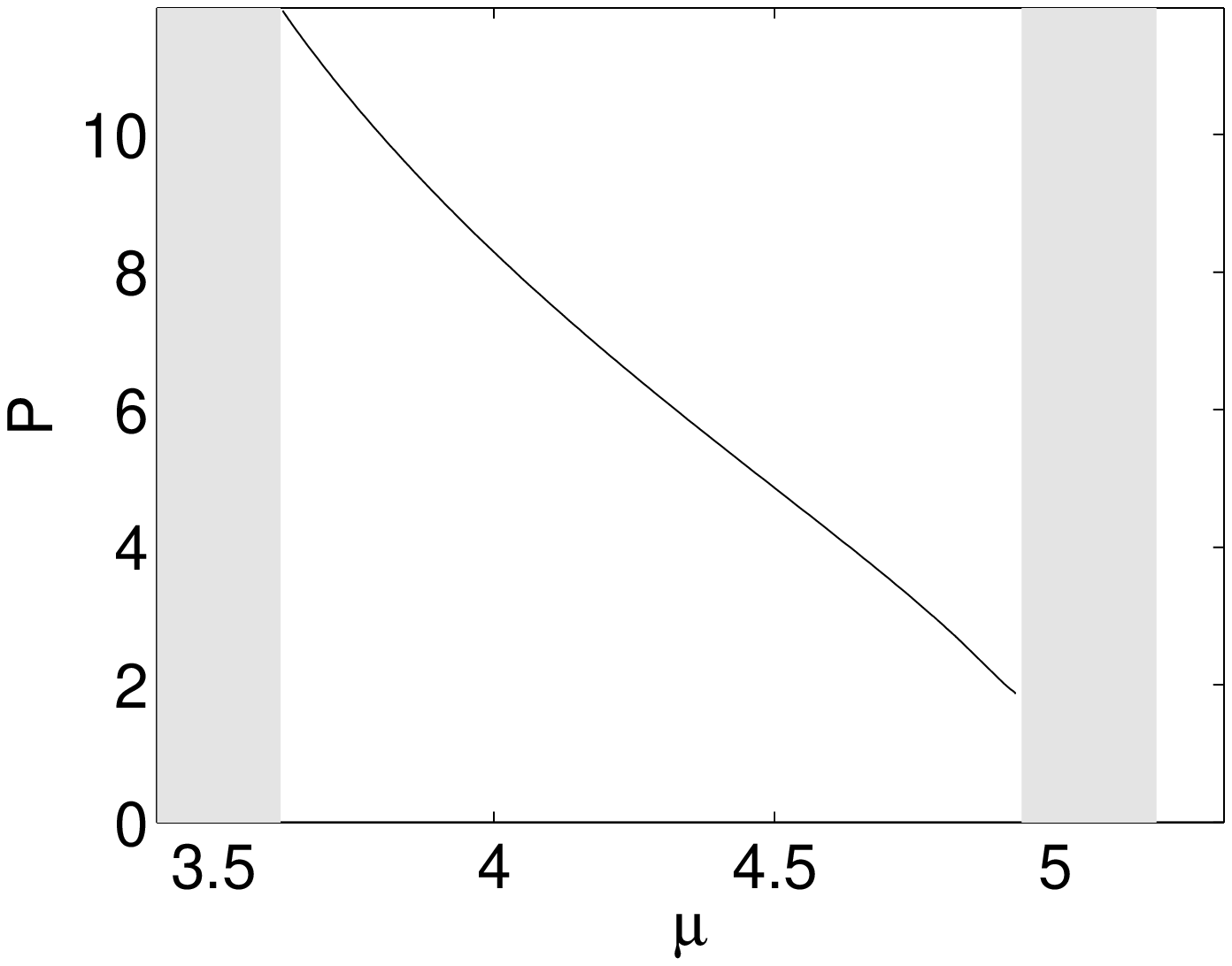}\\
\includegraphics[width=0.23\textwidth]{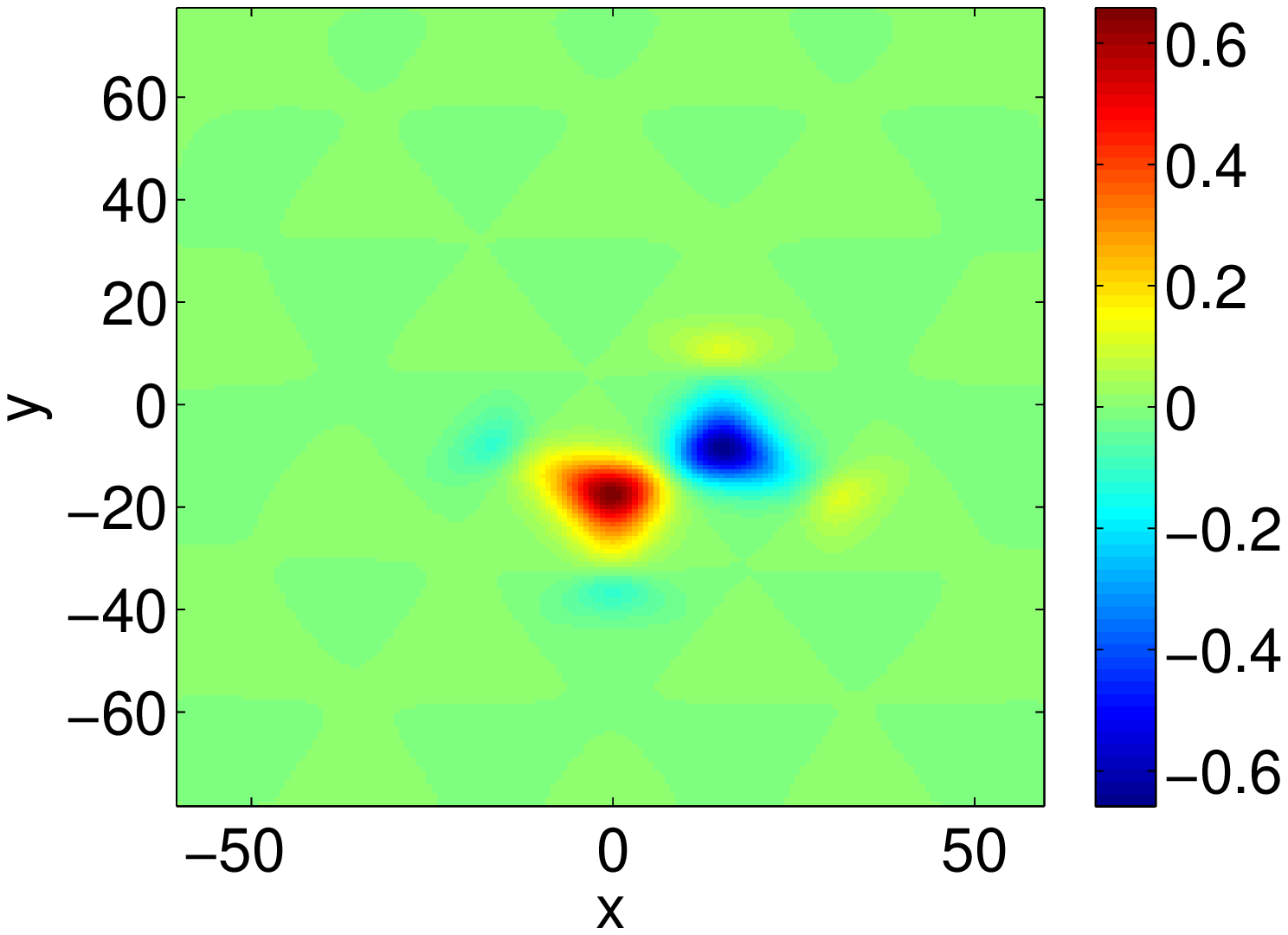}
\includegraphics[width=0.23\textwidth]{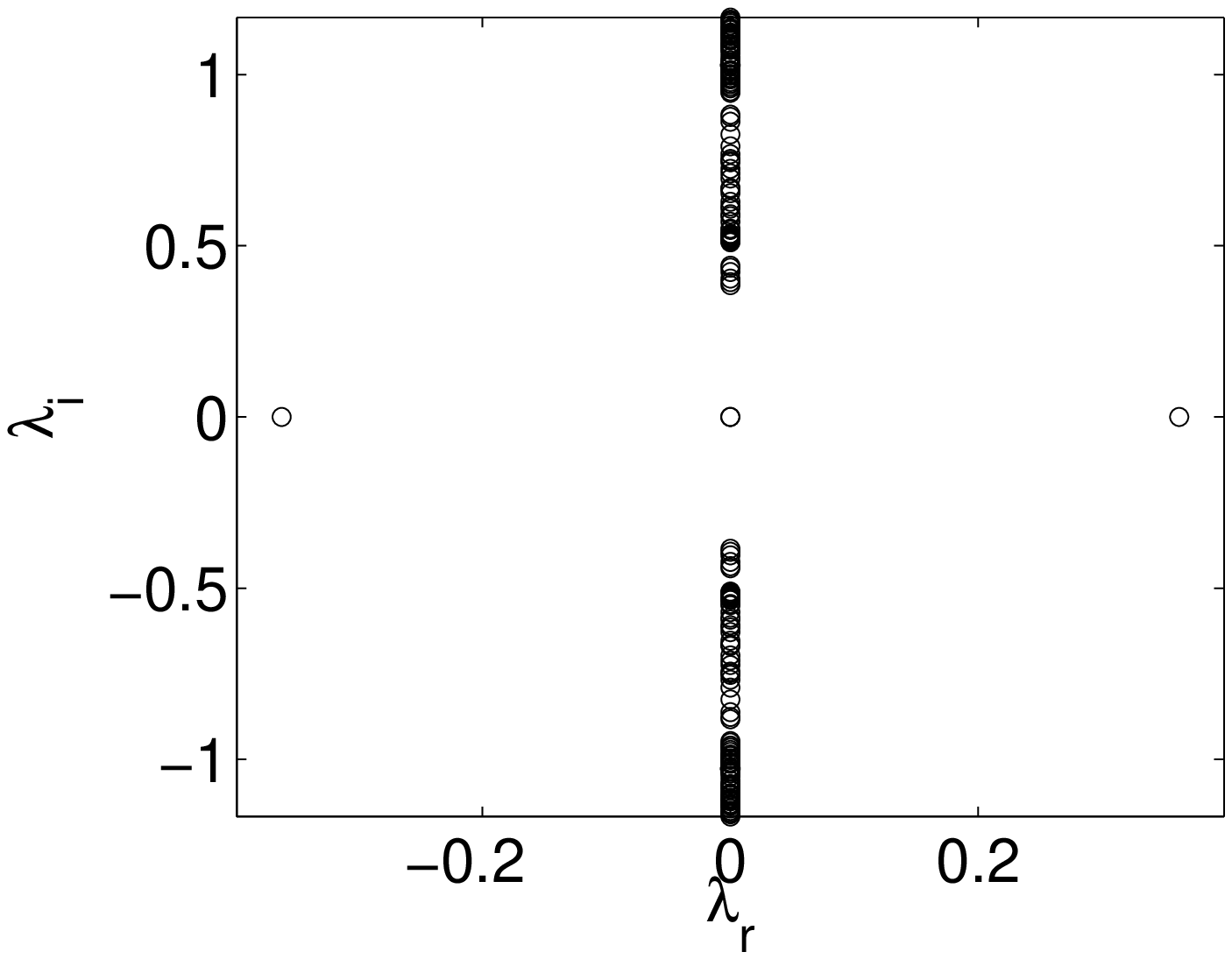}\\
\end{center}
\caption{(Color online) The top panels correspond to the same
panels of Fig.\ \ref{IPN}
but for OPN dipole solitons. The bottom panels show the
profile $u$ and the corresponding spectral plane of the
dipoles at $\mu=4$.}
\label{OPN}
\end{figure}

\begin{figure}[tbp!]
\begin{center}
\includegraphics[width=0.4\textwidth]{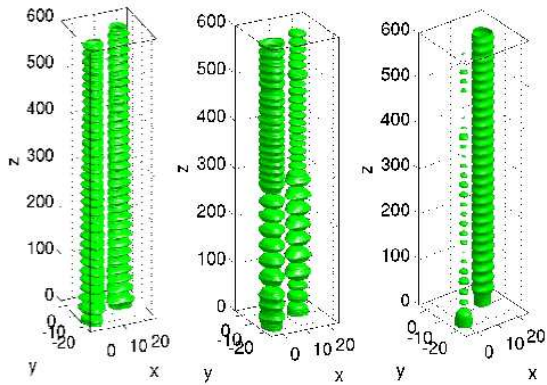}\\
\end{center}
\caption{The typical dynamical evolution of an unstable out-of-phase nearest
neighbor configuration from the family
presented in Fig\ \ref{OPN}. Depicted is the isosurface 
at half the maximum of the intial intensity amplitude. 
Notice that the OPN appears to oscillate between two sites and one
for the propagation constant $\mu=4$ (center) of the solution presented in 
the bottom panels of Fig.\ \ref{OPN}, as does it for smaller values
($\mu=3.6$, left), although for a larger value of $\mu=4.6$, 
right, the solution essentially transforms (due to
the instability) into a single site mode.  We hypothesize that the 
stronger effect of the nonlinearity on those solutions with smaller 
values of $\mu$ decreases the size of linear regime, causing these more 
linearly
unstable solutions to actually appear more stable in the full nonlinear
dynamical evolution, although it is away from the 
linear regime that the instability is manifested merely as
oscillations.}
\label{dyn_OPN}
\end{figure}

The middle left and right panels show the profile $u$ of the
in-phase nearest (IPN) neighbor dipole at
$\mu=4$ and the corresponding spectrum of linearization
eigenvalues $\lambda=\lambda_r+ i \lambda_i$ in the complex plane
$(\lambda_r,\lambda_i)$, respectively.
The corresponding profile and spectral plane for the saddle branch
(that eventually collides with the IPN solution)
at $\mu=4$ is shown in the bottom left and right panel, respectively,
of the same figure, illustrating the exponential instability of the latter.

We have simulated the dynamics of the solitary waves when they are unstable. The dipoles are perturbed by a random noise with
maximum intensity $2\times10^{-3}$. It is interesting to note that an unstable
IPN dipole turns out to be quite robust, even though it experiences only an
oscillatory instability. It is remarkable that up to $z=200$ we did not see
any signifant change in the configuration. Therefore, we do not depict our
evolution simulation here;
we simply note that this is consonant with the very weak growth rate of
the relevant oscillatory instability.

For the solution branch
 shown in the bottom panel of Fig.\ \ref{IPN}, we present its
dynamics in Fig.\ \ref{dyn_IPN}.
We found that the instability is strong as predicted above such that even after a relatively short propagation distance, the instability already sets in and
leads to recurrent oscillations (for the remainder of our dynamical evolution
horizon) between a dipole, two-site state and a three-excited-site state;
see Fig.\ \ref{dyn_IPN}.

\subsection{Out of Phase Nearest Neighbor Dipole Solitons}

We have also found OP dipoles arranged in nearest-neighboring lattice wells
which we refer to as OPN.
We summarize our findings in Fig.\ \ref{OPN} where one can see that the
solitons exist in the whole entire region of propagation constant
$\mu$ in the first Bragg gap, $\mu\in(3.62,4.94)$.
This smooth transition indicates that the OPN dipole solitons emerge out
of the Bloch band waves; see e.g. \cite{peli04} and \cite{shi07} for
a relevant discussion of the 1D and of the 2D problem respectively, in
the case of the cubic nonlinearity. The
OPN dipoles are
unstable due to a real eigenvalue pair, as expected from our above
theoretical predictions.

As the branch merges with the band edge, we observe an
interesting feature, namely that the configuration begins to
resemble a hexapole with a $\pi$ phase difference between each
well.
This can be an indication that these structures bifurcate out
of the Bloch band from the same bifurcation point.
We elaborate this further in our discussion at the end of
section VI.

In Fig.\ \ref{dyn_OPN} we present the unstable dynamics
of OPN dipole solitons perturbed by similar random noise perturbation
as in Fig.\ \ref{dyn_IPN}.
We display here three solutions for a range of chemical potentials 
to illustrate that the dynamical evolution of linearly 
unstable states is apparently correlated to
the power of the solution.
This type of dipoles is typically
more unstable than its IP counterpart, as is illustrated 
in the figure.
In particular, in all three
examples of unstable evolution given
the
instability already starts to manifest itself.
around $z=20$
However, for small values of $\mu$ (large power) the OPN continues 
oscillating between a single site structure
and a two site structure for the (longer) evolution distances 
investigated in this illustrative case, while for large enough 
$\mu$ (small enough power), one of the sites decays 
and the power is concentrated on a single site.


\section{Next Nearest Neighbor Dipole Solitons}

We have also obtained dipole solutions that are not oriented along the
two nearest-neighboring lattice wells, but rather where
the two humps of the structure are located at two
next-nearest-neighboring lattice sites. These humps can once again
have the same phase or a $\pi$ phase difference between them. We will
again use the corresponding IP and OP designations for these
next nearest neighbor waveforms.

\subsection{In Phase, Next Nearest Neighbor Dipole Solitons}

\begin{figure}[tbh]
\begin{center}
\includegraphics[width=0.23\textwidth]{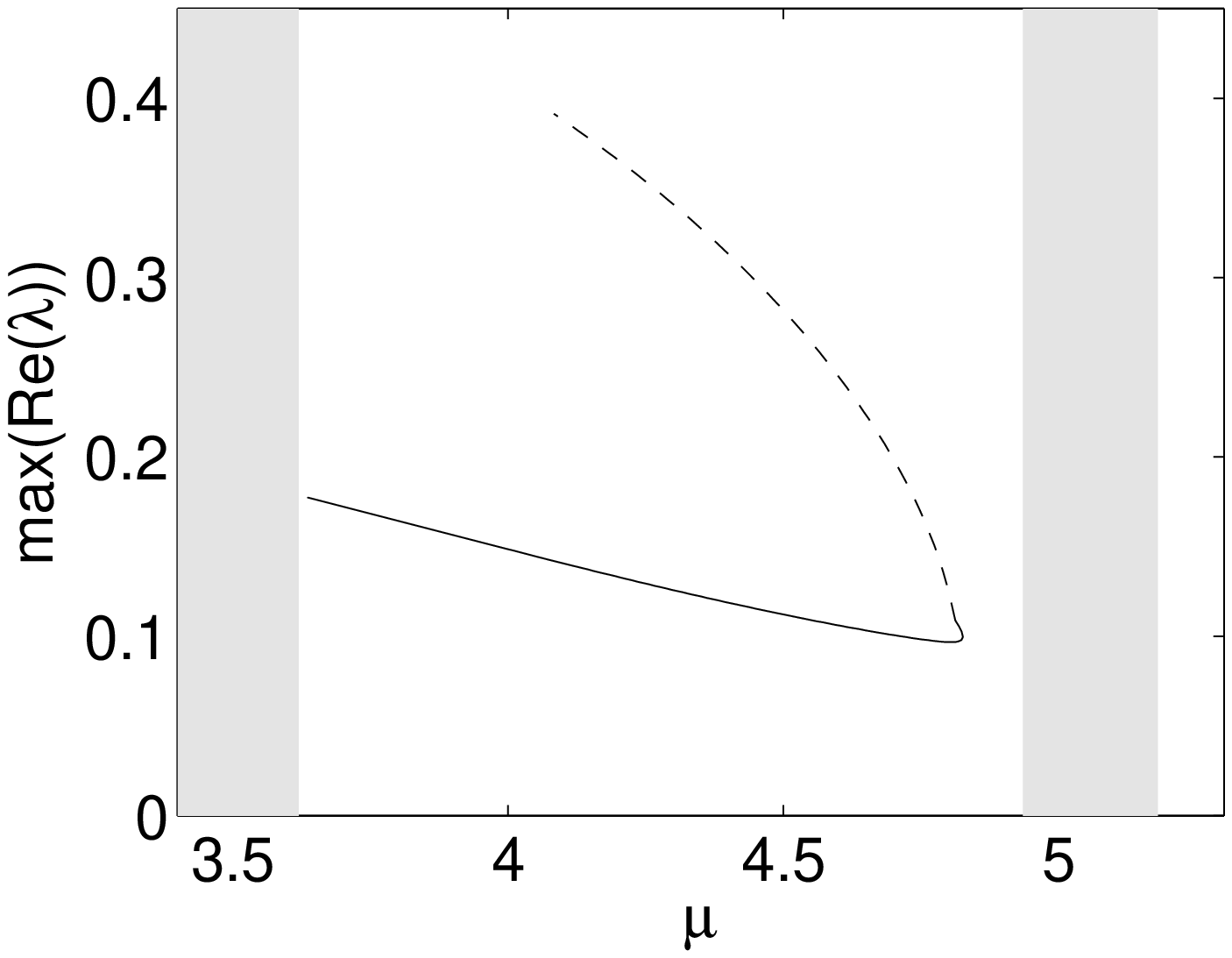}
\includegraphics[width=0.23\textwidth]{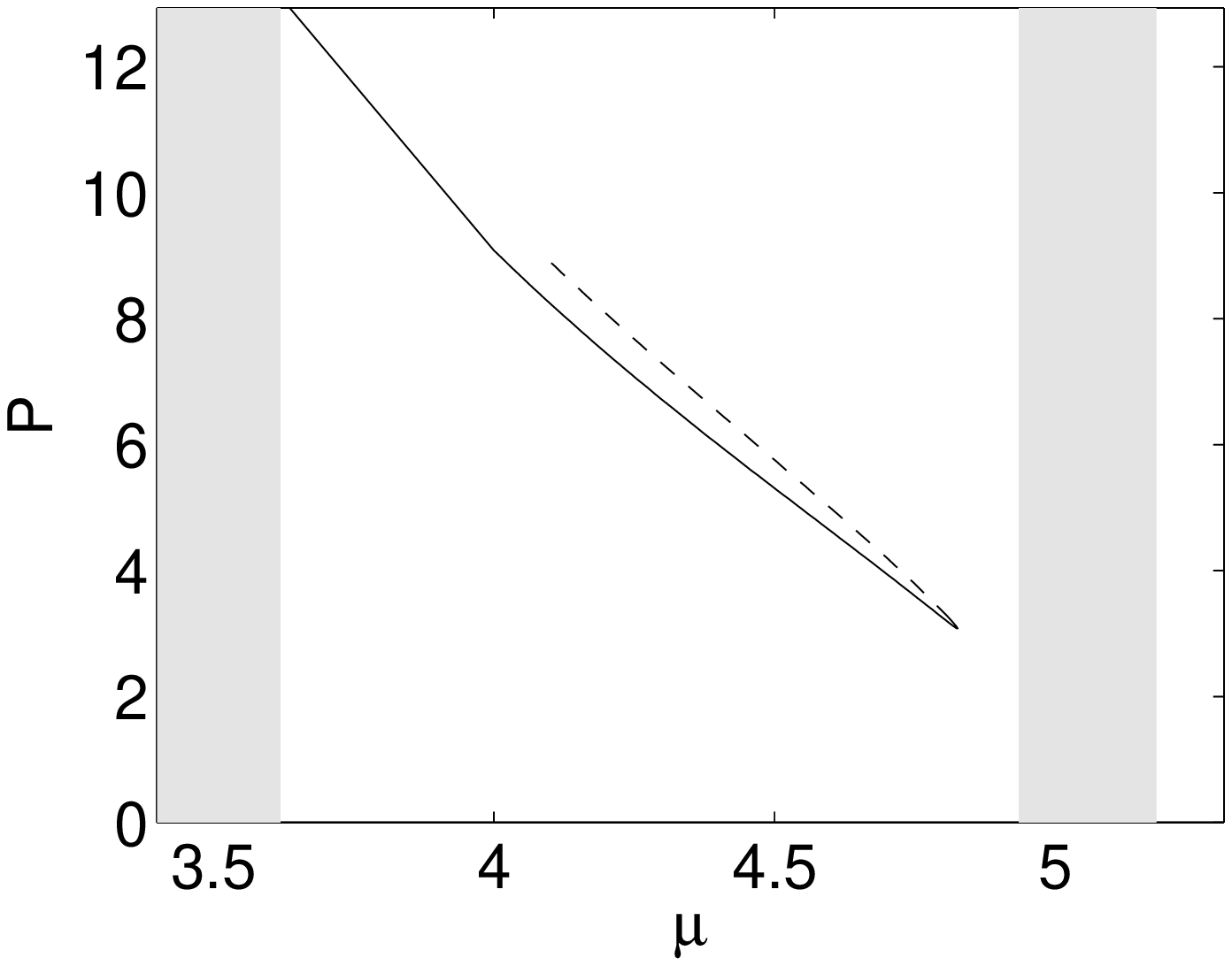}\\
\includegraphics[width=0.23\textwidth]{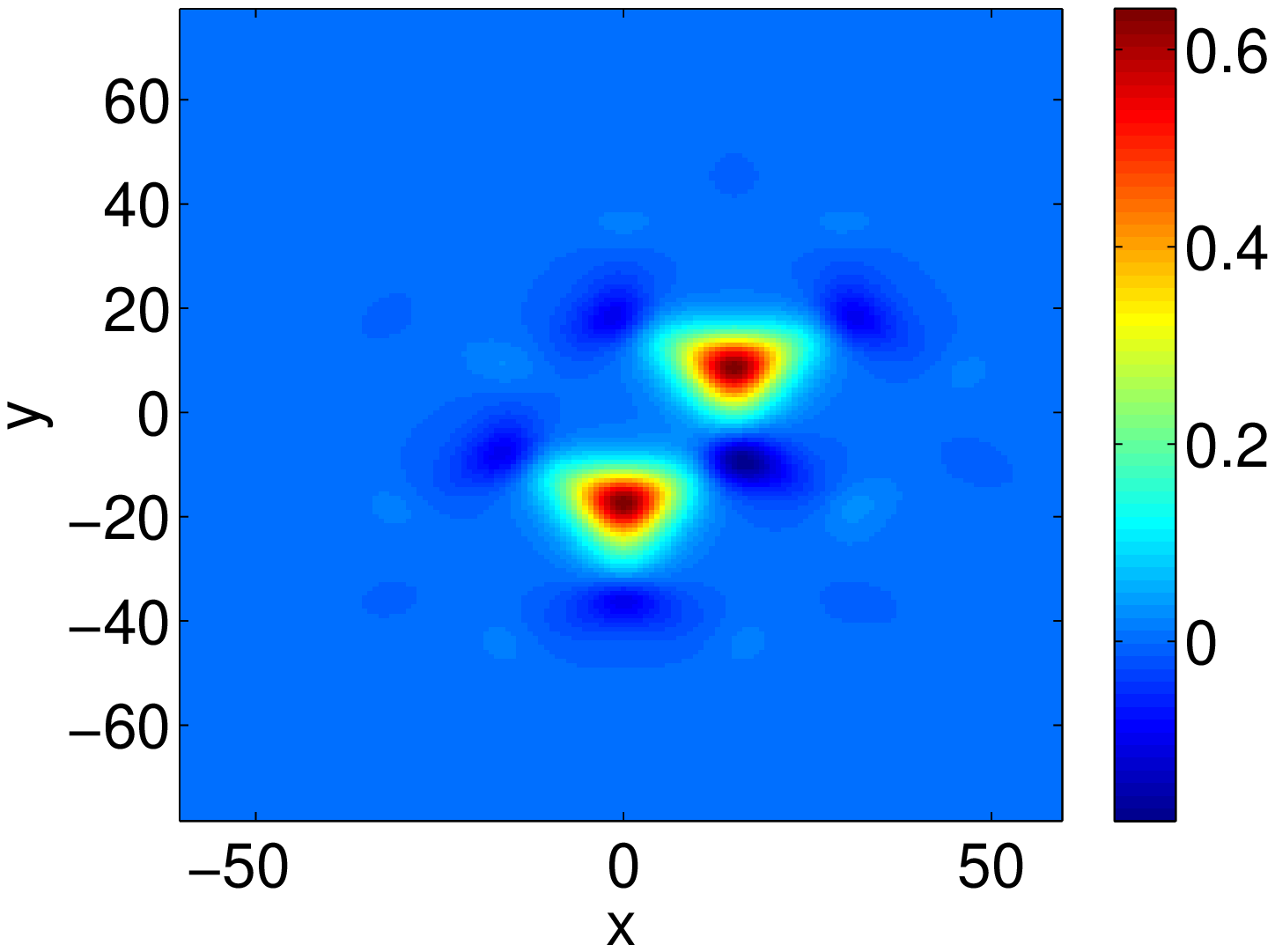}
\includegraphics[width=0.23\textwidth]{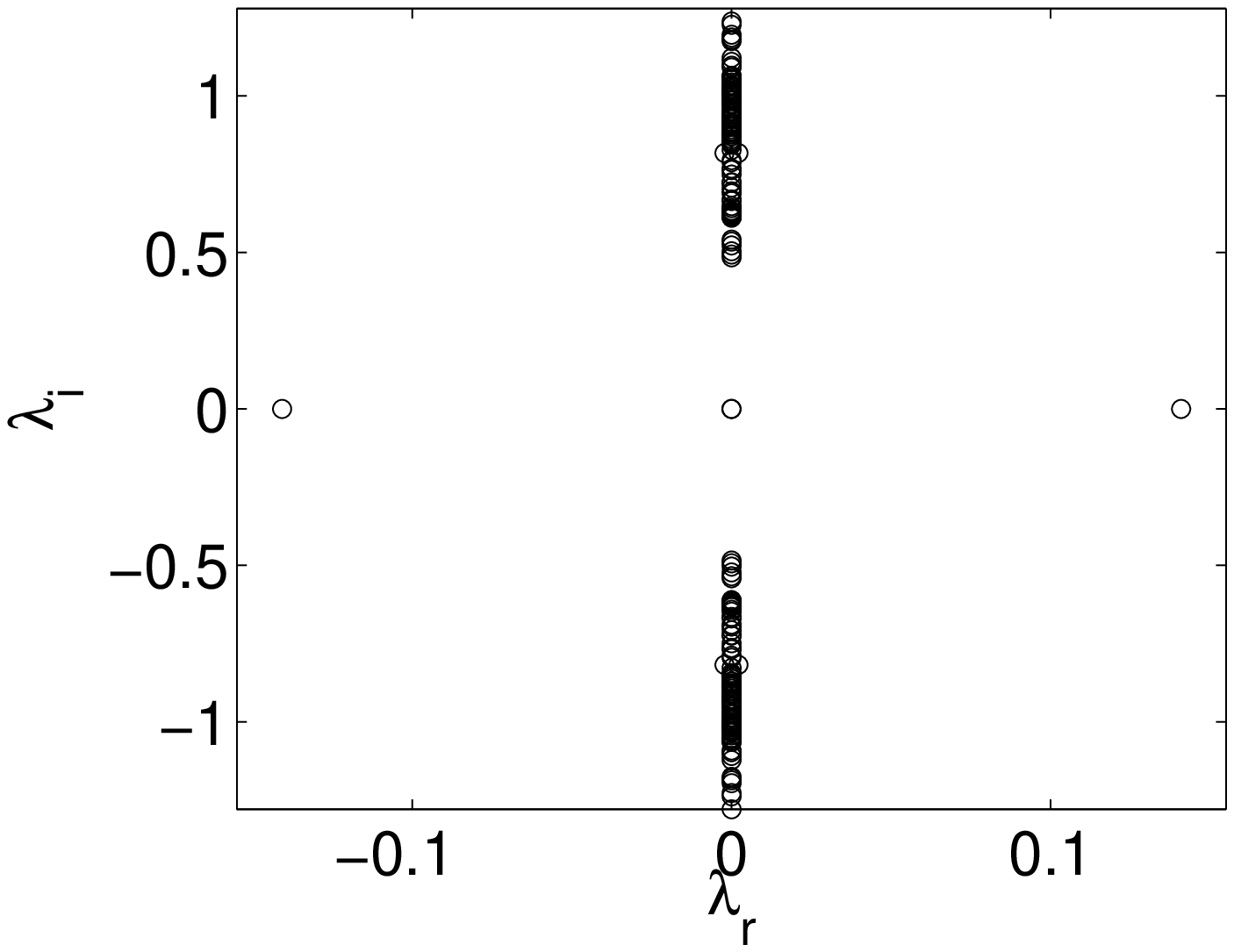}\\
\includegraphics[width=0.23\textwidth]{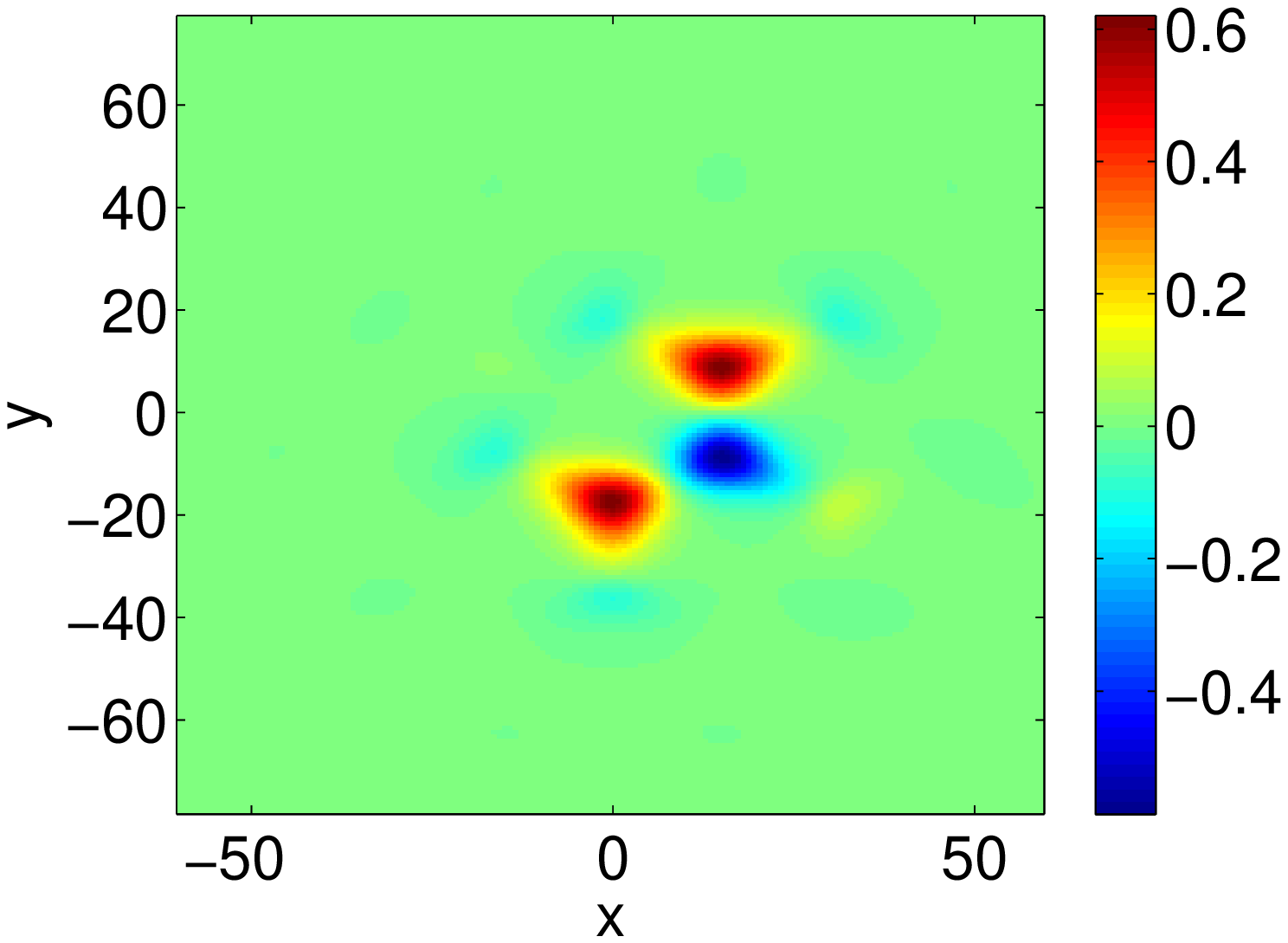}
\includegraphics[width=0.23\textwidth]{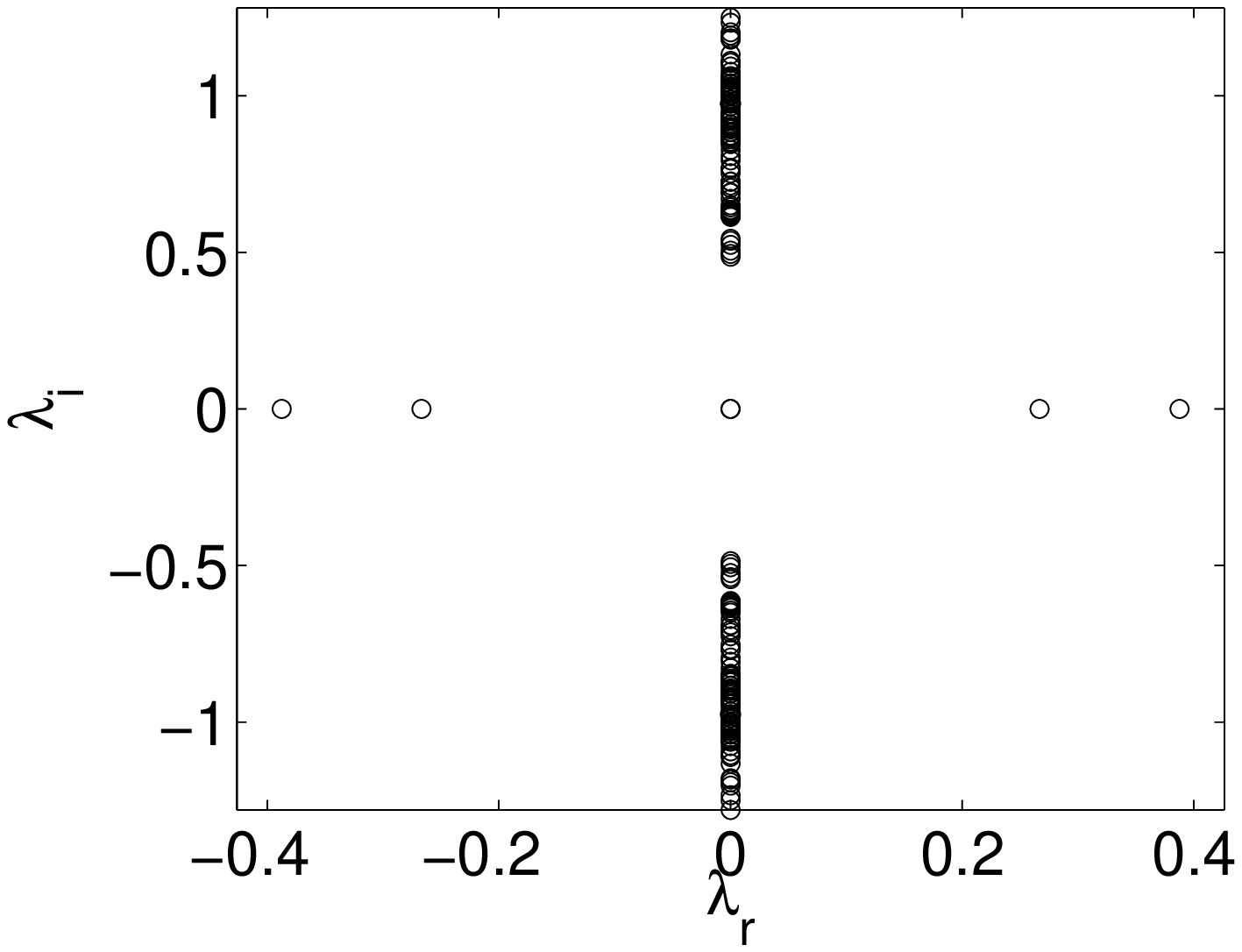}
\end{center}
\caption{(Color online) The top panels correspond to the same diagnostics
as in Fig.\ \ref{IPN} but
for IPNN dipole solitons. The middle panels show the profile
$U$ and the corresponding spectral plane of the IPNN dipole at $\mu=4.1$,
while the bottom row shows the same images for the solution branch
corresponding to the dashed line in the top panel, shown at
the same value of $\mu$.}
\label{IPNN}
\end{figure}

\begin{figure}[tbp!]
\begin{center}
\includegraphics[width=0.4\textwidth]{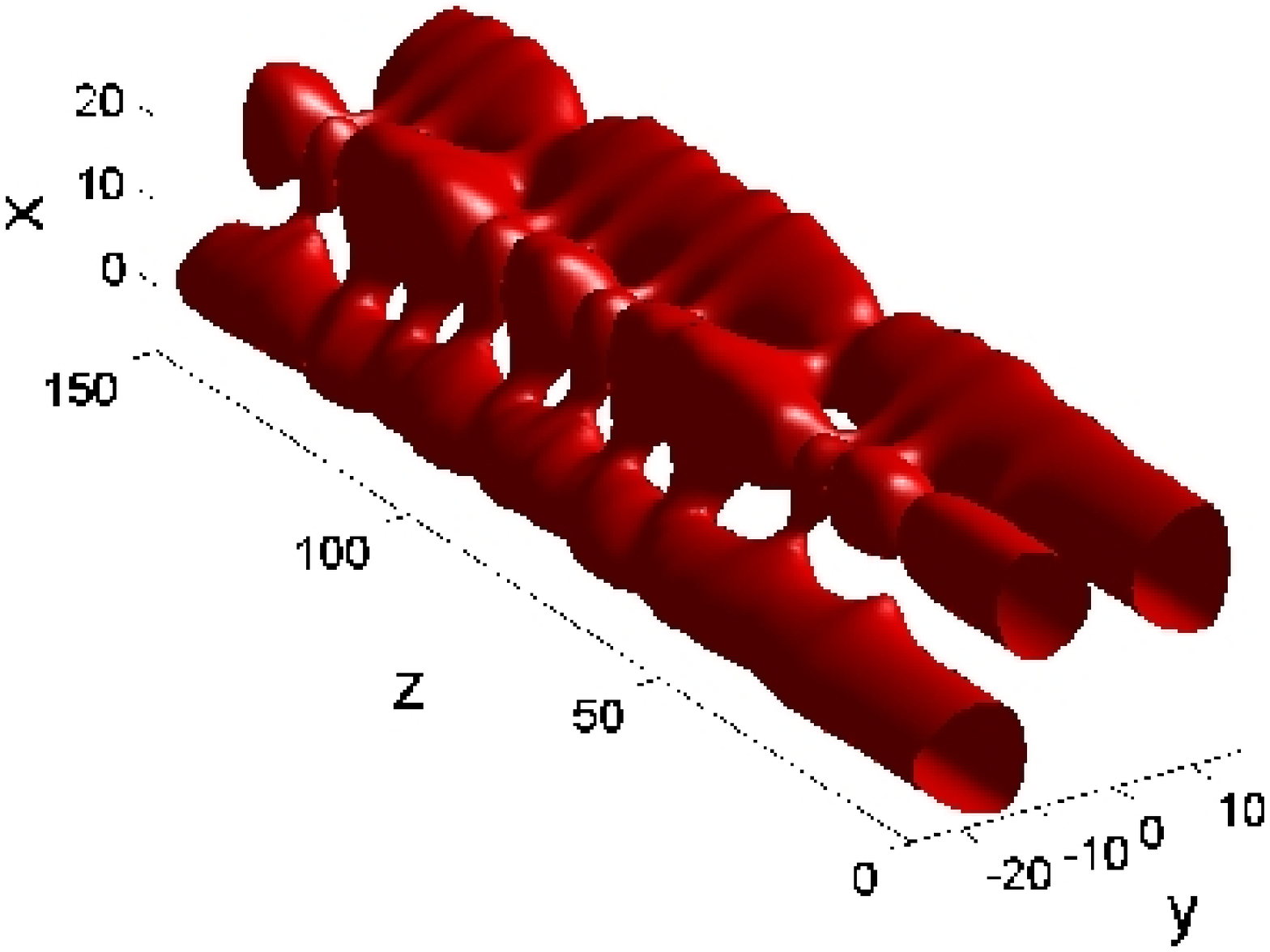}
\end{center}
\caption{The same figure as Fig.\ \ref{dyn_IPN},
but for the solution presented
in the bottom panel of Fig.\ \ref{IPNN}.
Depicted is the isosurface of height $0.05$.}
\label{dyn_IPNN}
\end{figure}

\begin{figure}[tbp!]
\begin{center}
\includegraphics[width=0.23\textwidth]{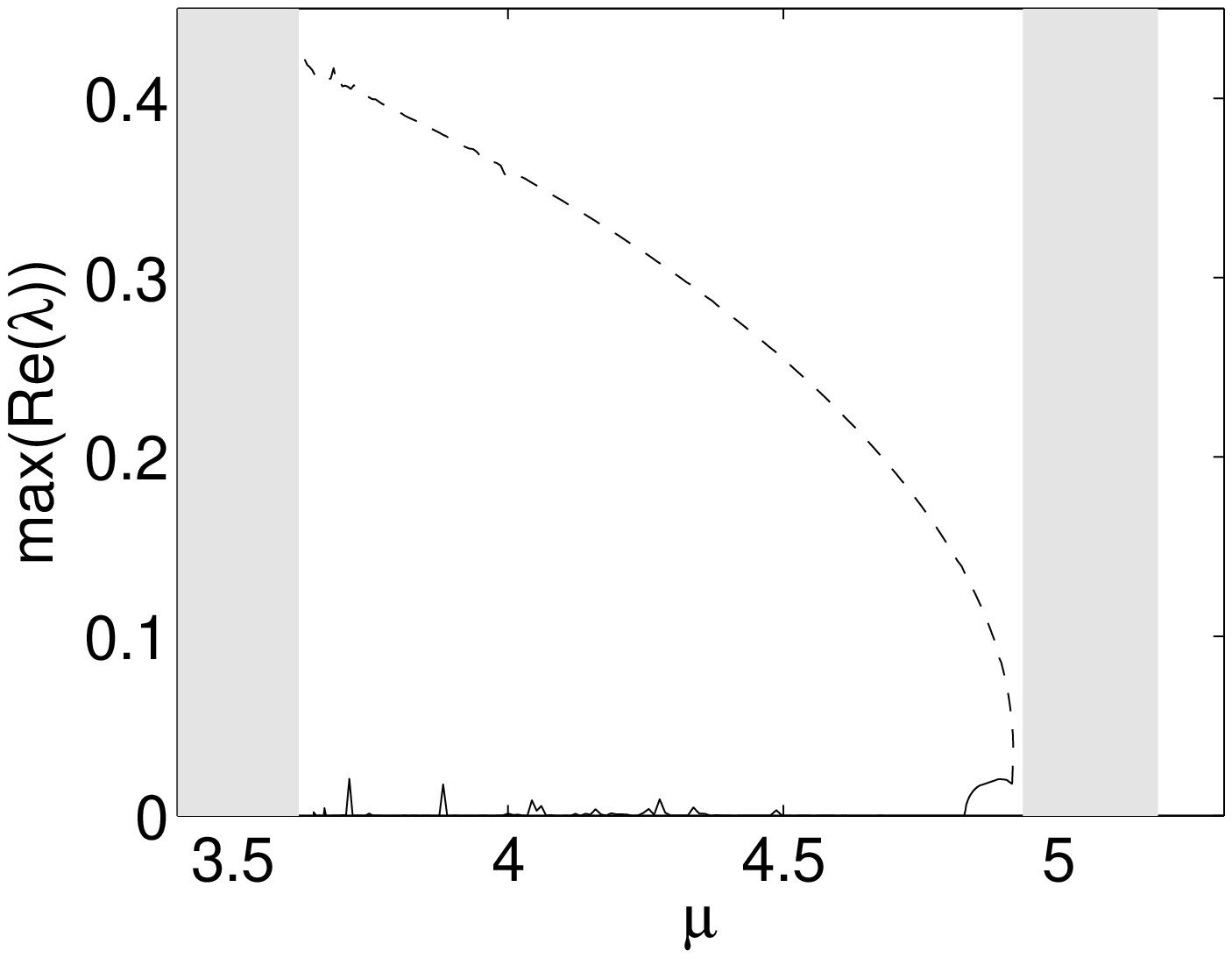}
\includegraphics[width=0.23\textwidth]{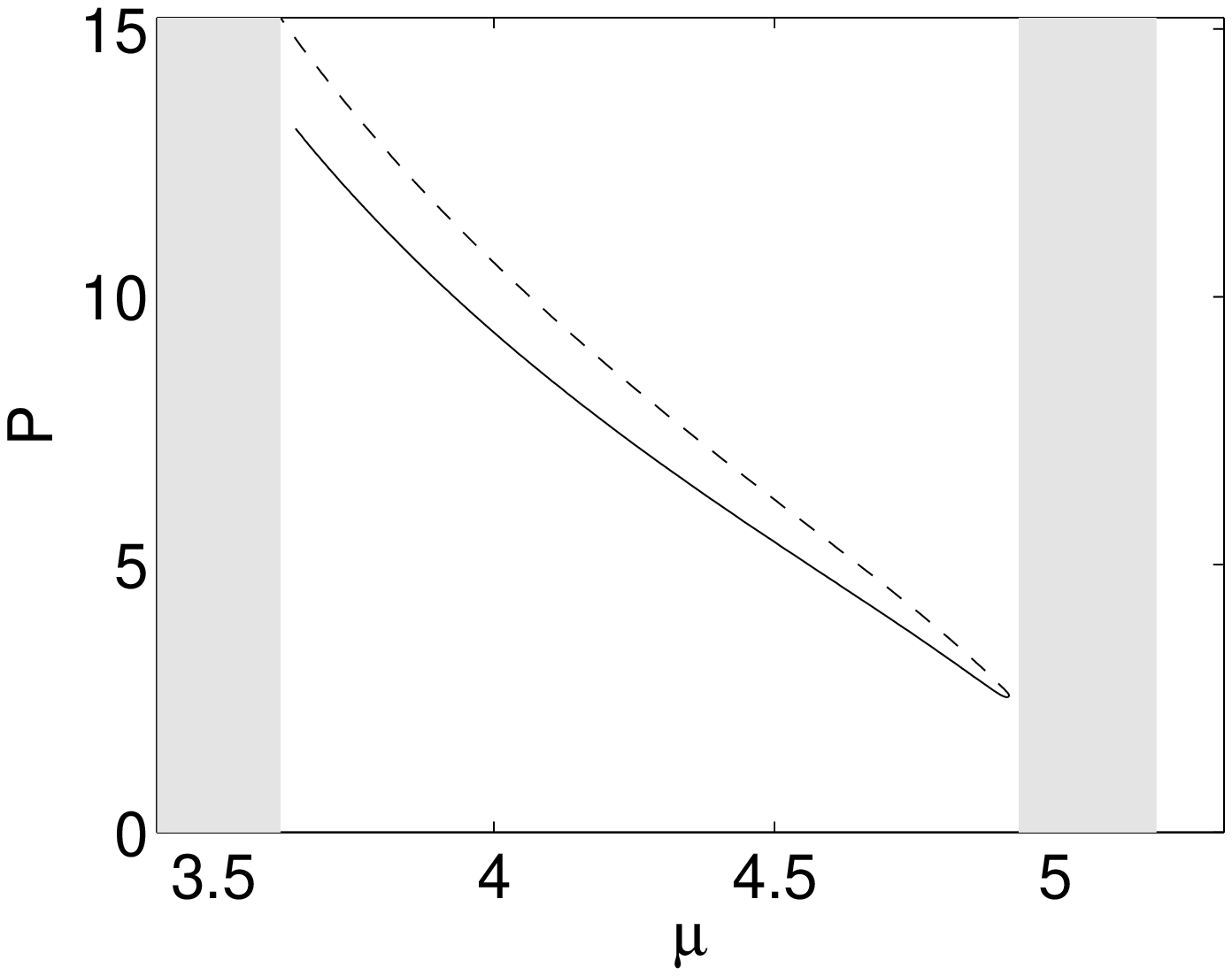}\\
\includegraphics[width=0.23\textwidth]{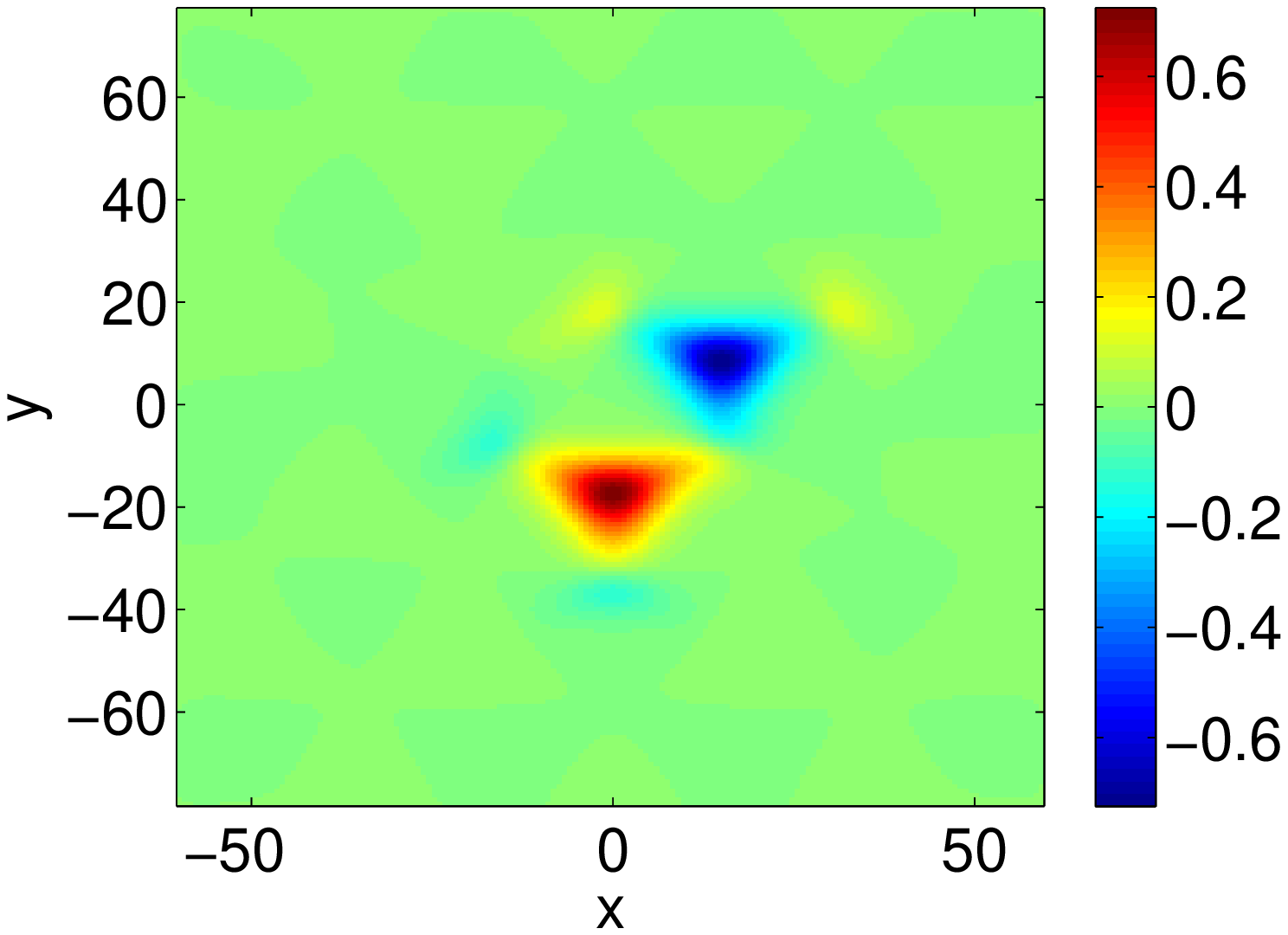}
\includegraphics[width=0.23\textwidth]{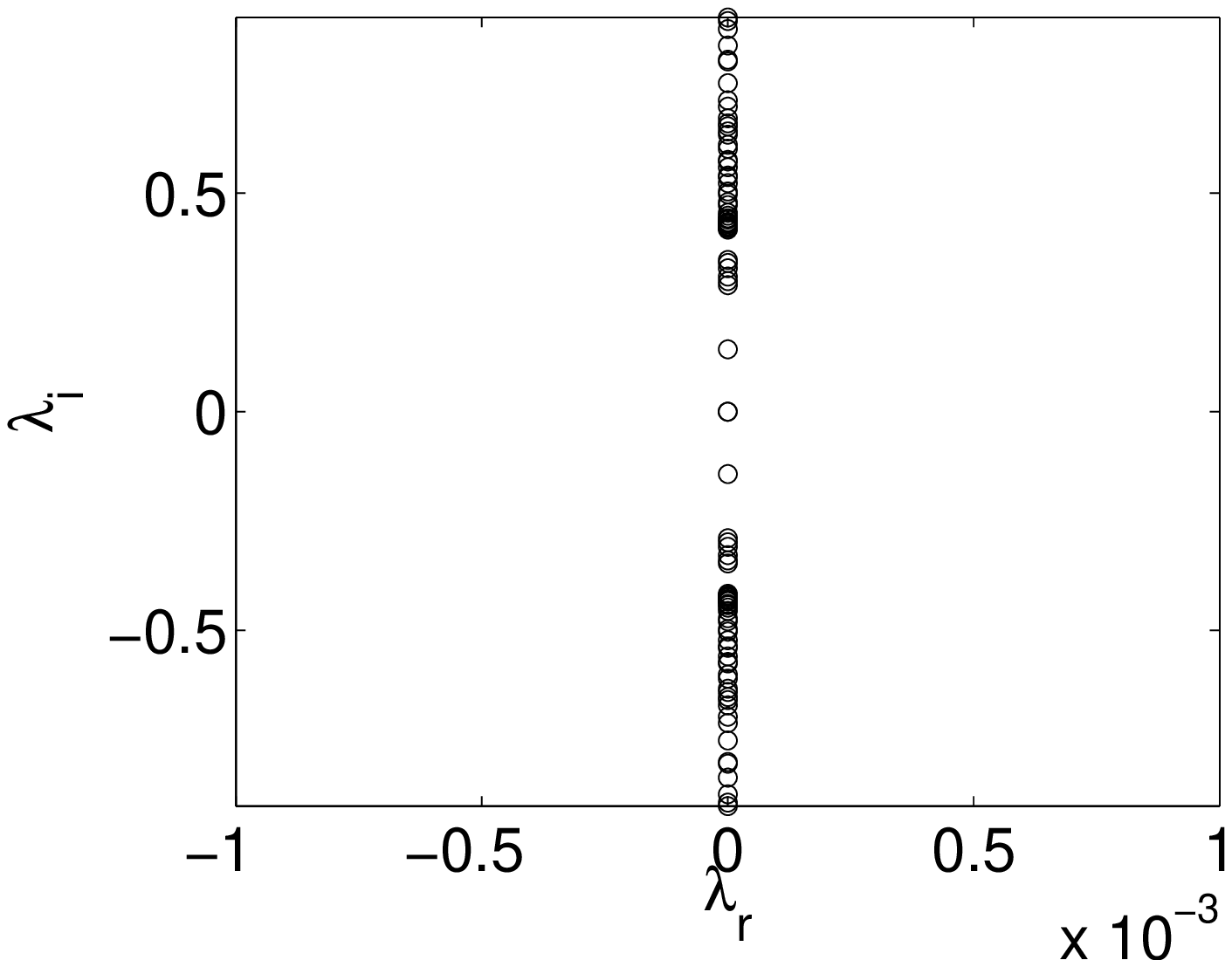}\\
\includegraphics[width=0.23\textwidth]{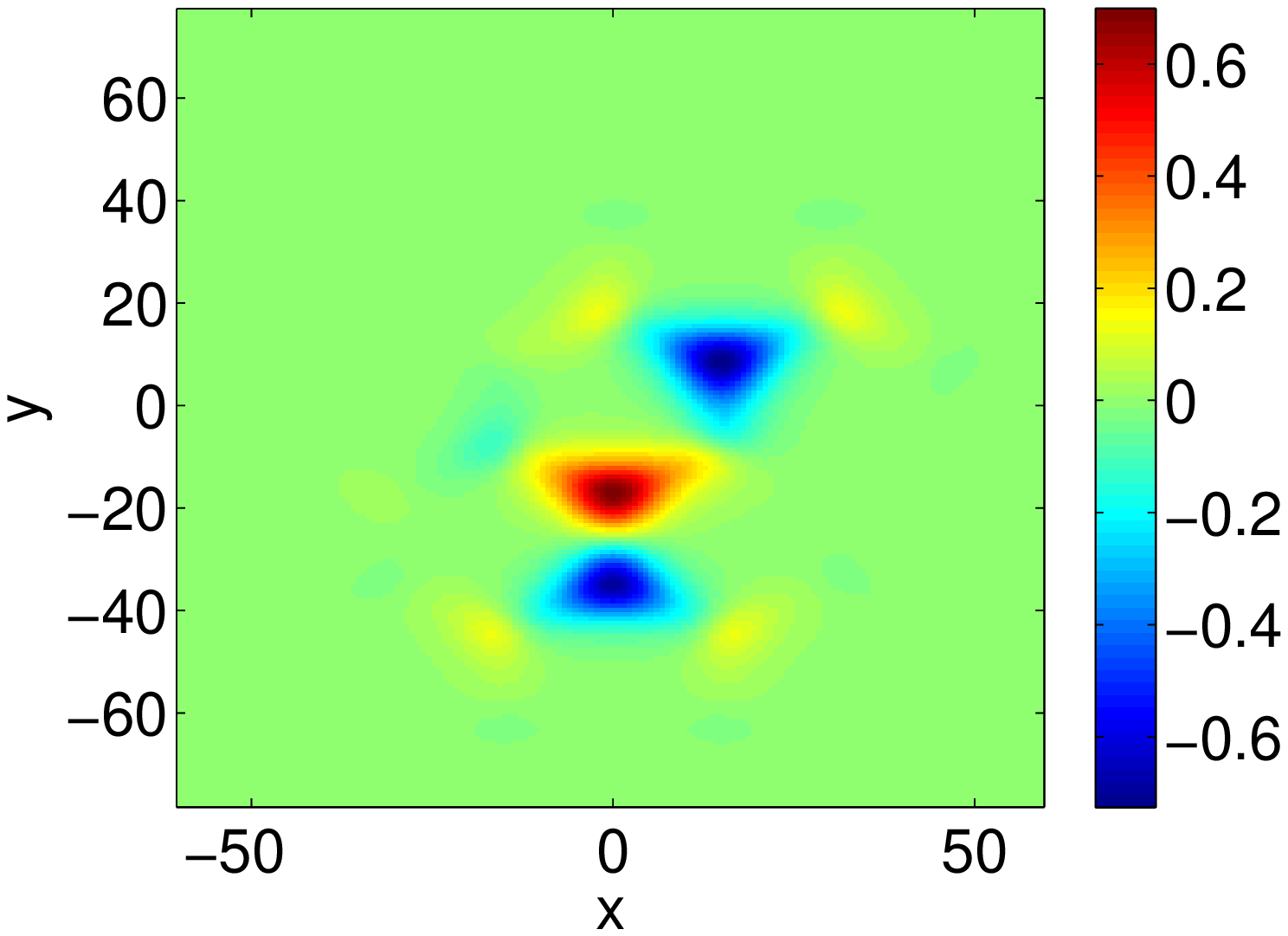}
\includegraphics[width=0.23\textwidth]{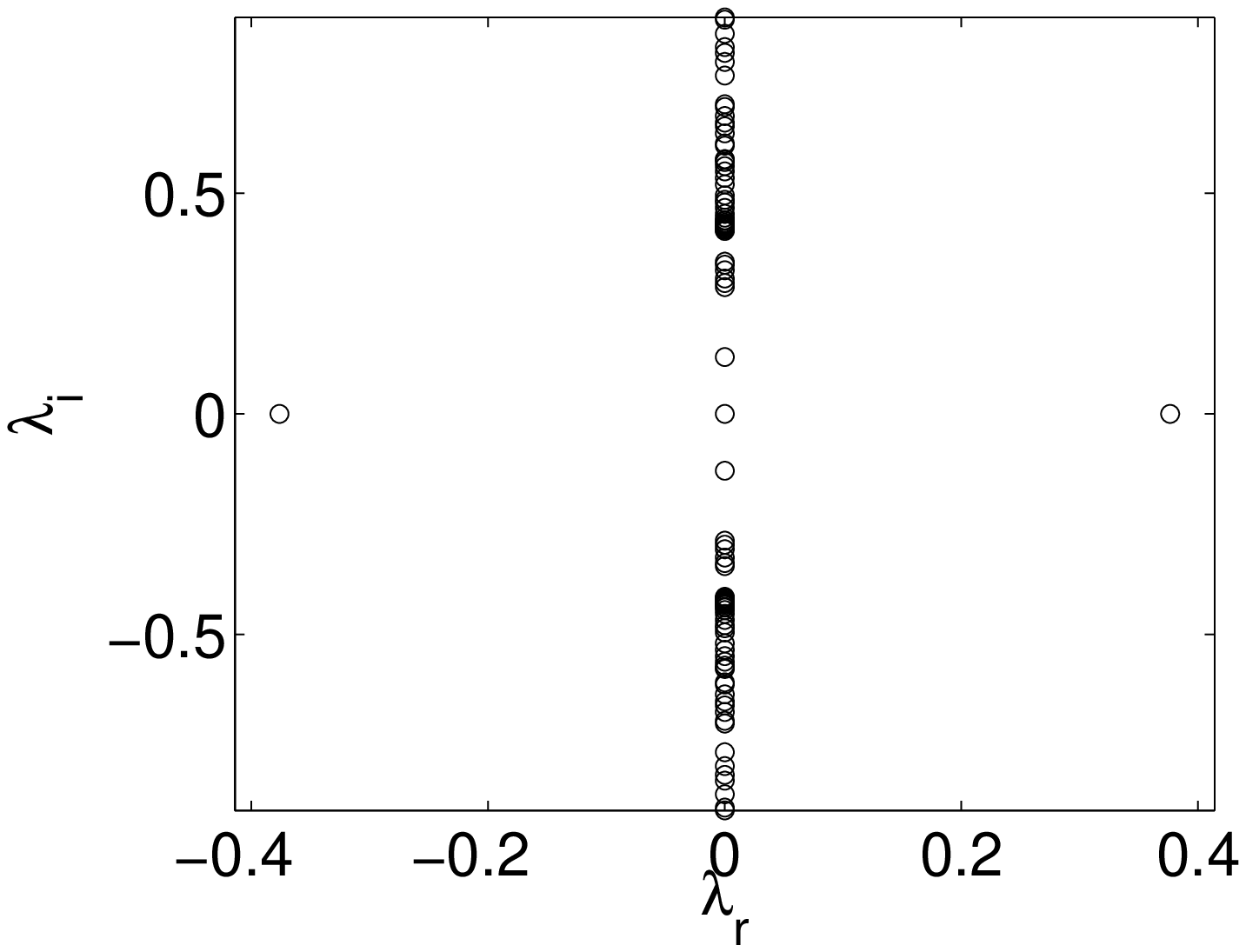}
\end{center}
\caption{(Color online) The top panels depict the largest
real part of the critical
eigenvalue, as well as the power of the OPNN dipole solitons. The middle panels show the profile $u$ and the
corresponding spectra in the complex plane of the dipole at $\mu=3.9$,
and the bottom is the unstable saddle configuration at the same value of $\mu$,
where one of the sites has merged with a neighbor out of phase and become an
OPN, accounting for the real eigenvalues.}
\label{OPNN}
\end{figure}

\begin{figure}[tbp!]
\begin{center}
\includegraphics[width=0.4\textwidth]{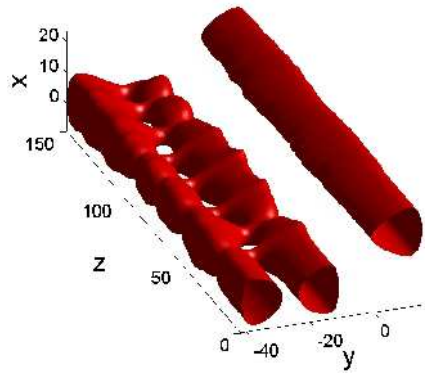}
\end{center}
\caption{The same figure as Fig.\ \ref{dyn_IPN}, but for the solution presented in the bottom panel of Fig.\ \ref{OPNN}.
Depicted is the isosurface of height $0.05$.}
\label{dyn_OOPNN}
\end{figure}

The in-phase next-nearest
(IPNN) neighbor solitons exist only up to a marginal distance from the
second band.
The stability and power of these dipoles are
shown in Fig.\ \ref{IPNN}. The stability is again consistent
with the theoretical discussion of Section II.
In particular, the IPNN configuration always possesses a real
eigenvalue pair; furthermore, the corresponding unstable ``saddle'' structure
with which it collides and terminates through a saddle-node bifurcation
has an additional such eigenvalue pair (two real eigenvalue pairs in
total for the solution branch indicated by dashed line
in Fig.\ \ref{IPNN}).

We have simulated also the dynamics of the unstable IPNN. Yet, we
do not present our simulation here as the typical
evolution of this configuration
is quite in resemblance to the dynamics of an unstable OPN
(see Fig.\ \ref{dyn_OPN})
in the fact that the configuration recurrently oscillates between
a two-soliton state and a one-soliton state.
Such an oscillation persists even up to $z=200$.

In Fig.\ \ref{dyn_IPNN}, we present the dynamical
evolution of the bifurcating solution
shown in the bottom panel of Fig.\ \ref{IPNN} under
similar random noise perturbation as above. One can note similarities in the typical evolution
of this configuration and the evolution of the bifurcating IPN solution shown in Fig.\ \ref{dyn_IPN}, one
of which is the recurrent oscillation between a pattern with three
pulses and one with just two peaks.

\subsection{Out of Phase Next Nearest Neighbor Dipole Solitons}

We have also obtained out-of-phase, next-nearest
(OPNN) neighbor dipole solitons. A typical profile of this
family of solutions for $\mu=3.9$ is shown in Fig.\ \ref{OPNN}.
The power diagram of these solitons is presented in the top panel of
Fig.\ \ref{OPNN}.
Typically these structures are stable
(as indicated again by the
comparison with the theoretical discussion and by
the numerical results shown in the middle right
panel of Fig.\ \ref{OPNN} ), suffering only windows of
oscillatory instability due to the presence of a single
eigenvalue with negative signature and
its collision with the spectral bands
In fact, we have found that a consistent stability
range for $E_0=8$ exists between $4.5\lesssim\mu\lesssim 4.85$.


For this solution we also observe that, similarly
to the IPNN dipoles, the solution disappears at non-zero intensity
because of the collision of this dipole with another (three-site)
configuration shown in
the bottom panels of Fig.\ \ref{OPNN} in a saddle-node bifurcation.
It is relevant to note that the point of the bifurcation is very close to
the edge of the Bloch band, i.e.,\ to $\mu \approx 4.94$.

The dynamics of the OPNN dipole do not manifest their very weak oscillatory
instability for the evolution distances considered herein. On the other hand,
the dynamics of the instability of
the 
three-site solution (with which the OPNN branch collides in the saddle-node
bifurcation) can be seen
in Fig.\ \ref{dyn_OOPNN}.
More specifically, the instability manifests itself
in the form of interactions between the closest out-of-phase pair of solitons
(leading to recurrent oscillations between a three-peak and a
two-peak state). Notice that the third peak
is almost not affected by these interactions.

\section{Opposite Dipole Solitons}

\begin{figure}[tbp!]
\begin{center}
\includegraphics[width=0.23\textwidth]{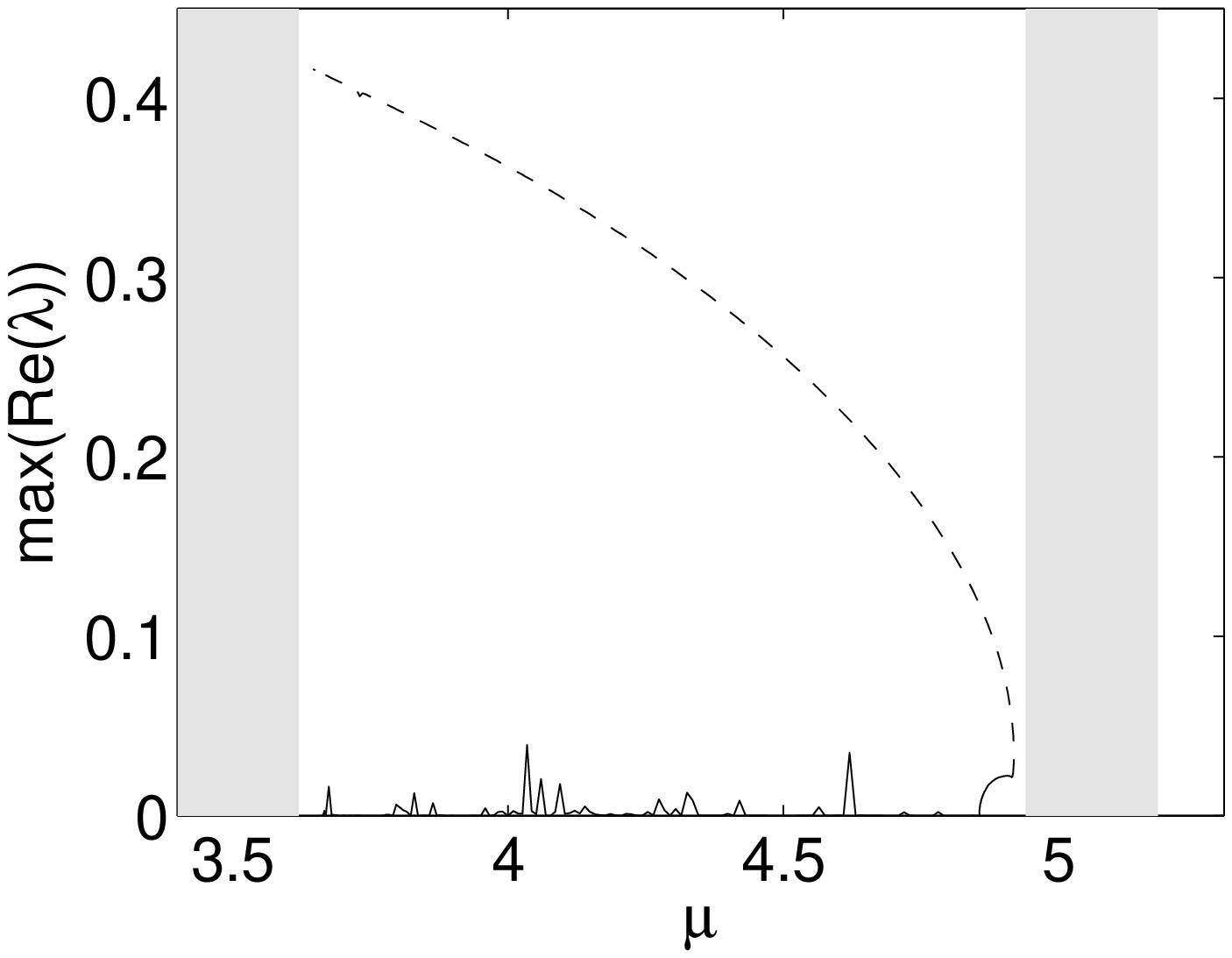}
\includegraphics[width=0.23\textwidth]{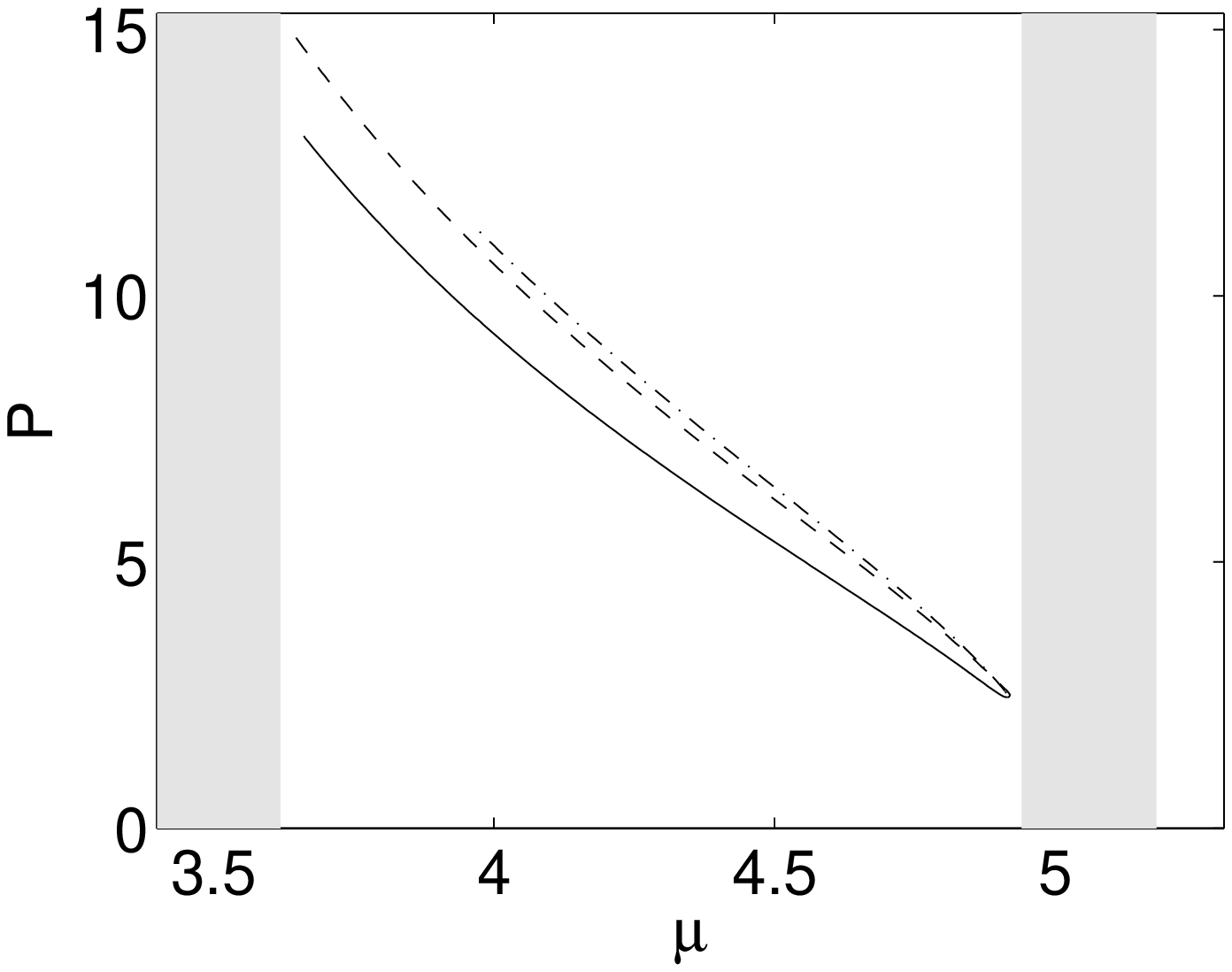}\\
\includegraphics[width=0.23\textwidth]{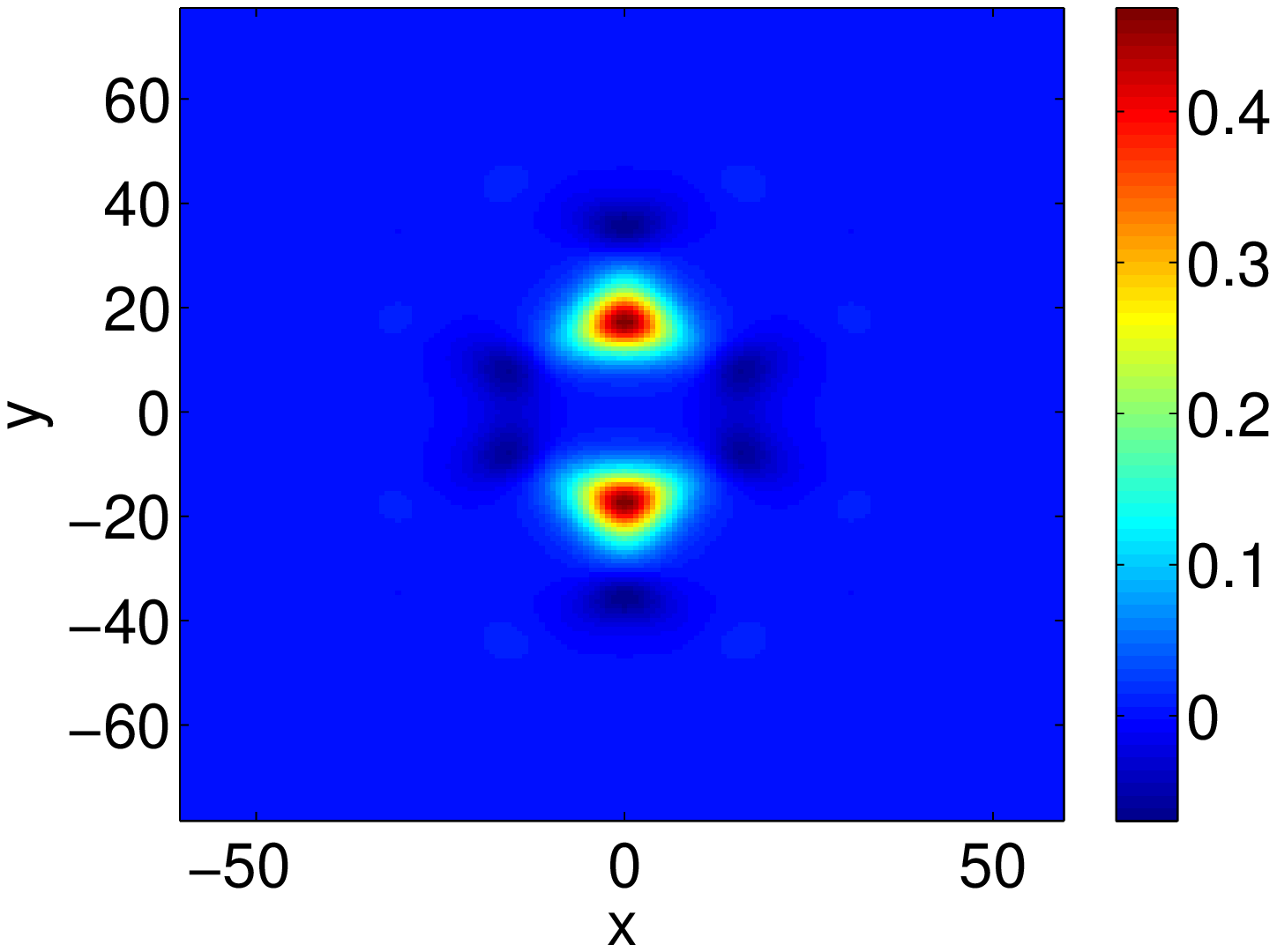}
\includegraphics[width=0.23\textwidth]{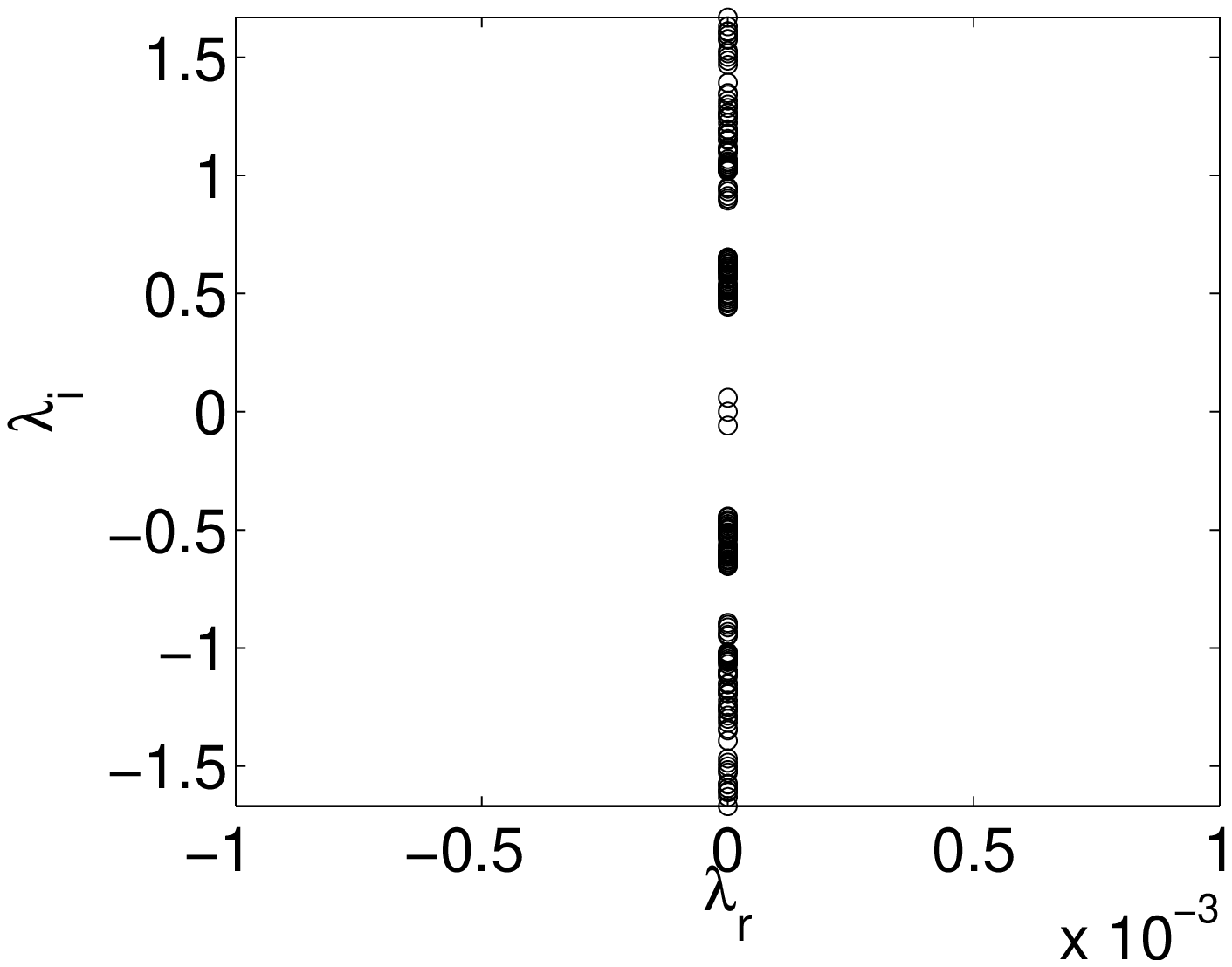}\\
\includegraphics[width=0.23\textwidth]{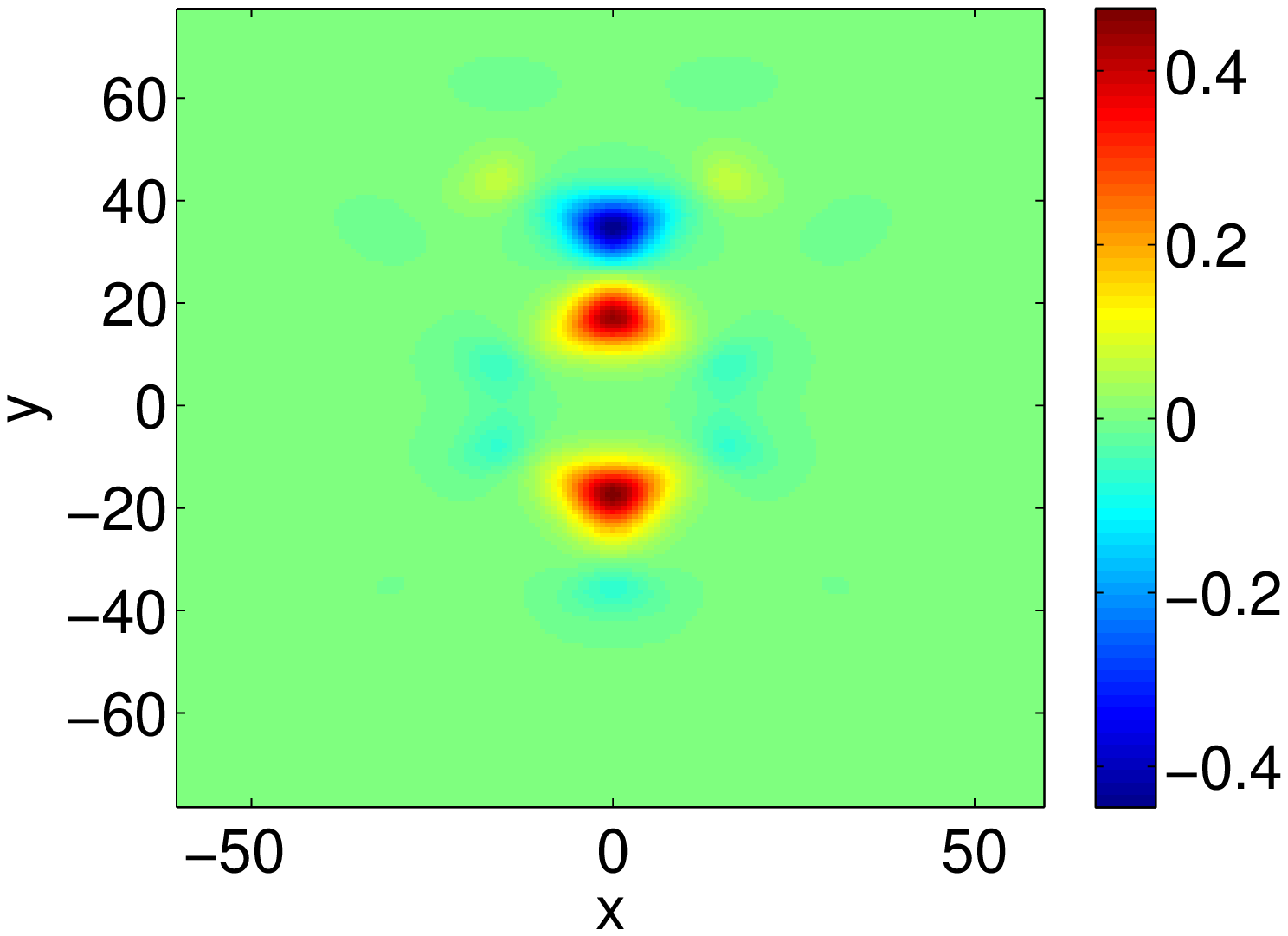}
\includegraphics[width=0.23\textwidth]{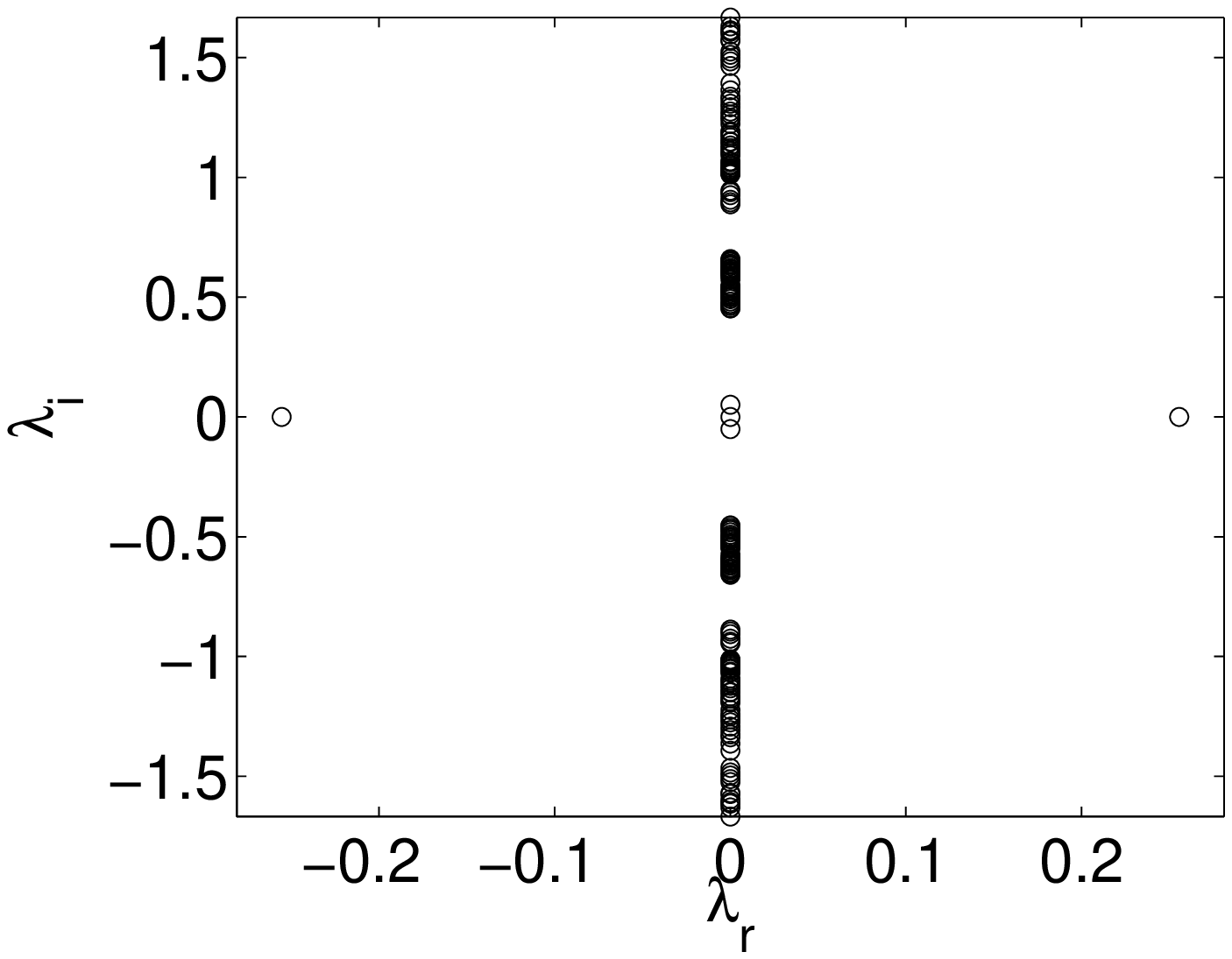}\\
\includegraphics[width=0.23\textwidth]{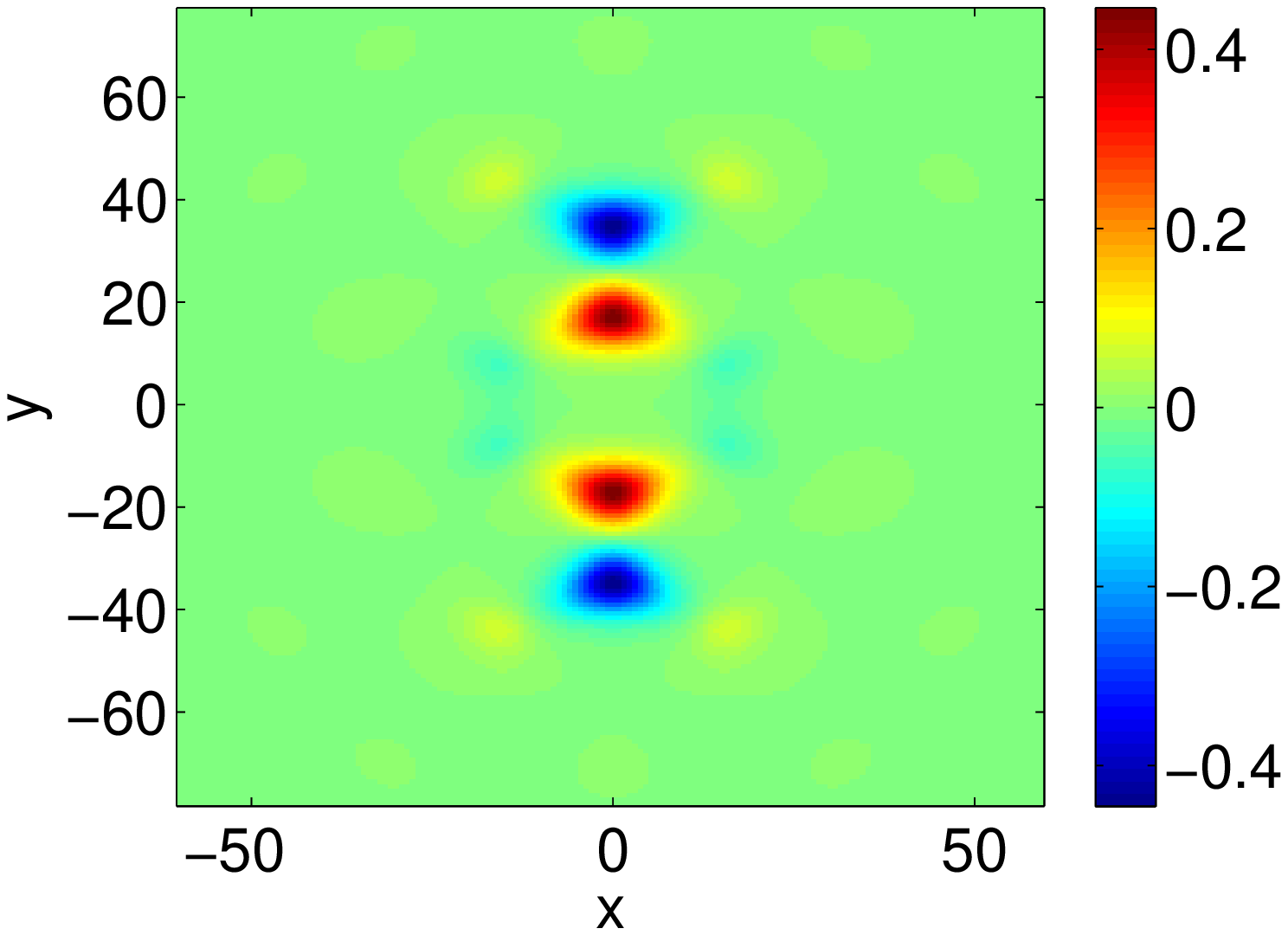}
\includegraphics[width=0.23\textwidth]{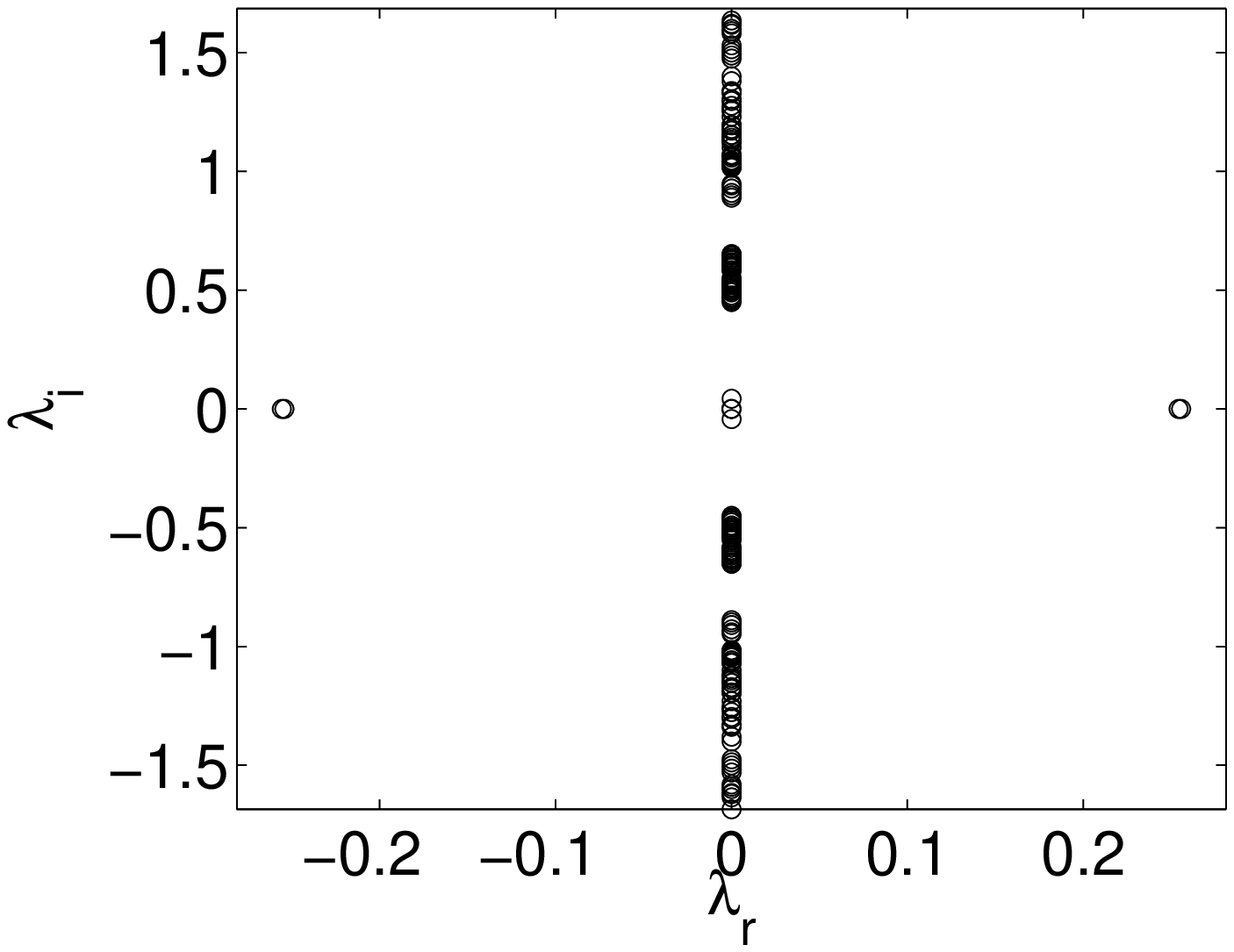}
\end{center}
\caption{(Color online) The top panels depict the largest
real part of the critical
eigenvalue, as well as the power and the peak intensity of
the IPO dipole solitons.
The panels in the second row show the profile $u$ and the
corresponding spectra in the complex plane of the dipole at $\mu=4.5$,
the third row shows the same images at the same value of
$\mu$ for the middle branch (dashed line) of the bifurcation diagram
and the bottom row is a
solution along the top branch (dash-dotted line) at the same value.}
\label{IPO}
\end{figure}

\begin{figure}[tbp!]
\begin{center}
\includegraphics[width=0.4\textwidth]{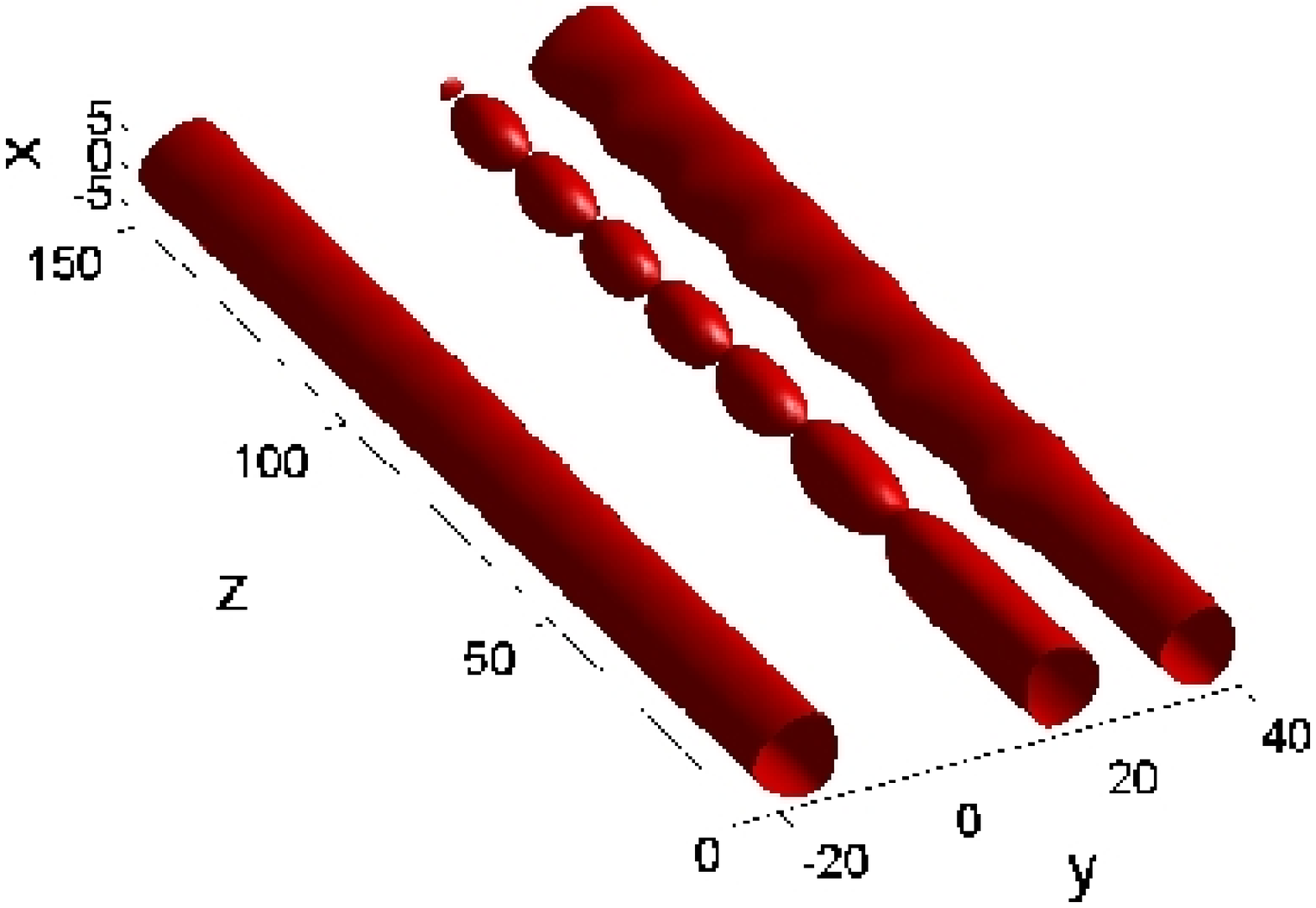}
\end{center}
\caption{The same figure as Fig.\ \ref{dyn_IPN}, but for the solution presented in the middle panel of Fig.\ \ref{IPO}.
Depicted is the isosurface of height $0.05$.}
\label{dyn_IPO}
\end{figure}

\begin{figure}[tbp!]
\begin{center}
\includegraphics[width=0.23\textwidth]{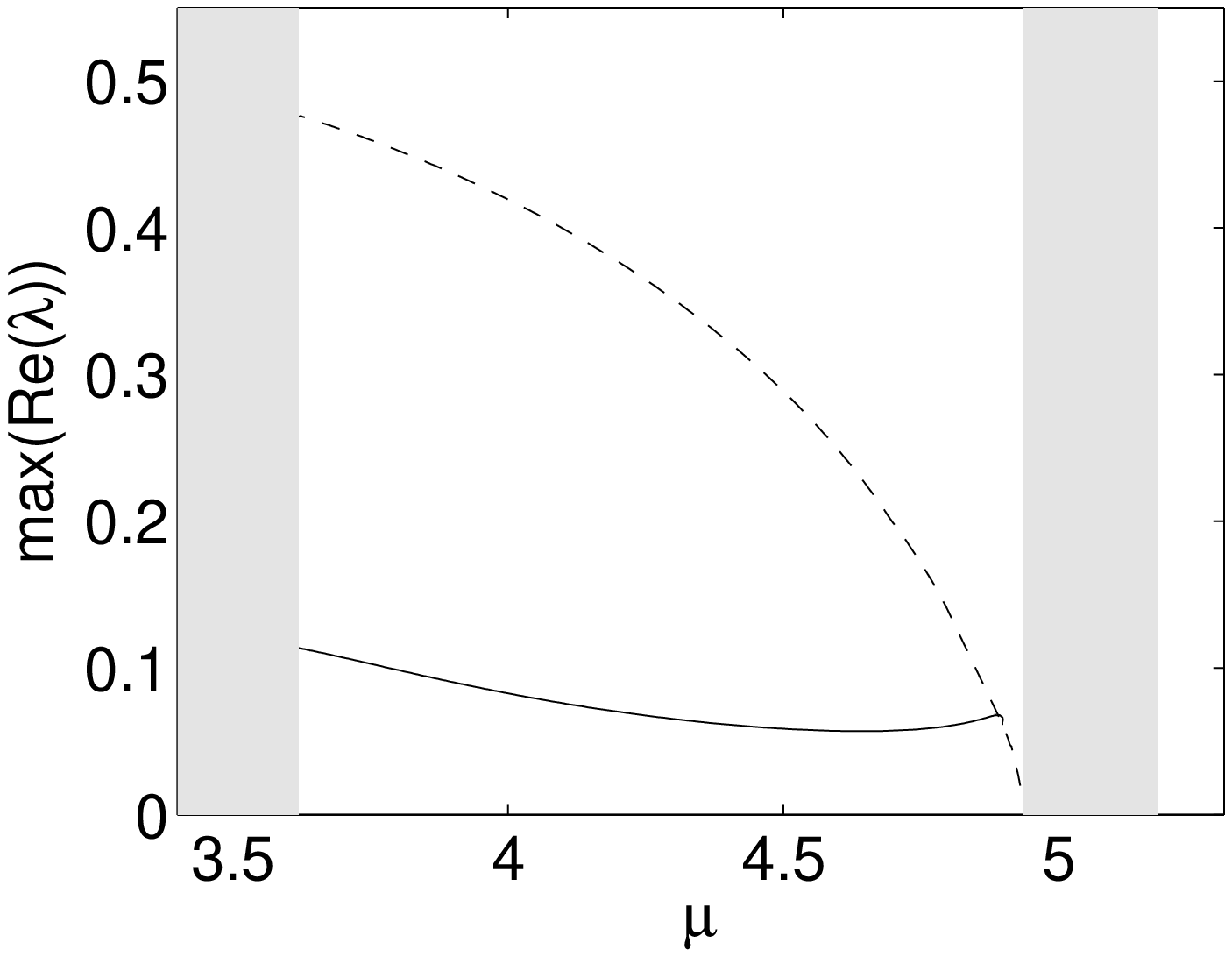}
\includegraphics[width=0.23\textwidth]{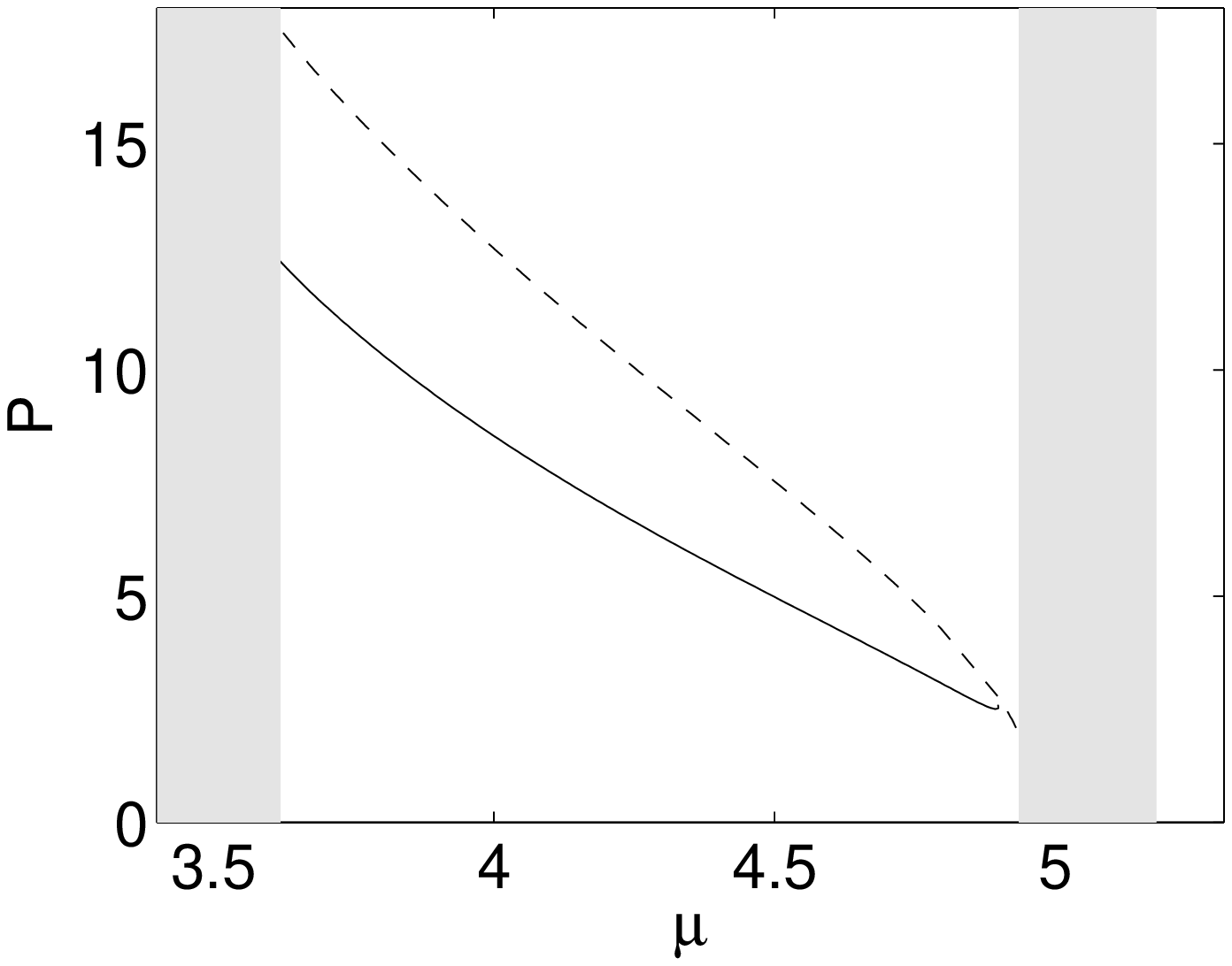}\\
\includegraphics[width=0.23\textwidth]{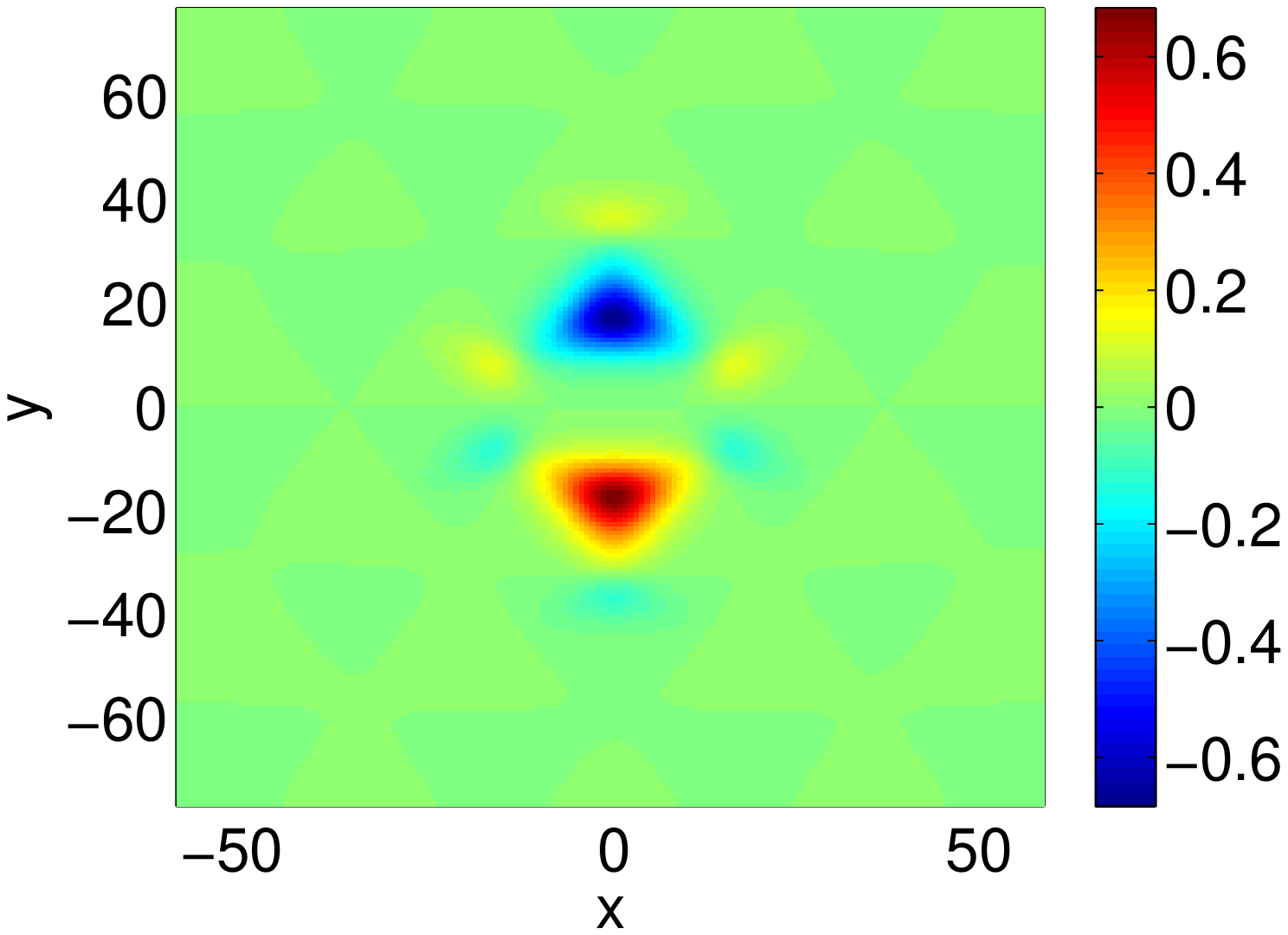}
\includegraphics[width=0.23\textwidth]{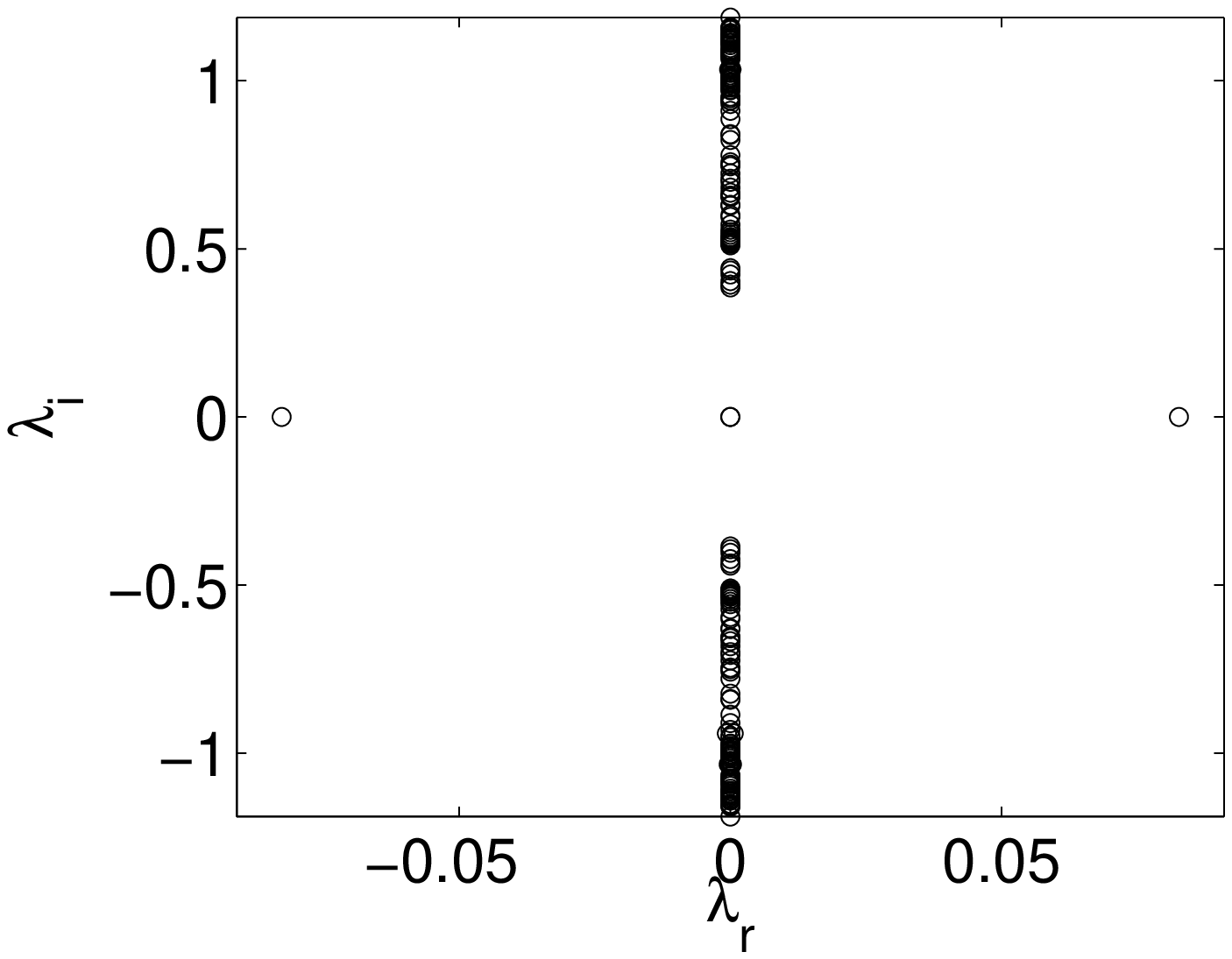}\\
\includegraphics[width=0.23\textwidth]{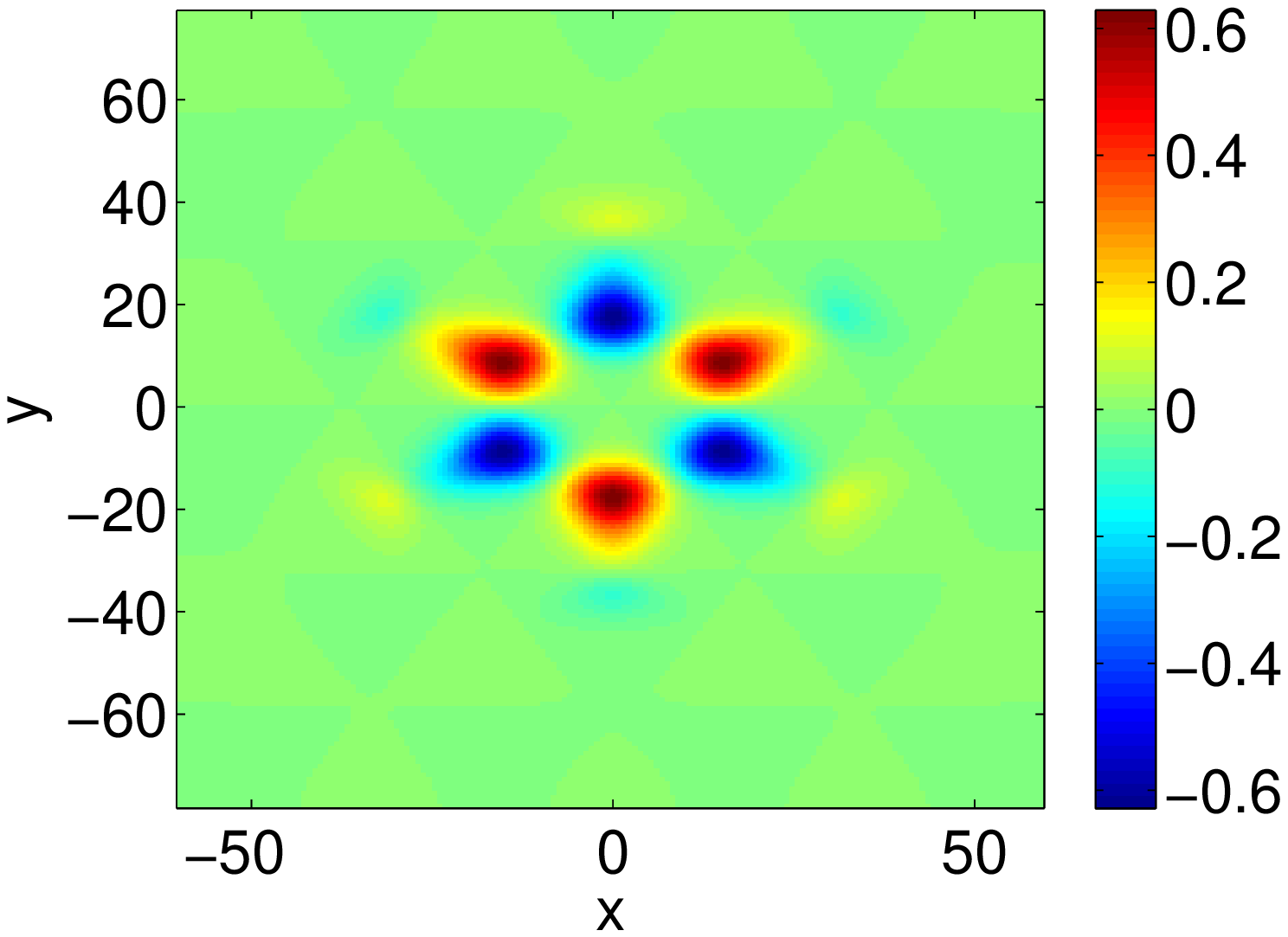}
\includegraphics[width=0.23\textwidth]{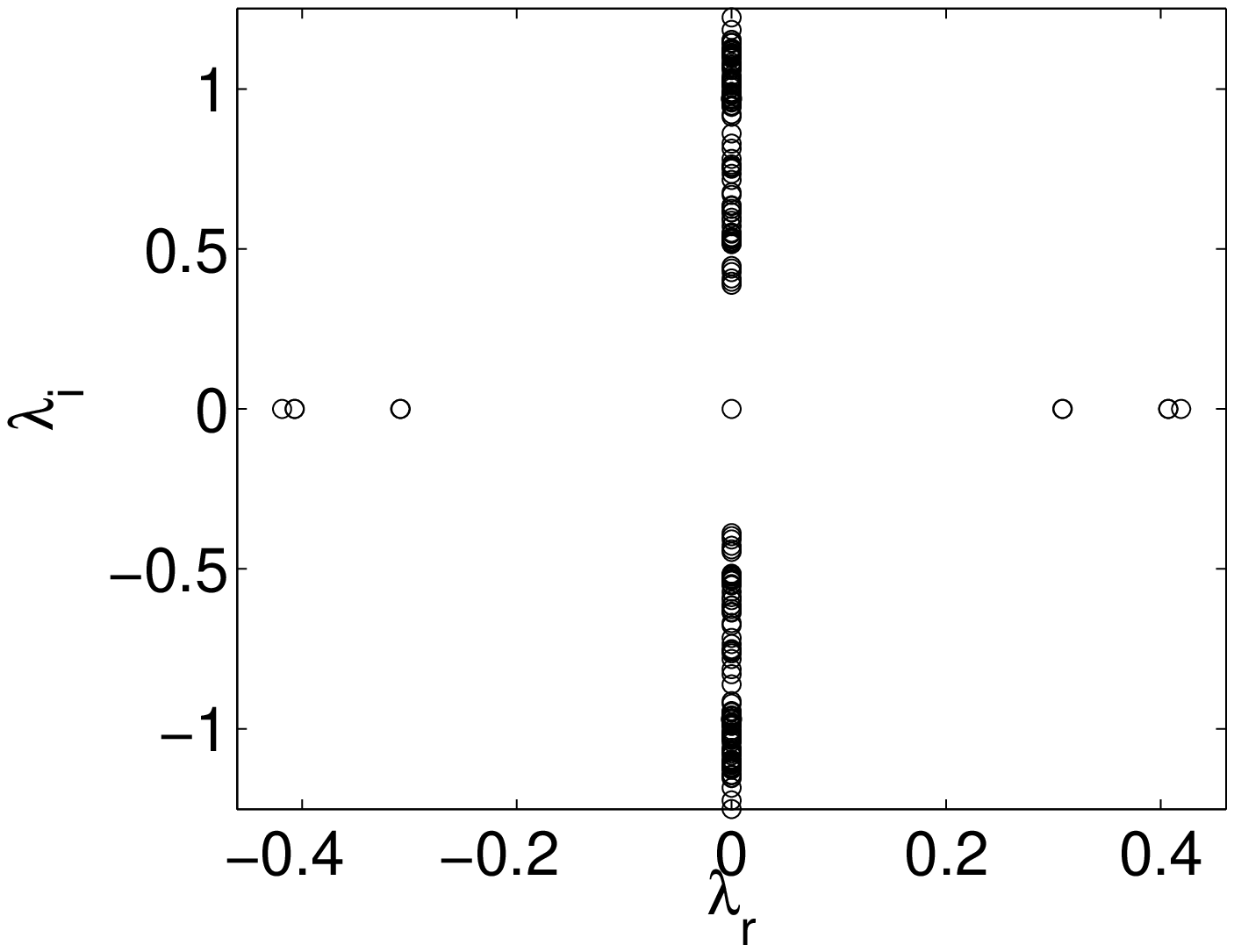}
\end{center}
\caption{(Color online) The top panels depict the largest
real part and the power of the OPO dipole solitons. The middle panels
again show the profile $u$ and spectra at $\mu=4$,
and the bottom is the more unstable saddle configuration, consisting
this time of a hexapole configuration constructed out of three such OPOs.}
\label{OPO}
\end{figure}

\begin{figure}[tbp!]
\begin{center}
\includegraphics[width=0.4\textwidth]{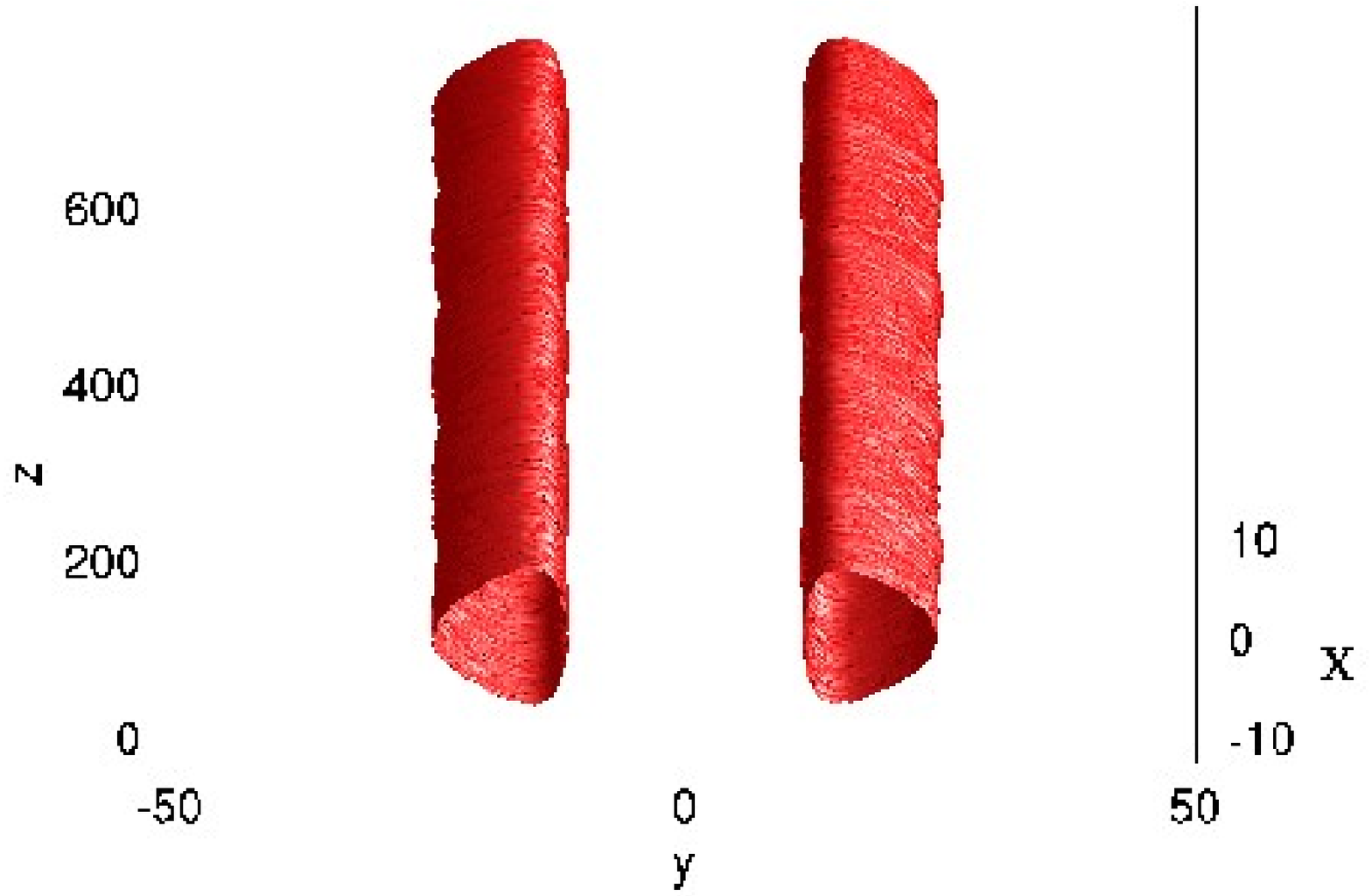}\\
\includegraphics[width=0.4\textwidth]{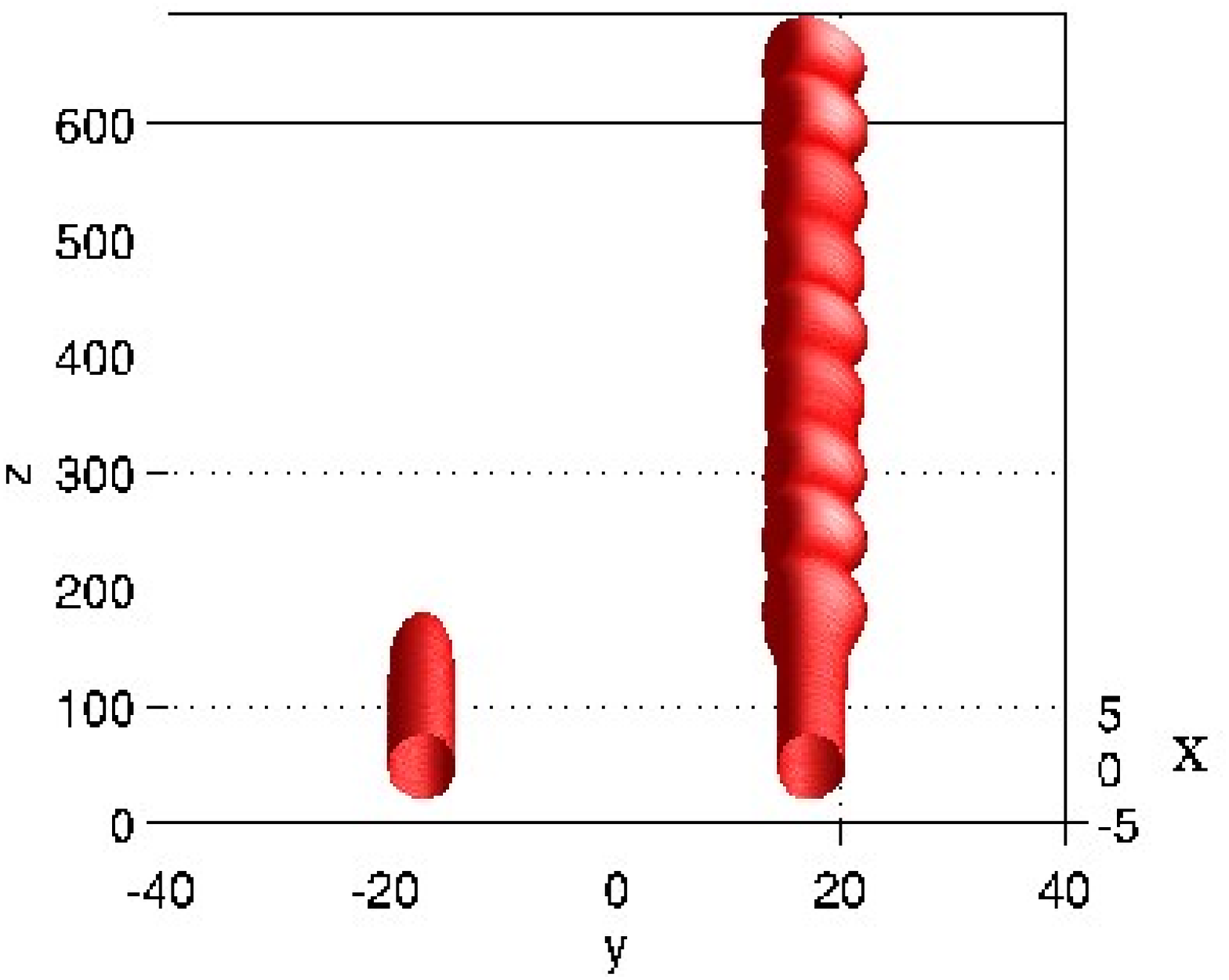}
\end{center}
\caption{The top panel is the same as Fig.\ \ref{dyn_IPN}, but for the 
solution presented in the middle panel of Fig.\ \ref{OPO}, 
with isosurface of height $0.1$.  On the other hand, the bottom panel 
illustrates again (c.f. Fig. \ \ref{dyn_OPN}) 
that the 
linear stability analysis is more predictive of
the nonlinear dynamics for 
the solution with the larger value of 
$\mu=4.88$ (and accordingly smaller amplitude) close to the 
intersection with the extended OP quadrupole branch (the isosurface is 
taken at half the maximum of the initial intensity amplitude). The 
growth rates for each solution are comparable, while the 
dynamical evolutions differ drastically.}
\label{dyn_OPO}
\end{figure}

We now address opposite (O)
dipole solitons
residing at the two sites along a diameter of a
local maximum of the lattice.  This is
the final type of dipole configuration for a
symmetric triangular lattice, exhausting
the possibilities up to phase and rotational invariances.
Again, we partition our considerations into in-phase and out-of-phase
cases.

\subsection{In Phase Opposite Dipole Solitons}

We have found in-phase opposite (IPO)
solitons
throughout the first gap in the linear
spectrum. Our numerical findings are presented in Fig.\ \ref{IPO}.


Again, the solution branch is largely stable with
small windows of Hopf quartets and again a
saddle node bifurcation occurs as the branch approaches
the first spectral band.  Also, interestingly,
the configuration 
with which this branch collides when it disappears
resembles an OPN (or two pairs of OPNs-- see the third and fourth row
of the figure). The latter branches are naturally unstable due to
one (or more) real pair of eigenvalues.

The dynamics of one of the bifurcating solutions,
i.e.\ the configuration with a single OPN structure,
is presented in Fig.\ \ref{dyn_IPO},
where one can see that, as usual,
only the pair of out-of-phase
nearest neighbor dipole interacts, while the other
soliton is almost uninfluenced.

Using the same reasoning, one can deduce
as well that the dynamics of the other
bifurcating solution, presented
in the bottom panel of Fig.\ \ref{IPO},
will be similar, except the fact that
now there are two pairs of OPN interacting
among themselves.

\subsection{Out of Phase Opposite Solitons}

Lastly, as regards dipoles,  we consider
the out of phase opposite (OPO)
dipole.
The first interesting characteristic
of the OPO is its strong instability stemming from a real pair of eigenvalues,
seen in the top left and middle rows of Fig. \ \ref{OPO}.
Once again the direct instability of this mode follows from
our theoretical considerations of Section II.
On the other hand, the figure also reveals an interesting
bifurcation structure in this case.  The branch actually merges
with a hexapole made of three copies of itself close to the band, when
solutions start becoming extended.  This hexapole then intersects with the
linear spectrum shortly thereafter and the solution transforms itself
into a  fully
extended ``checkerboard"-like configuration of all adjacent wells excited
out-of-phase.
As can be seen in the top left and the bottom right of Fig.\ \ref{OPO},
the hexapole configuration is significantly more unstable, possessing
five real eigenvalue pairs.


We have numerically monitored the full evolution to observe
the dynamics of the unstable OPO dipoles. It is interesting to note that
even though the state has a pair of real eigenvalues, our simulation reveals
that the
instability is 
barely detectable for the state depicted in the middle rows of 
Fig. \ \ref{OPO}, presumably because of the
spatial separation of the populated sites (top row of Figure \ \ref{dyn_OPO}); 
the solution oscillations are very mild (and almost indetectable) 
between similar structures with mass concentrated 
in one site or another.  On the other hand, for significantly
smaller power (larger $\mu$) as seen in the bottom panel of 
Figure \ \ref{dyn_OPO}, one site decays fairly rapidly and a robust
single site remains.

\begin{figure}[tbp!]
\begin{center}
\includegraphics[width=0.4\textwidth]{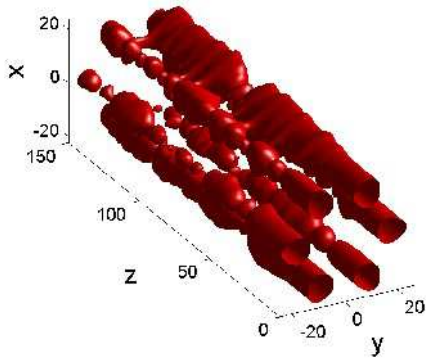}
\end{center}
\caption{The same dynamical evolution
figure as Fig.\ \ref{dyn_IPN}, but for the
out-of-phase hexapole depicted in the bottom panel of Fig.\ \ref{OPO}.
Depicted is the isosurface of height $0.1$.}
\label{dyn_OOP_nck}
\end{figure}

\begin{figure}[tbp!]
\begin{center}
\includegraphics[width=0.23\textwidth]{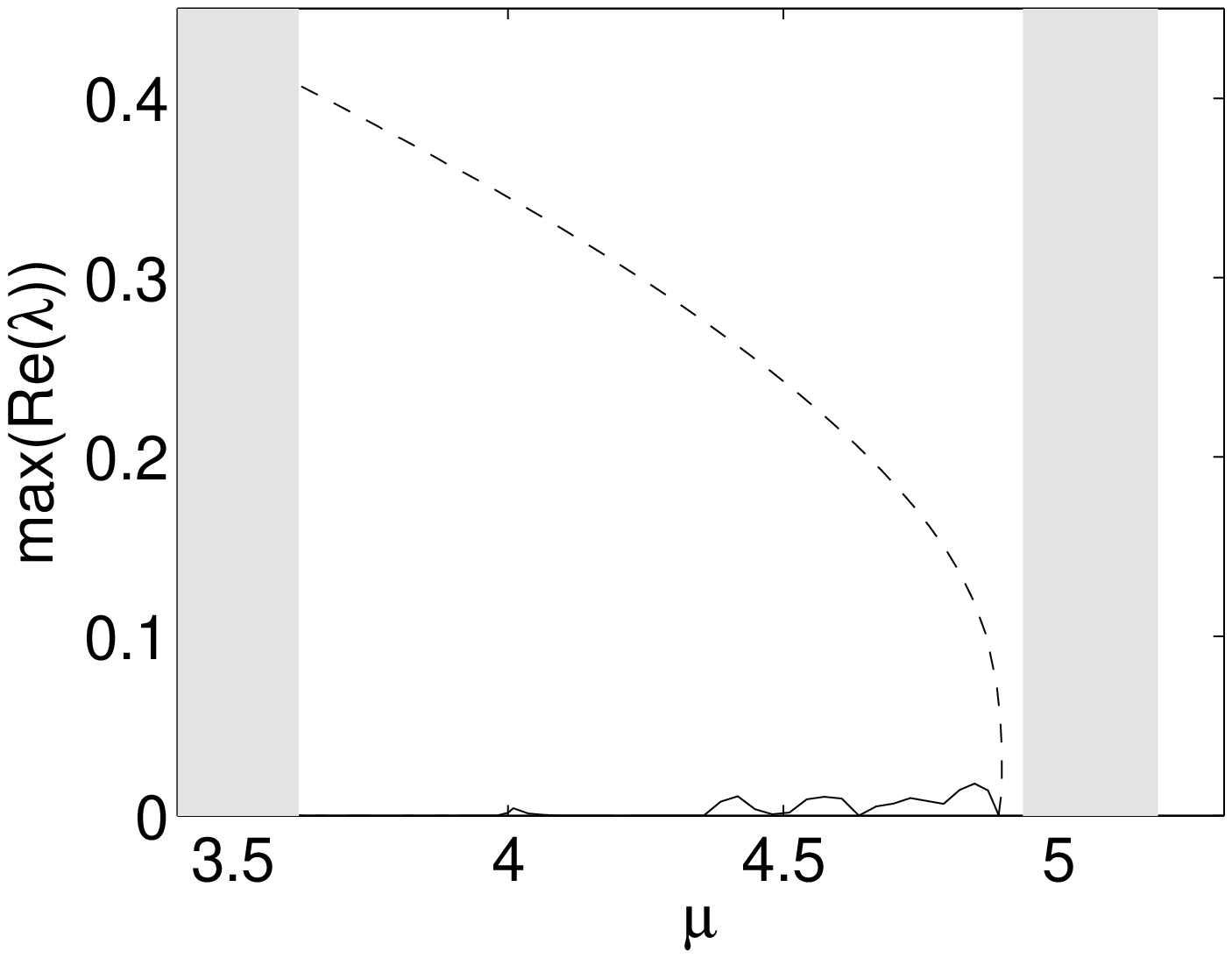}
\includegraphics[width=0.23\textwidth]{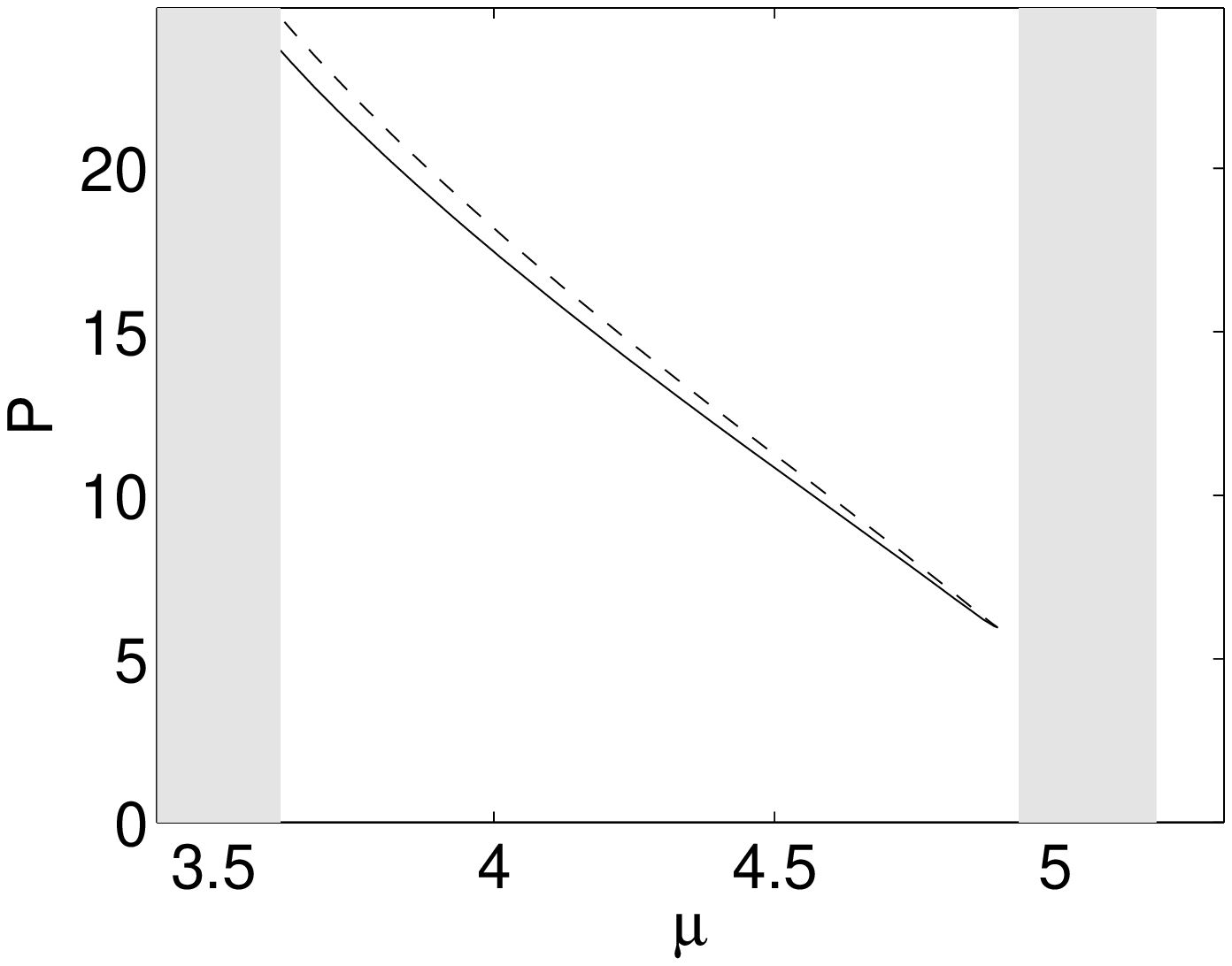}\\
\includegraphics[width=0.23\textwidth]{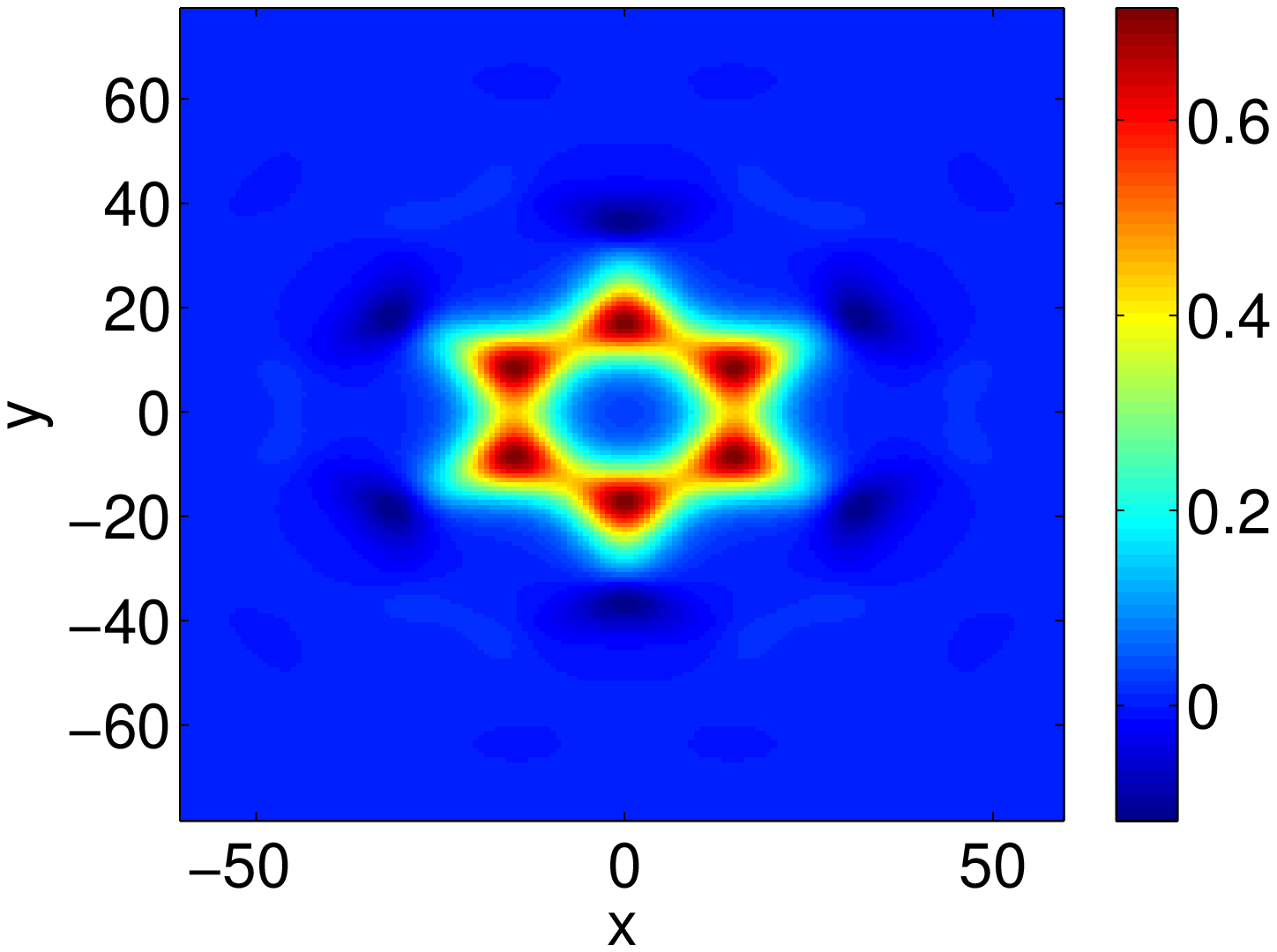}
\includegraphics[width=0.23\textwidth]{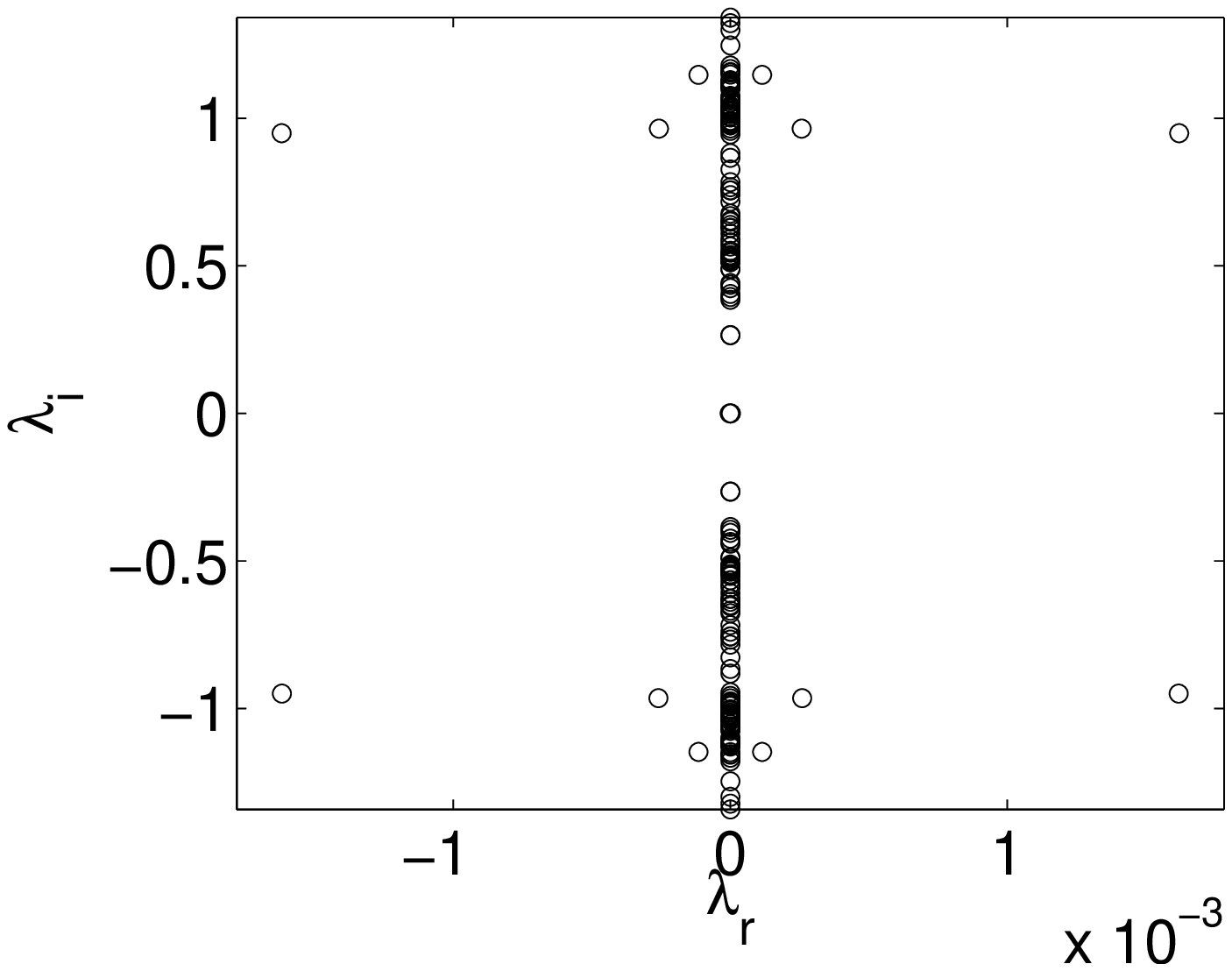}\\
\includegraphics[width=0.23\textwidth]{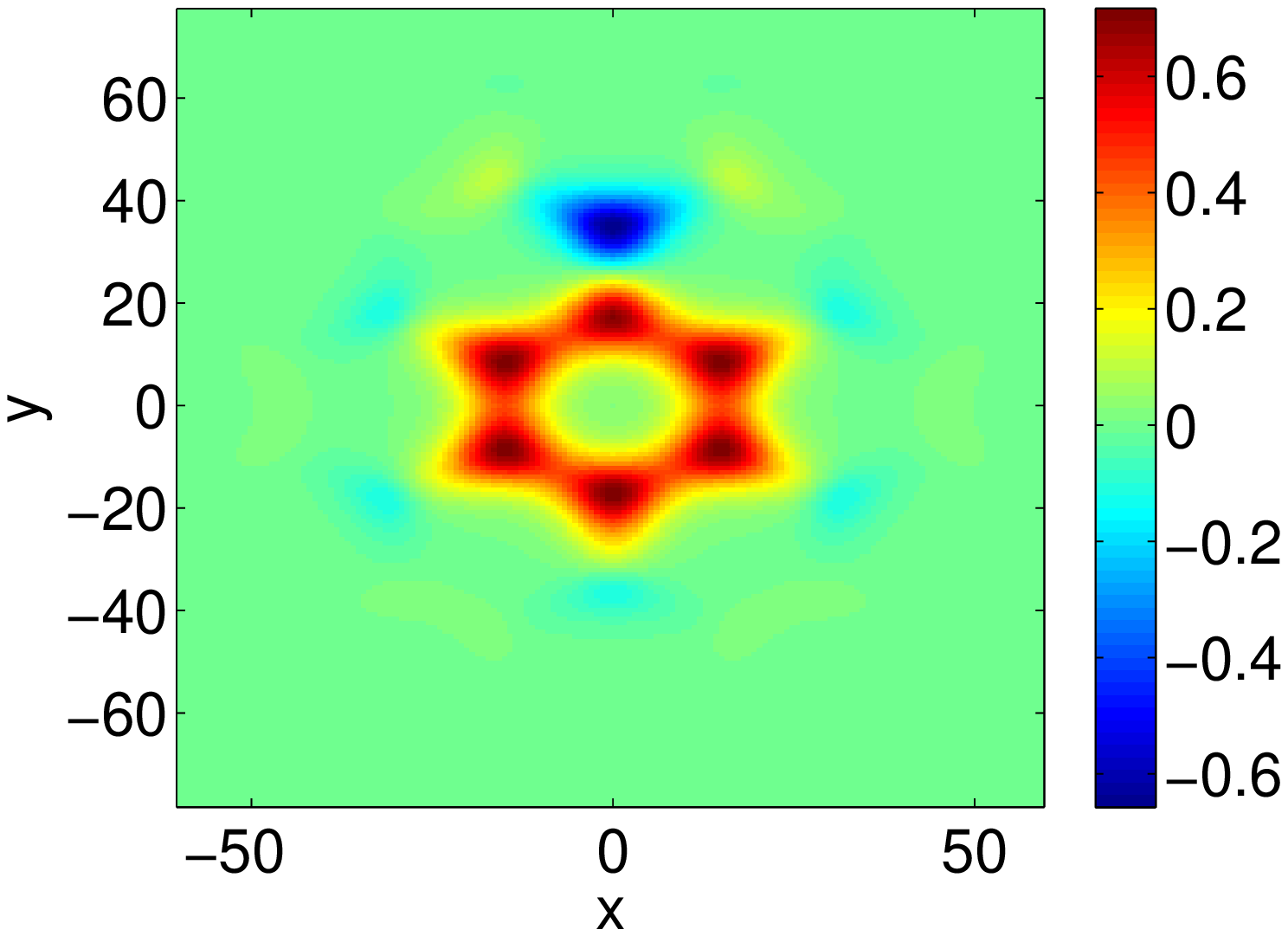}
\includegraphics[width=0.23\textwidth]{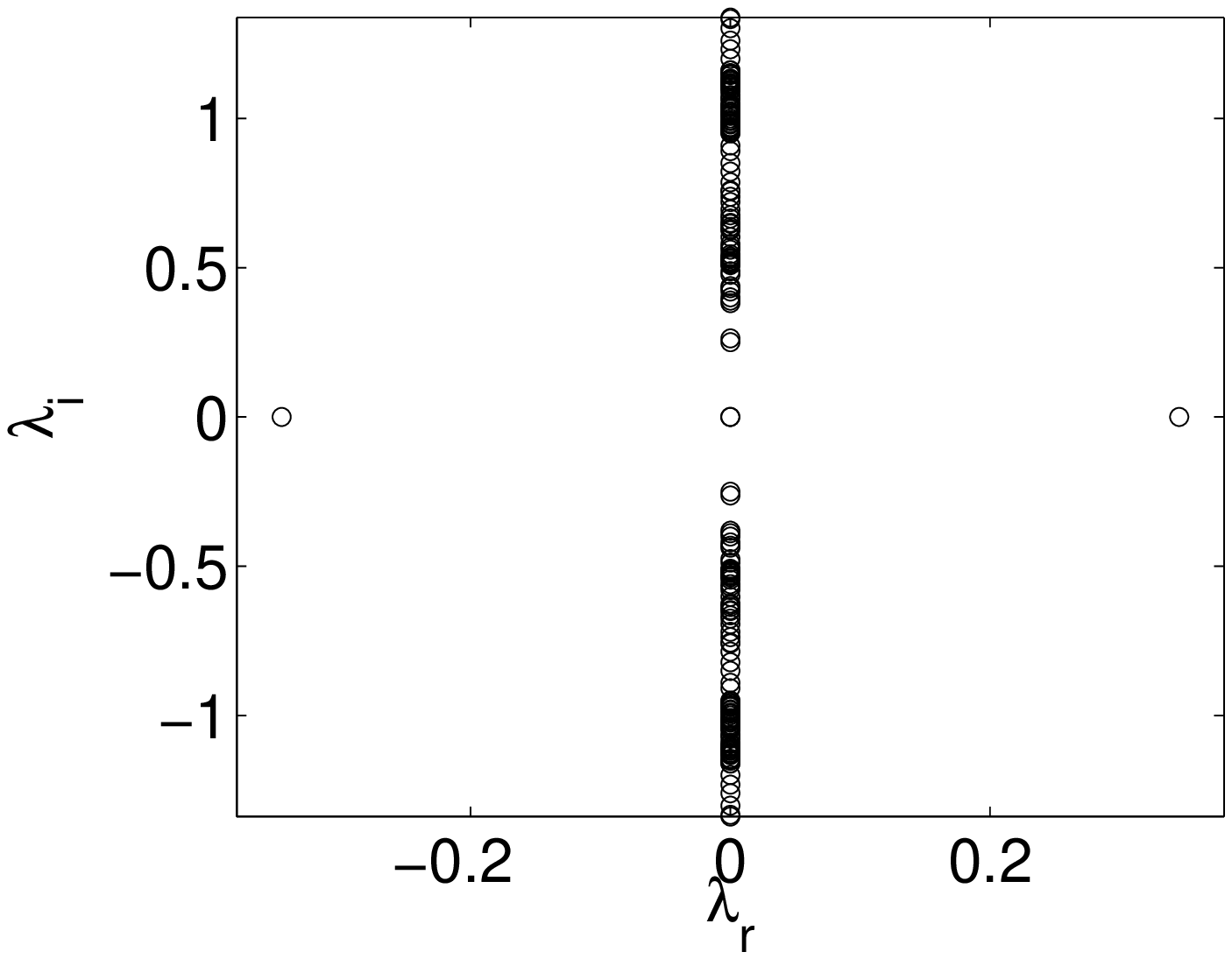}
\end{center}
\caption{(Color online) The top panels depict the largest
real part and the power of the IP hexapole solitons. The middle panels
again show the profile $u$ and spectra at $\mu=4$,
and the bottom is the more unstable saddle configuration,which features
our expected OPN sidekick.}
\label{HEX}
\end{figure}

\begin{figure}[tbp!]
\begin{center}
\includegraphics[width=0.4\textwidth]{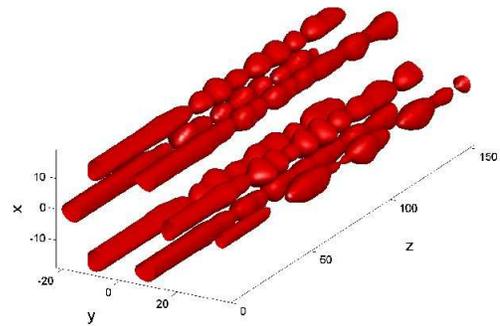}
\end{center}
\caption{The same figure as Fig.\ \ref{dyn_IPN}, but for the in-phase hexapole depicted in the bottom panel of Fig.\ \ref{HEX}.
Depicted is the isosurface of height $0.3$.}
\label{dyn_IP_nck}
\end{figure}

Regarding the bifurcating solution, which is an
out-of-phase hexapole, we will explore it as well as
the other hexapole configurations in more detail
in the following section.

\section{Hexapole Solitons and Vortex Necklaces}

\begin{figure*}[tbp!]
\begin{center}
\includegraphics[width=0.3\textwidth]{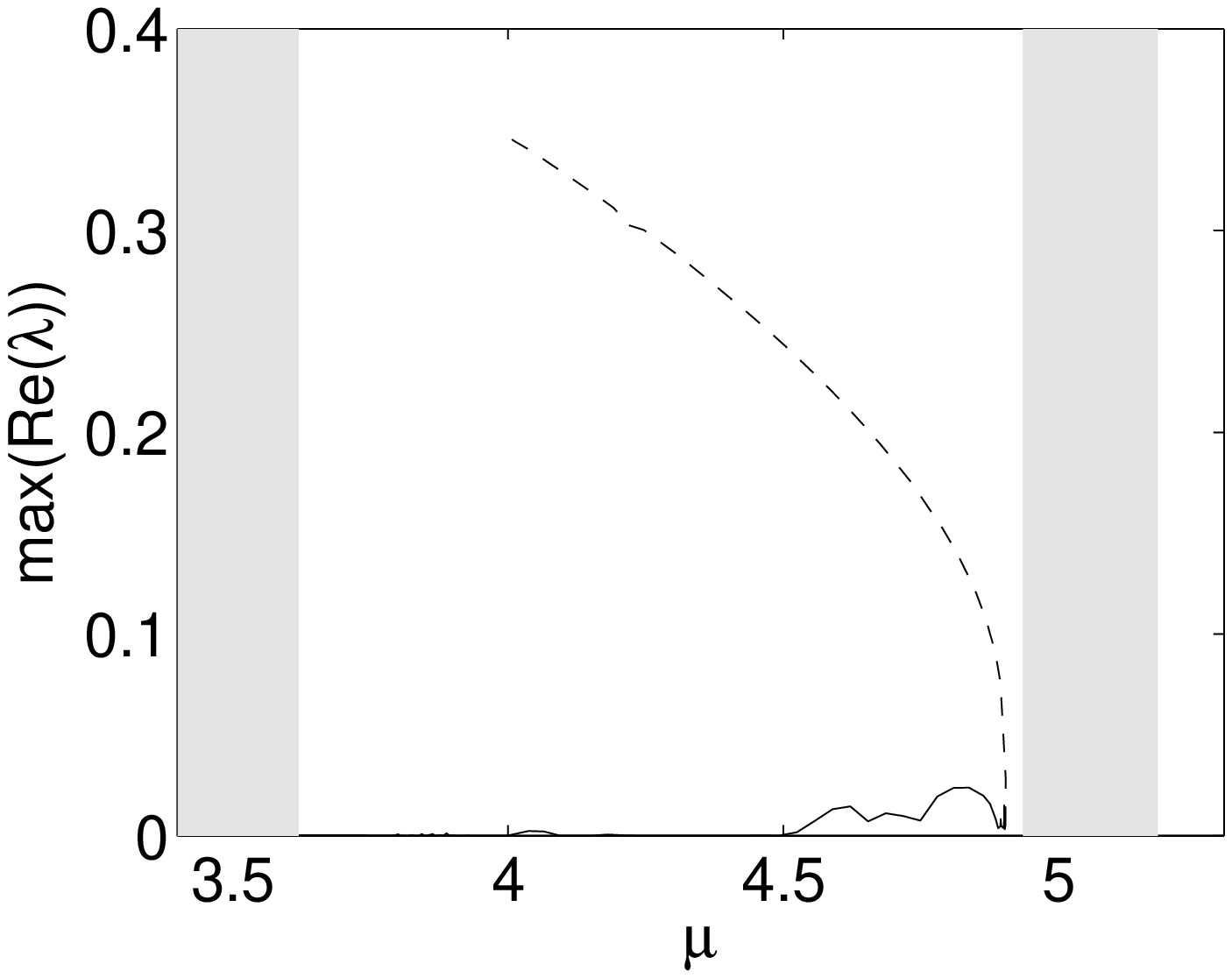}
\includegraphics[width=0.3\textwidth]{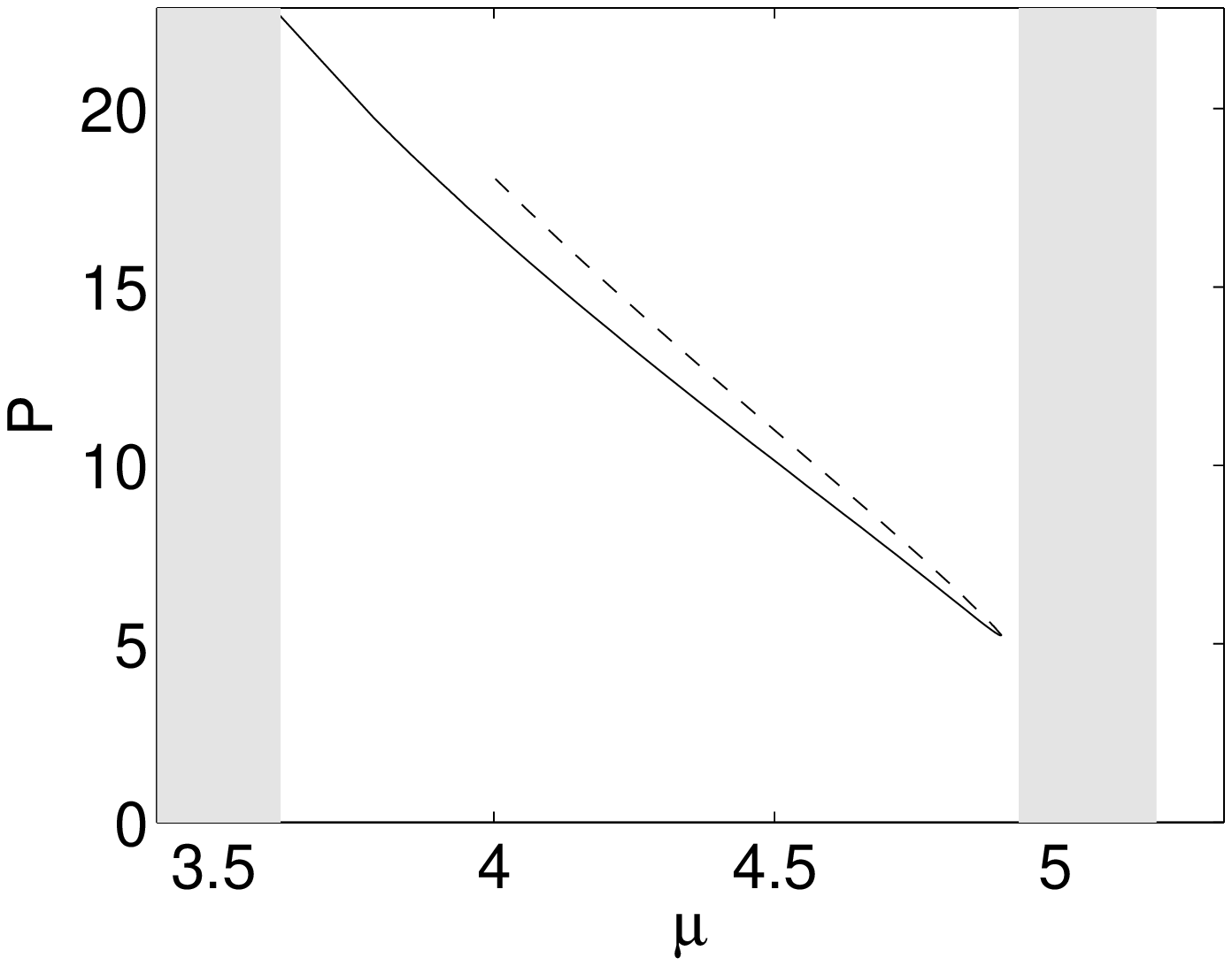}\\
\includegraphics[width=0.3\textwidth]{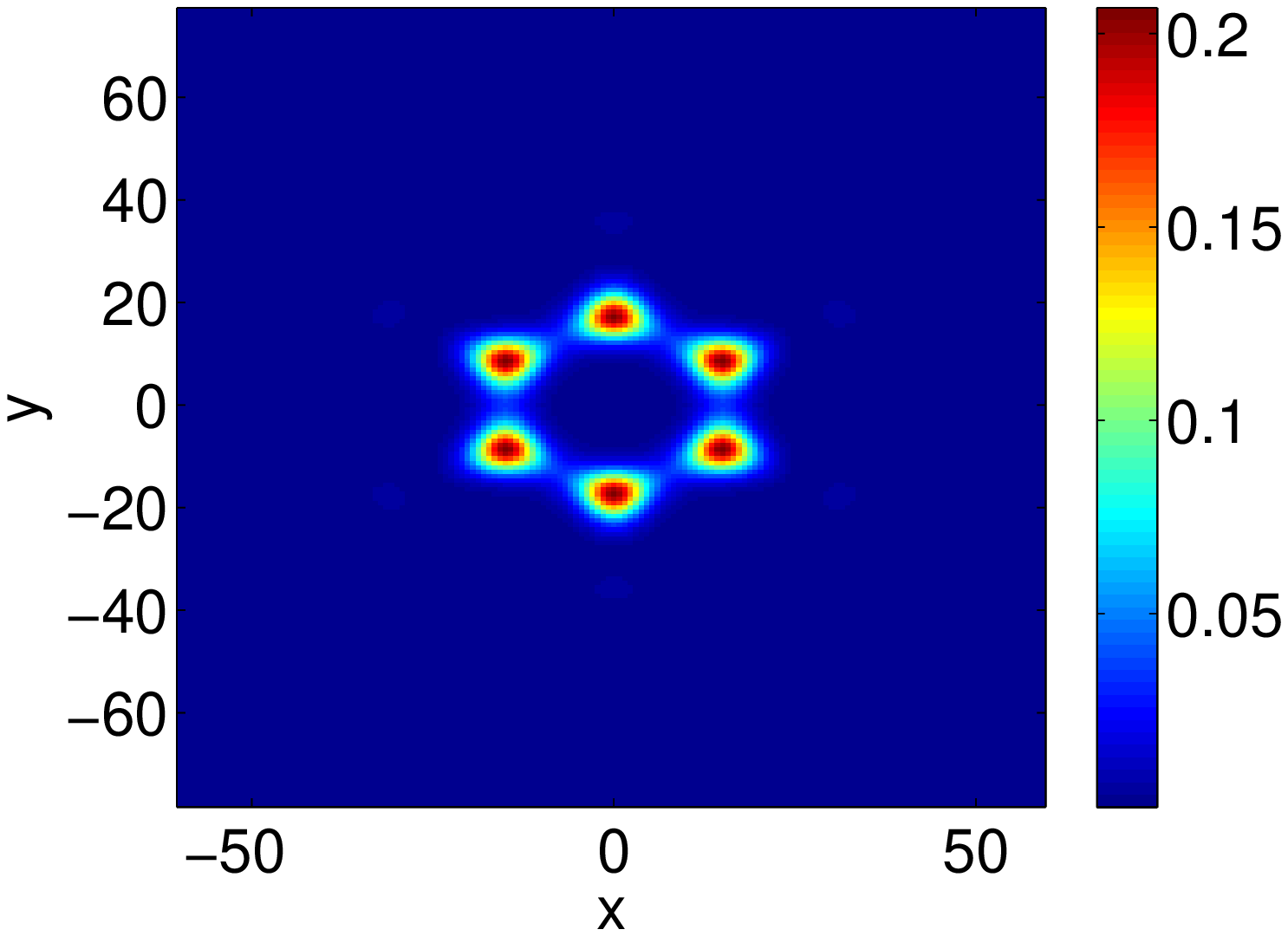}
\includegraphics[width=0.3\textwidth]{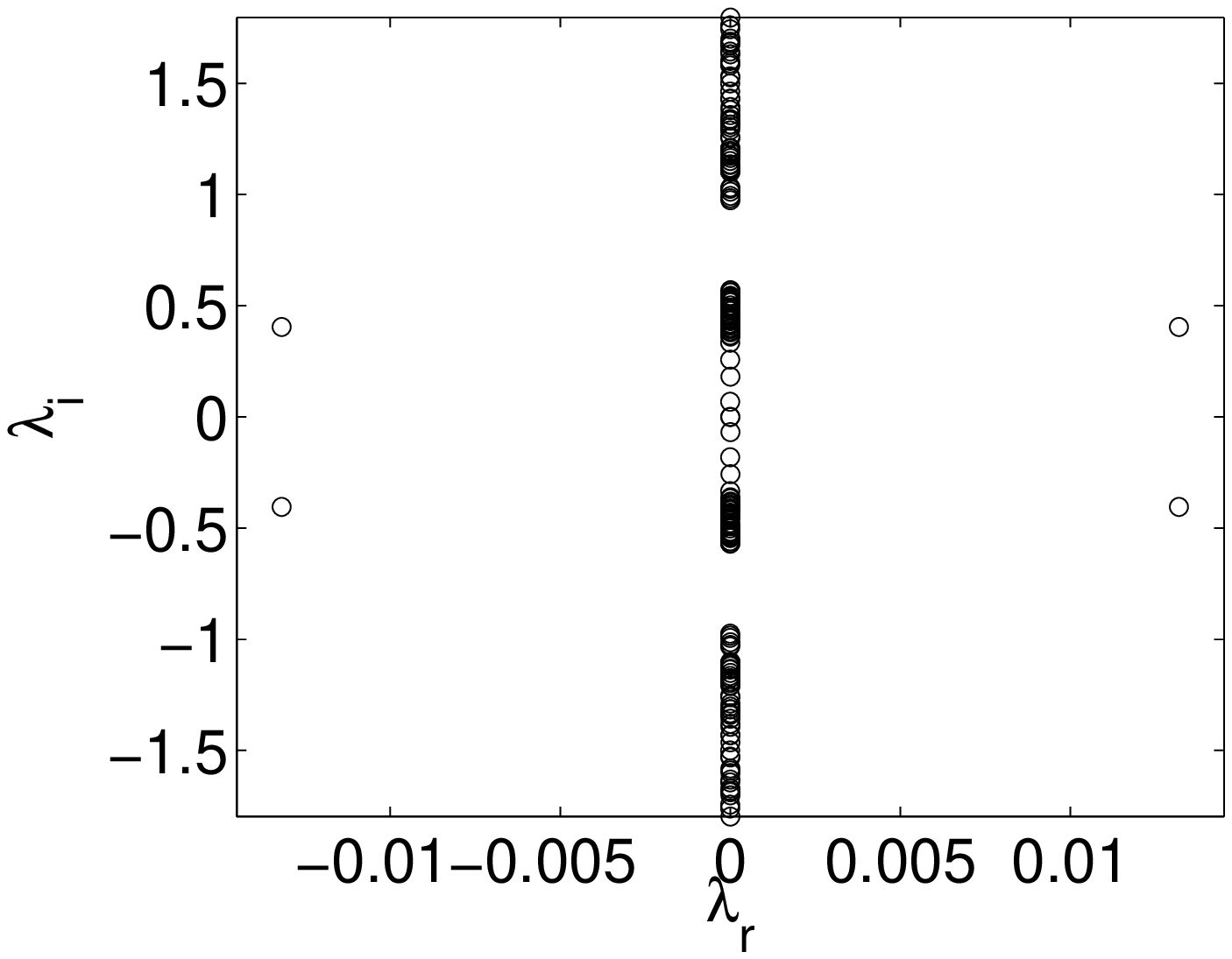}\\
\includegraphics[width=0.3\textwidth]{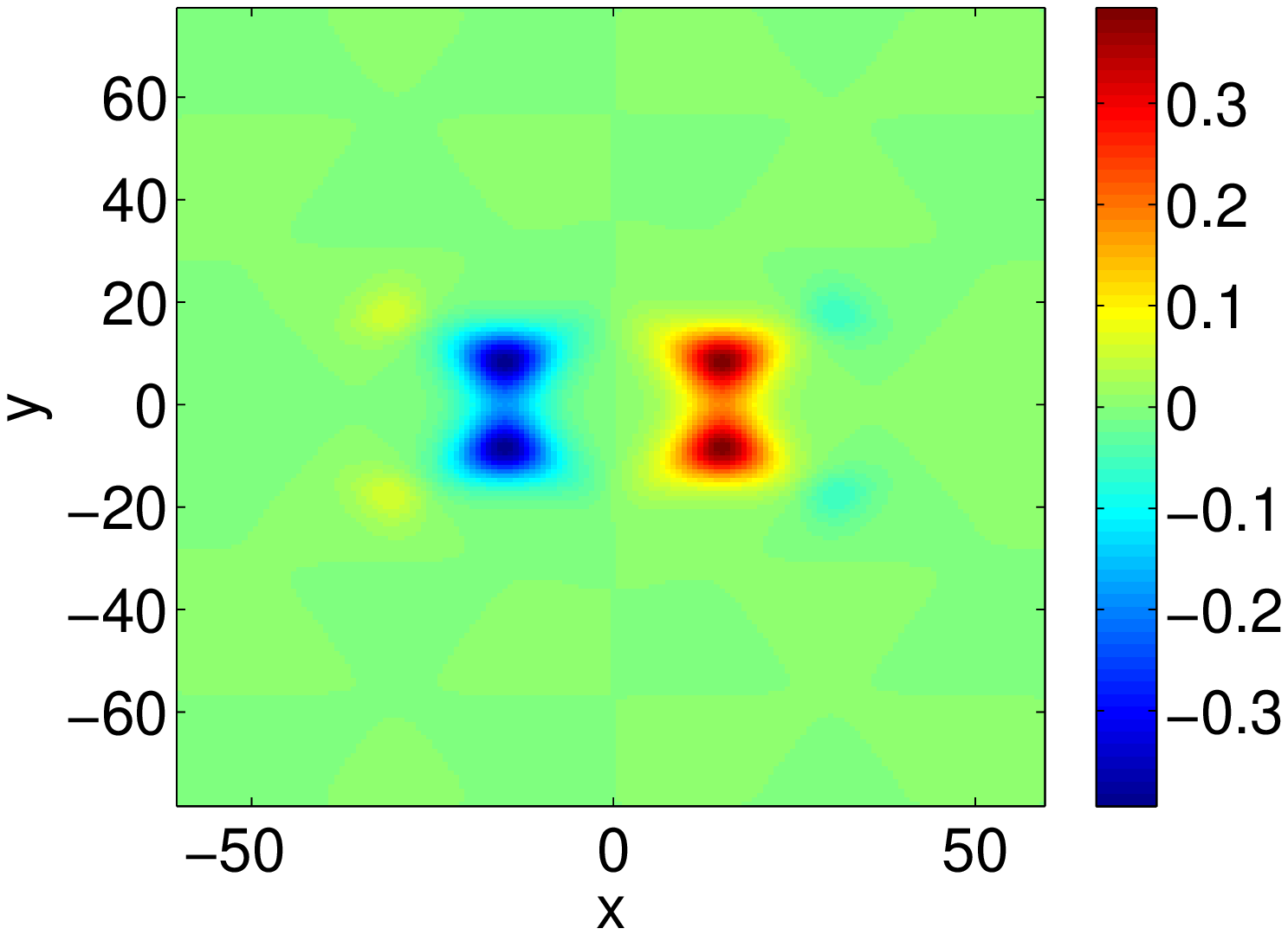}
\includegraphics[width=0.3\textwidth]{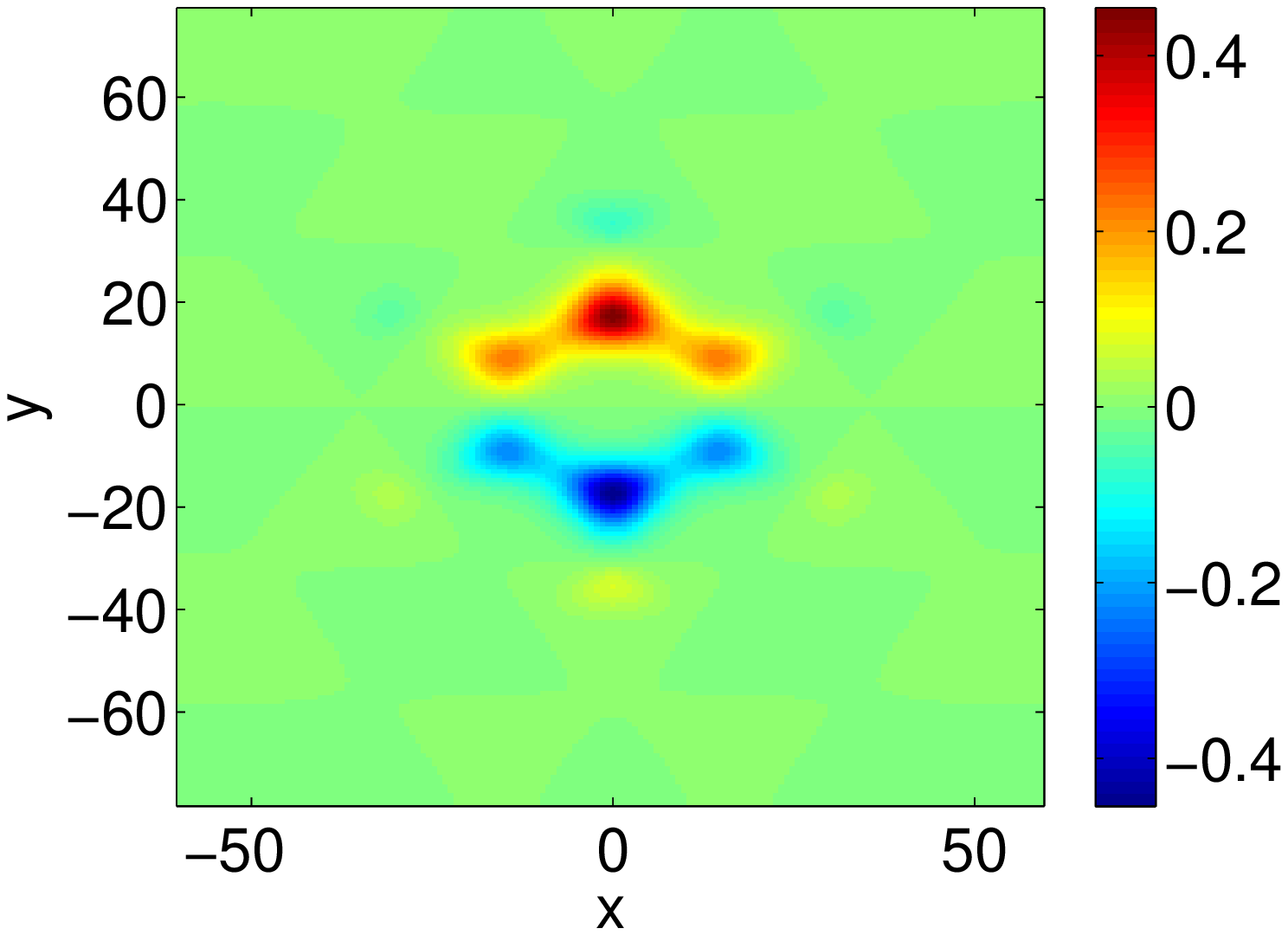}
\includegraphics[width=0.3\textwidth]{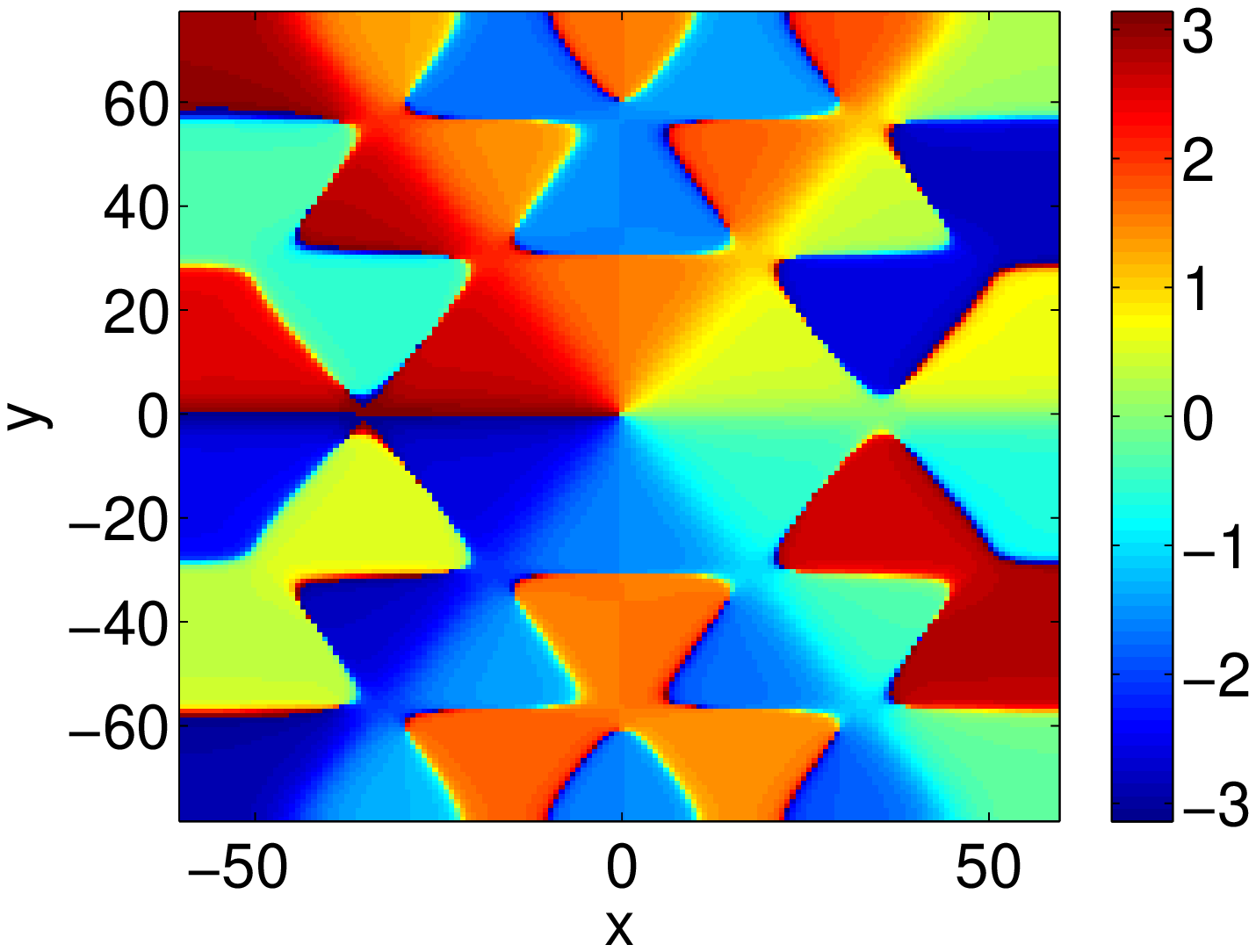}\\
\includegraphics[width=0.3\textwidth]{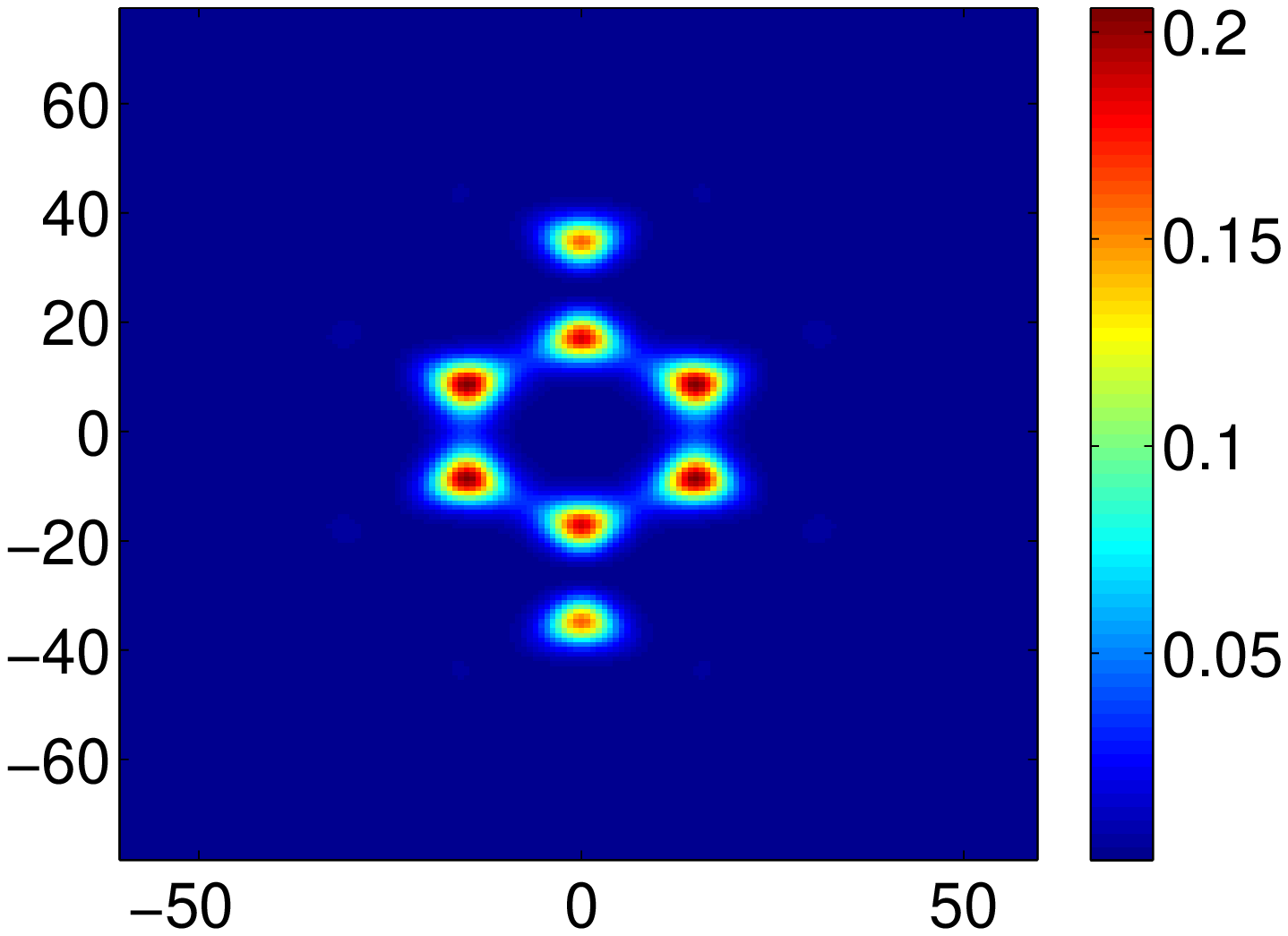}
\includegraphics[width=0.3\textwidth]{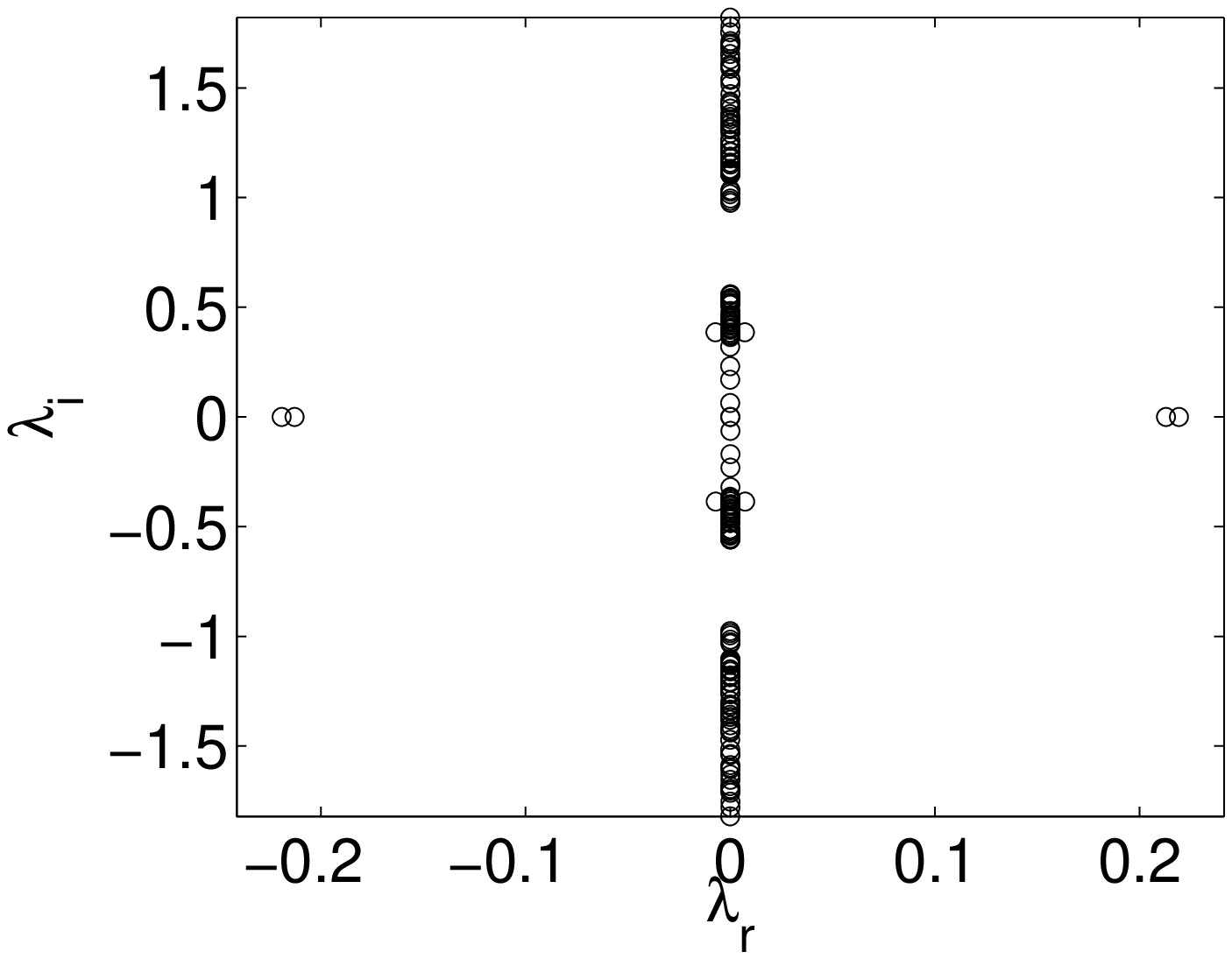}\\
\includegraphics[width=0.3\textwidth]{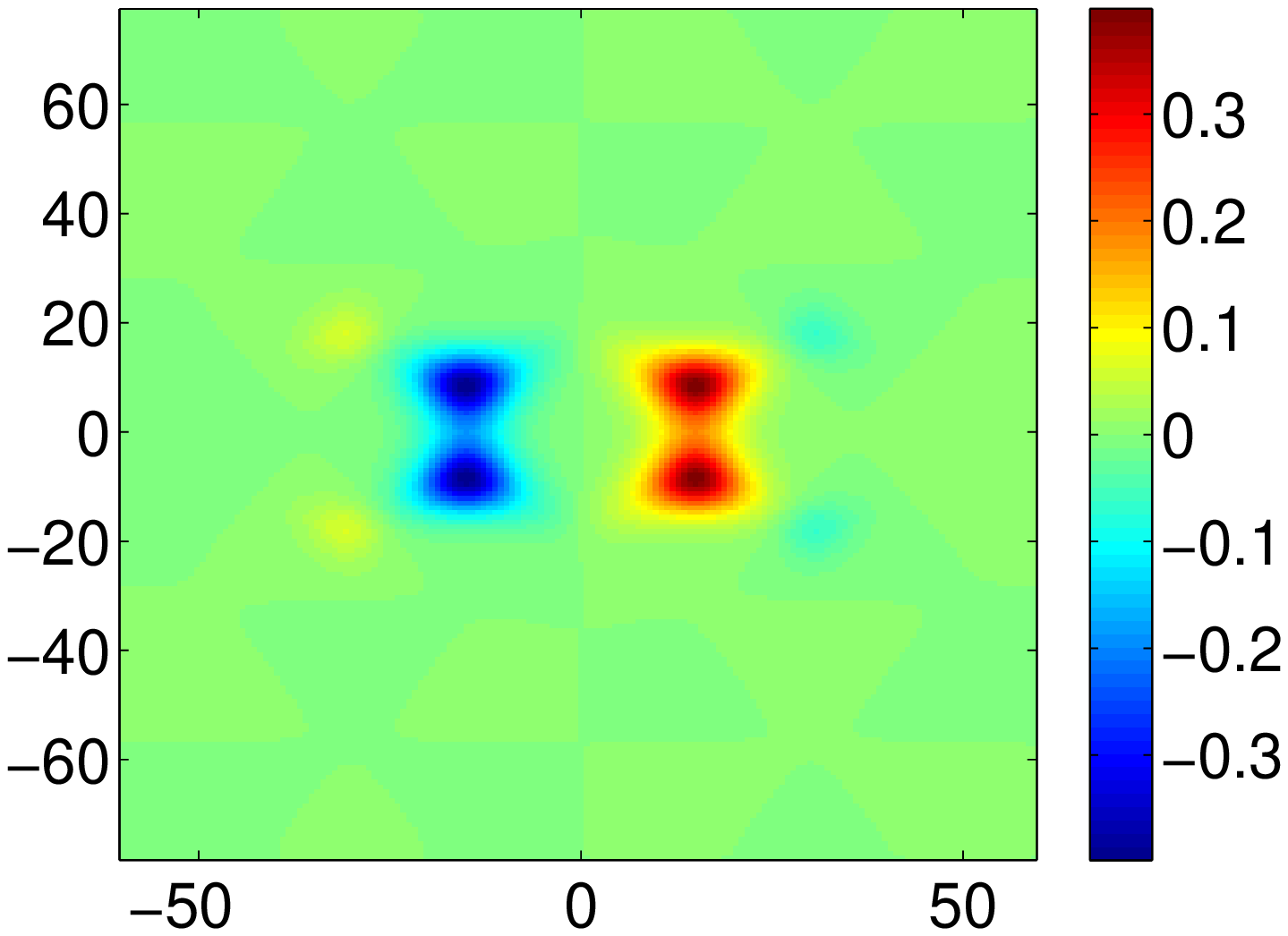}
\includegraphics[width=0.3\textwidth]{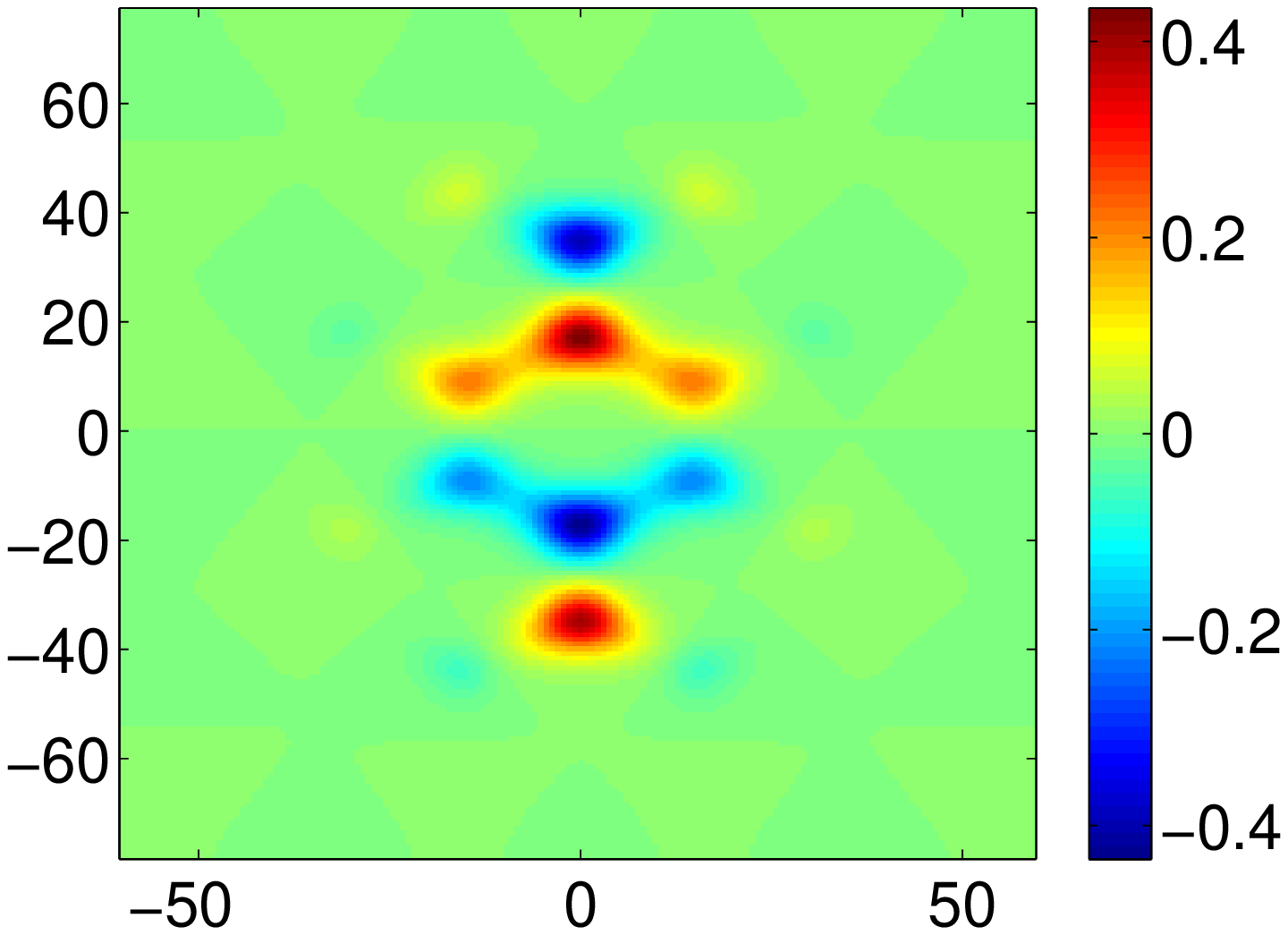}
\includegraphics[width=0.3\textwidth]{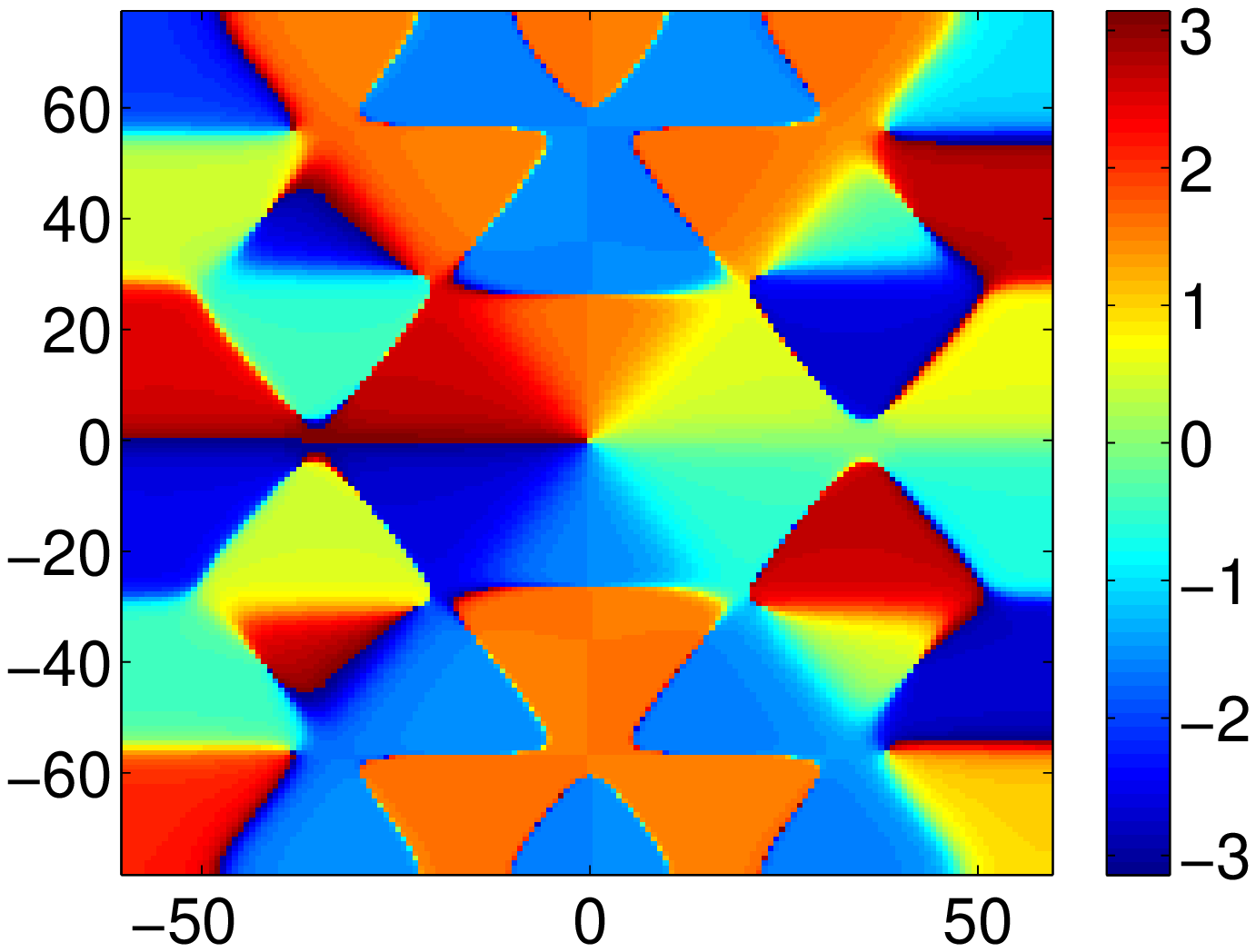}
\end{center}
\caption{(Color online) The top panels depict the largest
real part and the power of the vortex necklace hexapole configuration.
The second row depicts the modulus of the solution and corresponding
spectrum when $\mu=4.6$.  The third row illustrates the real and
imaginary components of the field and the phase (from left to right).
The fourth and fifth rows show the same properties as the second and
third but for the unstable eight-site configuration of the dashed line in the
top panels.}
\label{VTX}
\end{figure*}

\begin{figure}[tbp!]
\begin{center}
\includegraphics[width=0.4\textwidth]{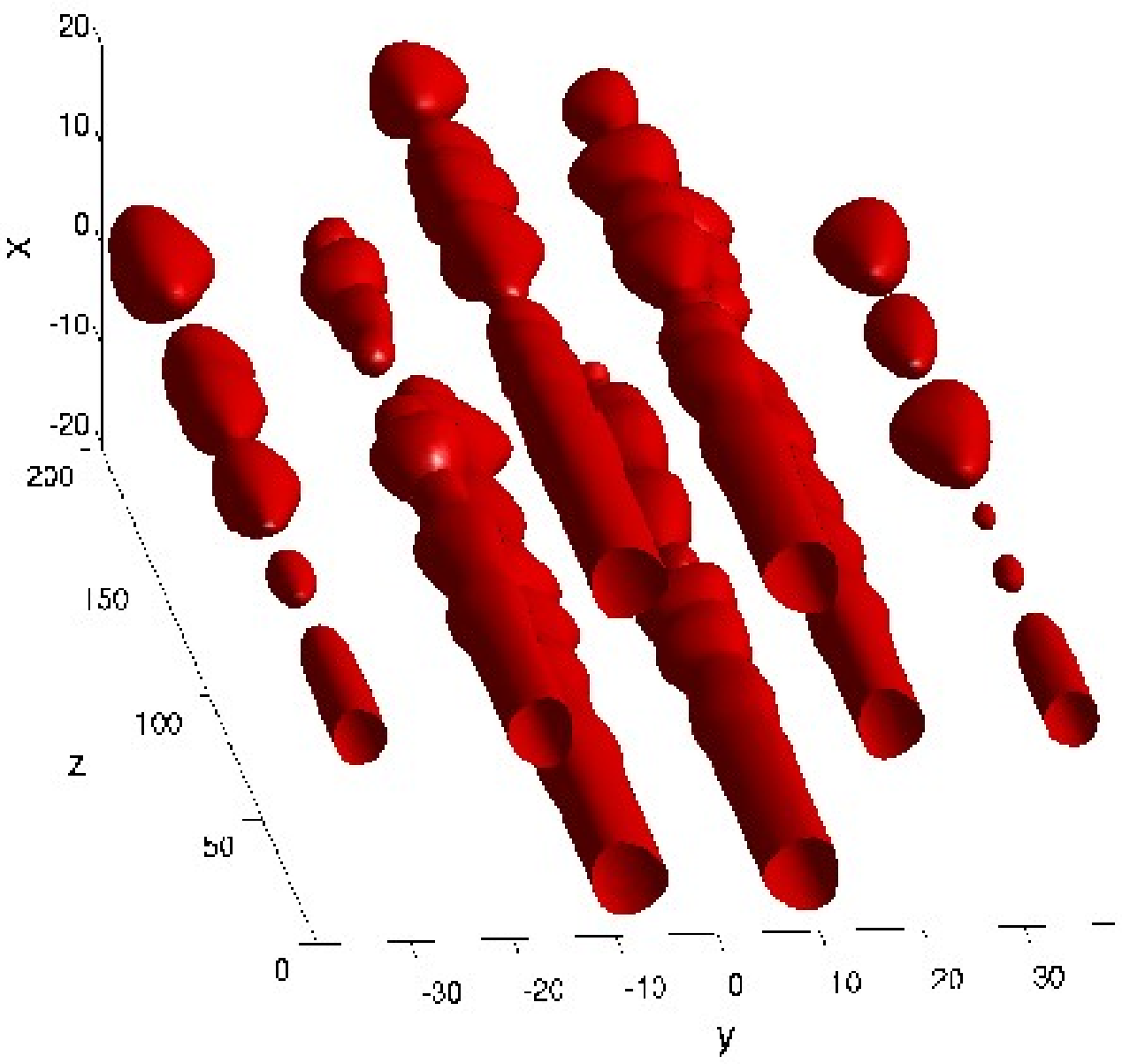}
\end{center}
\caption{The same figure as Fig.\ \ref{dyn_IPN}, but for the vortex necklace
with eight lobes depicted in the bottom panel of Fig.\ \ref{VTX}.
Depicted is the isosurface of height $0.1$.}
\label{dyn_vtx}
\end{figure}

First, we consider the out-of-phase hexapole. The existence and
the stability of this configuration has been described in
the preceding section. As the state has multiple pairs of real eigenvalues,
it is natural to expect that it should be prone to break up
under the instability's dynamical evolution.
A typical example of such a numerical experiment
is presented in Fig.\ \ref{dyn_OOP_nck}.

We found IP hexapole configurations as well, which, in
accordance with our considerations in Section II, turn
out to chiefly be stable within the first gap, although they
may possess
weak oscillatory instability inducing eigenvalue quartets.

This configuration also suffers a saddle-node bifurcation with
an OPN-type pair
emanating off of one of its lobes, when a
neighboring well becomes populated
out of phase near the first band.
The latter configuration is unstable always
possessing a real eigenvalue pair
in its linearization spectrum.
We note in passing that this is among
any of the six equivalent symmetric versions of
this configuration.

As for the dynamics of the instability, the solution
along the main lower branch is quite robust to strong perturbation.
Even though the solution suffers from an oscillatory instability,
a random perturbation with a maximum intensity almost $10^{-1}$ cannot
lead to a breakup of the configuration until propagation distances
of the order of $z=200$. On the other hand, the oscillatory dynamics
leading to the break up of the configuration of the bottom panel of
Fig.\ \ref{HEX} is shown in Fig.\ \ref{dyn_IP_nck}.


Finally, we investigate the complex-valued hexapole
configuration for which each lobe has the same modulus
and their phase increases counterclockwise in phase
increments of $\pi/6$, yielding
a vortex-necklace configuration.
This configuration turns out to be stable for the
most part within the first gap as well, with minor Hamiltonian 
Hopf-bifurcation
induced oscillatory instabilities.
We also found that this solution undergoes a saddle-node
bifurcation near the first band, in which it
collides with a waveform with two pairs of OPNs.
The stability
of the latter configuration in the presence
of these additional OPN dipoles is consistent with that
of their real counterparts from the previous sections,
each appearing to contribute one real pair, rendering the
relevant configuration quite unstable.

Similar to the case of in-phase hexapoles,
even though vortex necklaces may be unstable,
they are quite robust to perturbation, given the weak nature
of the relevant oscillatory instabilities. We therefore only
depict the dynamics of the
solutions
which have eight lobes as shown in the fourth and fifth
row panels of Fig.\ \ref{VTX}. The typical evolution of
this state is shown in Fig.\ \ref{dyn_vtx}, showcasing the oscillatory
breakup of this structure into one with a smaller number of lobes.

\section{Conclusions}

In this communication, we examined in detail theoretically and
numerically the existence, stability
and dynamics of multipole lattice solitons excited with a
saturable defocusing photorefractive nonlinearity in a
triangular geometry.
We have obtained a wide array of relevant
structures, including different types of dipoles and hexapoles,
as well as vortices. For the
dipole configurations we examined the
different possible phase configurations (in phase, and
out phase profiles), as well as cases where the excited
sites are separated by 0, 1, or 2 intermediate lattices sites.  For
hexapoles, we examined in phase and out of phase, and we also studied
the monotonic increasing phase of a discrete vortex necklace.

We have found good  agreement with the
general guidelines, explained in section II, stemming from the
theoretical analysis of the discrete model.
This intuition led to the illustration of a wide variety
of potentially stable solutions (although
they may incur oscillatory instabilities)
such as the in phase, nearest neighbor dipole,
the out of phase, next nearest neighbor dipole, and the in phase
opposite dipole. We have also identified
those solutions including e.g., the out of phase nearest neighbor,
in phase next-nearest neighbor, and out of phase opposite dipoles
which are typically unstable due to exponential instabilities
and real eigenvalues.
By the same considerations, the in-phase hexapole was proposed
and was indeed found to be typically stable, while the out-of-phase
one was predicted and observed to be quite unstable, due to multiple
real eigenvalue pairs. Finally, we have seen that the discrete vortex
structure is also potentially stable.

Furthermore, we have also identified an interesting set
of bifurcations that are associated with the parametric continuation
and termination of some of the above branches.
The dynamical instabilities encountered in the present work have been
monitored through direct integration of the relevant dynamical
equation. The result of evolution in every case involved oscillations
between the original configuration and one with fewer sites
which is more stable, such as a single site solitary wave, and
sometimes degeneration to such a configuration.
Solutions with smaller power
tend to decay into a single site solitary wave for certain solutions 
investigated, while those with larger power tend to oscillate.
This connection is beyond the scope of the present work, but is
currently being investigated further.

Since the framework of defocusing equations has been studied
far less extensively than their focusing counterparts, it would
be particularly interesting to extend the present considerations
to other structures. Perhaps the most interesting example would
be the study of multiple charge vortices in this context
which would be an interesting endeavor
both from a theoretical, as well as from an experimental point of
view.
Such studies are currently in progress and
will be reported in future publications.

\vspace{5mm}

{\it Acknowledgements}.
PGK acknowledges support from NSF-DMS-0505663,
NSF-DMS-0619492 and NSF-CAREER. W. Krolikowski is gratefully
acknowledged for numerous informative discussions on the theme
of this work.

\end{document}